\documentclass[article]{siamart220329}
\usepackage[caption=false]{subfig}
\usepackage{multirow}
\title{Comparison of lubrication theory and Stokes flow in corner geometries with flow separation}

\author{Sarah Dennis\thanks{Department of Mathematics, Brandeis University, Waltham MA (\email{sarahdennis@brandeis.edu})} \and Thomas G. Fai\thanks{Department of Mathematics and Volen Center for Complex Systems, Brandeis University, Waltham MA (\email{tfai@brandeis.edu})}}

\begin{document}
\maketitle
\begin{abstract}
The Reynolds equation from lubrication theory and the Stokes equations for zero Reynolds number flows are distinct models for an incompressible fluid with negligible inertia. Here we investigate the sensitivity of the Reynolds equation to large surface gradients, and explore flow recirculation in corner geometries in comparison to the Stokes equation.
We compare the solutions for the Reynolds and Stokes equations in the backward facing step (BFS), the regularized BFS, and the lid-driven triangular cavity. For the BFS variations listed above, we compute the error in terms of the average pressure drop through the channel and show how the error increases with increasing expansion ratio and with increasing magnitude of surface gradients. We further investigate the phenomenology of corner flow recirculation that arises in the Stokes solutions. In particular, we observe that occluding the corner separated region in the Stokes solution to the BFS does not disrupt the bulk flow characteristics.
\end{abstract}

\section{Introduction}
As two distinct fluid models in which inertia is considered negligible, it is interesting to contrast the behaviors of lubrication theory and Stokes flows. In particular, comparing these two models provides insights into the connections between different limiting regimes of the Navier-Stokes equations.

Previous research has demonstrated the sensitivity of lubrication theory to large gradients in the height of the fluid film \cite{brown_applicability_1995,biswas_backward-facing_2004,dobrica_about_2009,shyu_numerical_2018,julian_influence_2023}. When the lubrication assumptions of a thin film and a small scaled Reynolds number break down, the pressure drop modeled by the Reynolds equation is typically underestimated \cite{dobrica_reynolds_2005,dobrica_about_2009}. Moreover, lubrication theory is generally unable to capture flow separation and features including the so-called Moffatt eddies that appear in corner flows \cite{armaly_experimental_1983,biswas_backward-facing_2004,shyu_numerical_2018}. Here we compare the solutions of lubrication theory and Stokes flow for various geometries featuring large and/or discontinuous surface gradients, i.e. corner geometries, and further assess the extent to which lubrication theory is applicable in these examples.

The backward facing step (BFS) is a foundational example in lubrication theory and exhibits clear discrepancies between the Stokes and Reynolds solutions. In the Stokes and Navier-Stokes solution to the BFS, significant cross film pressure variation at the step discontinuity induces flow separation and flow recirculation in the concave corner \cite{matsui_separation_1975,uchida_numerical_1975,armaly_experimental_1983,julian_influence_2023}. The corner recirculation region approaches a constant size in the Stokes limit \cite{biswas_backward-facing_2004}. The Reynolds equation on the other hand does not capture this flow recirculation, and by construction has no cross-film pressure gradient. The step discontinuity in the BFS presents a maximal surface gradient such that we can examine the effect of varying the length scale ratio in isolation. Furthermore, the BFS example showcases how neglecting the cross-film pressure gradient in the lubrication approximation yields solutions without the patterns of flow separation and corner recirculation observed in the Stokes approximation. 

Corner recirculation in Stokes flow has been revealed with formal analysis of the Stokes equation $\nabla^4\psi=0$, where $\psi$ is the stream function. In terms of polar coordinates $(r,\theta)$, solutions of the form $\psi(r,\theta) = r^\lambda f_\lambda(\theta)$ to Stokes flow between rigid boundaries intersecting at an angle $\theta \in [-\alpha,\alpha]$ have complex exponent $\lambda$ whenever $2\alpha \le 146^\circ$ \cite{dean_steady_1949}. Such solutions reveal a sequence of recirculation zones, also known as Moffatt eddies, receding into the corner with rapidly decreasing size and intensity \cite{moffatt_viscous_1964}. The size of each recirculation zone is characterized by the points of flow separation: the half-saddle points of the stream function located along the rigid boundaries. Meanwhile, as a result of the thin film assumption and neglecting the cross-film pressure gradient, solutions in lubrication theory do not capture the patterns of flow separation and corner recirculation seen in Stokes flow. 

We consider solutions to the Reynolds equation of lubrication theory and Stokes equations for several variations of corner geometries. First, we examine the classical BFS; then two further variations on the BFS. The first variation focuses on occluding the corner flow recirculation region in the BFS, the second variation focuses on smoothing the slope of sudden expansion at the step. For the BFS variations, we characterize the error in average pressure drop between the Reynolds and Stokes equations, and determine that error increases with increasing expansion ratio and with increasing magnitude of surface variation. We also consider a lid-driven triangular cavity, which displays the classic sequence of corner recirculation zones, or Moffatt eddies. For the BFS variations and the lid-driven triangular cavity, we characterize the corner flow recirculation in the Stokes solutions and determine the points of flow separation. We observe that corner recirculation in the BFS can be minimized through occluding the corner separated region from the domain, and find that this does not otherwise disrupt the bulk flow.

Source code is available at \href{https://github.com/sarah-dennis/flow-separation}{https://github.com/sarah-dennis/flow-separation}.

\section{Lubrication Theory}
The Navier-Stokes equations,
\begin{align}\label{n-s_x}
 \frac{\partial p}{\partial x} &= \eta \Big(\frac{\partial^2 u}{\partial x^2} + \frac{\partial^2 u}{\partial y^2}\Big) - \rho\Big(  \frac{\partial u}{\partial t}+u\frac{\partial u}{\partial x} + v \frac{\partial u}{\partial y}\Big),\\
 \label{n-s_y} \frac{\partial p}{\partial y} &= \eta \Big(\frac{\partial^2 v}{\partial x^2} + \frac{\partial^2 v}{\partial y^2}\Big) - \rho\Big(  \frac{\partial v}{\partial t}+u\frac{\partial v}{\partial x} + v \frac{\partial v}{\partial y}\Big),
\end{align}
subject to incompressibility,
\begin{equation}\label{incompressibility}
 \frac{\partial u}{\partial x} + \frac{\partial v}{\partial y} = 0,\end{equation}
serve as the governing equations for a two-dimensional incompressible fluid \cite{leal_advanced_2007}. The pressure is $p$, the velocity is $(u,v)$, and the kinematic viscosity is $\nu=\eta/\rho$, where $\eta$ is the bulk viscosity and $\rho$ is the constant density.

We consider the two-dimensional fluid domain $[x_0, x_L] \times [0,h(x)]$ with the no-slip boundary condition at the fluid-surface interfaces $y=0$ and $y=h(x)$. We assume, without loss of generality, that the relative velocity $\mathcal{U}$ is imposed at the lower surface $y=0$ and the upper surface $y=h(x)$ is stationary,
\begin{align}\label{bc_u}
&& u (x,0) = \mathcal{U} && u(x,h(x)) = 0, \\ \label{bc_v}
&& v (x,0) = 0&& v(x,h(x)) = 0 .
\end{align}
We prescribe the constant flux $\mathcal{Q}$ such that the velocity satisfies, 
\begin{equation} \label{flux} 
\mathcal{Q} = \int_0^{h(x)} u(x,y) \,dy.
\end{equation}
We also prescribe a fixed outlet pressure $p(x_L,0) = 0$. Because the flux and outlet pressure are prescribed, the average pressure drop, 
\begin{equation}\label{dp}
    \Delta p = \frac{1}{h(x_0)}\int_0^{h(x_0)} p(x_0,y)\,dy - \frac{1}{h(x_L)}\int_0^{h(x_L)} p(x_L,y)\,dy,
\end{equation}
is determined by the solution. 

 We introduce the characteristic length scales $L_x = x_f - x_0$ and $L_y = \max h(x) > 0$, and the length scale ratio $\varepsilon = L_y/L_x$. For a prescribed constant flux $\mathcal{Q}\ne 0$, define the characteristic velocities $U_* = \mathcal{Q}/L_y$ and $V_* = \mathcal{Q}/L_x$. Or, in the case of $\mathcal{Q}= 0$ and $\mathcal{U}\ne 0$, let $U_*=\mathcal{U}$ and $V_*= U_*L_y/L_x$. The Reynolds number is given by $\text{Re}=\rho U_*{\color{black}L_x}/\eta$, the characteristic time is $T_* = L_x/U_*$, and the characteristic pressure is $P_*=\eta U_*L_x/L_y^2$. Upon nondimensionalization and applying the lubrication assumptions $\varepsilon^2 \ll 1$ and $\varepsilon^2 \text{Re} \ll 1$, the incompressible Navier-Stokes equations \cref{n-s_x,n-s_y} are reduced to the governing equations of lubrication theory:
\begin{align}\label{gap_x} 
 \frac{\partial p}{\partial x}&=\eta \frac{\partial^2 u}{\partial y^2}, \\ \label{gap_y}
 \frac{\partial p}{\partial y}&=0,
\end{align}
subject to incompressibility \cref{incompressibility}. 

The velocity $u(x,y)$ is determined through integration of \cref{gap_x} and applying the boundary conditions \cref{bc_u},
\begin{equation}
 \label{reyn_u} u(x,y)=\frac{1}{2\eta}\frac{d p}{d x} \Big(y^2-h(x)y\Big) +\frac{\mathcal{U}}{h(x)} \Big(h(x)-y\Big).\end{equation} The velocity $v(x,y)$ is then determined from incompressibility \cref{incompressibility} and applying the boundary conditions \cref{bc_v},
\begin{equation}\label{reyn_v}
v(x,y) =\frac{-1}{6\eta}\frac{d^2 p}{d x^2}y^3 + \frac{1}{2}\Bigg(\frac{1}{2\eta}\bigg(\frac{d^2 p}{d x^2}h(x) + \frac{d p}{d x}\frac{dh}{dx}\bigg) -\frac{\mathcal{U}}{[h(x)]^2}\frac{dh}{dx}\Bigg)y^2.\end{equation} 
The condition of constant flux,
\begin{equation}\label{reyn_flux}\mathcal{Q} = \int_{0}^{h(x)} u(x,y) \, dy =\frac{-1}{12\eta}\Bigg(\Big[h(x)\Big]^3\frac{dp}{dx} -6\eta\mathcal{U}h(x)\Bigg),\end{equation}
and the boundary condition $v(x,h)=0$, are satisfied exactly when $p(x)$ satisfies the Reynolds equation,
\begin{equation}\label{reynolds}
 \frac{d}{dx}\Bigg[\Big[h(x)\Big]^3\frac{dp}{dx}\Bigg] = 6\eta\,\mathcal{U}\frac{dh}{dx}.
\end{equation}

To solve the Reynolds equation given a prescribed flux, we use a mixed boundary condition for pressure,
\begin{align}\label{reyn_bc_p}&& \frac{dp}{dx}\bigg|_{x_0} = \frac{-12\eta \mathcal{Q}}{\big[h(x_0)\big]^3} +\frac{6\eta\mathcal{U}}{\big[h(x_0)\big]^2}, &&p (x_L) =0. &&\end{align}
That is, the flux $\mathcal{Q}$ and boundary velocity $\mathcal{U}$ determine the Reynolds pressure gradient according to \cref{reyn_flux}. Without loss of generality, the pressure gradient is prescribed at the inlet and the fixed pressure is taken at the outlet.

In our previous work \cite{dennis_fast_2026}, we presented a linear time solver for the Reynolds equation, which gives the exact solution for piecewise linear heights as considered here. A brief outline of the method is provided in the appendix. 

\section{The Stokes equation}
The Stokes equation is obtained from the incompressible Navier-Stokes equations in the limit of zero Reynolds number \cite{leal_advanced_2007}. 
With the standard dimensionless variables, $X=x/L_x$, $Y=y/L_x$, $U=u/U_*$, $V=v/U_*$,  $T = t/T_*$ and $P=p/P_*$, the Navier-Stokes equations \cref{n-s_x,n-s_y} are expressed in dimensionless terms as, 
\begin{align}\label{n-s_x_dimless_Stokes}
 \frac{\partial P}{\partial X} &= \Big(\frac{\partial^2 U}{\partial X^2} + \frac{\partial^2 U}{\partial Y^2}\Big) - \text{Re}\Big( \frac{\partial U}{\partial T}+U\frac{\partial U}{\partial X} + V \frac{\partial U}{\partial Y}\Big),\\
 \label{n-s_y_dimless_Stokes} \frac{\partial P}{\partial Y} &=  \Big(\frac{\partial^2 V}{\partial X^2} + \frac{\partial^2 V}{\partial Y^2}\Big) - \text{Re}\Big( \frac{\partial V}{\partial T}+U\frac{\partial V}{\partial X} + V \frac{\partial V}{\partial Y}\Big).
\end{align}
Compared with lubrication theory, this nondimensionalization is performed using a single length scale $L_x$ and velocity scale $U_*$. The characteristic pressure is now $P_* =\eta U_*/L_x$. Consequently, \cref{n-s_x_dimless_Stokes,n-s_y_dimless_Stokes} are not sensitive to the length scale ratio $\varepsilon$ that appears in lubrication theory.
 
Through introduction of the stream function $\psi(x,y)$ satisfying,
\begin{align} \label{stream}
 \hfill&& u = \frac{\partial\psi}{\partial y} \hspace{3em} v = -\frac{\partial \psi}{\partial x}.&&\hfill
\end{align}
and assuming a steady state flow, the dimensionless Navier-Stokes equations \cref{n-s_x_dimless_Stokes,n-s_y_dimless_Stokes} may be expressed in
stream-velocity form,
\begin{equation}\label{n-s_psivel}
 \nabla^4\Psi = \text{Re}\big(V\nabla^2 U - U\nabla^2 V).
\end{equation}
where $\Psi = \psi/(U_*L_x)$ and the Reynolds number $\text{Re} = \rho U_* L_x/\eta$ controls the inertial terms. When $\text{Re} = 0$, the stream-velocity formulation \cref{n-s_psivel} reduces to the biharmonic Stokes equation $\nabla^4 \Psi=0$, equivalently $\nabla^4 \psi = 0$ in dimensional variables.

To solve the Stokes equation $\nabla^4 \psi = 0$, we require additional boundary conditions for the velocity at the inlet and outlet, and corresponding boundary conditions for the stream function. At the inlet and outlet, we assume a fully developed laminar flow. For velocity, the inlet and outlet boundary conditions are,
\begin{align}\label{n-s_bc_vel}
 u(x_0,y) &= u_\text{Reyn}(x_0,y) &&\frac{\partial u}{\partial x}\Big|_{x_L,y} = 0,\\
 v(x_0,y) &= 0 && v(x_L,y) = 0,
\end{align}
where $u_\text{Reyn}(x,y)$ is the velocity \cref{reyn_u} of lubrication theory.
At the surfaces $y=0$ and $y=h(x)$, we use the no-slip boundary conditions \cref{bc_u,bc_v}. Then for the stream function,
\begin{align}\label{n-s_bc_stream}
 \psi(x_0,y) &= \int_{0}^y u_\text{Reyn}(x_0,\hat{y})d\hat{y} &&\frac{\partial \psi}{\partial x}\Big|_{x_L,y} = 0,\\
 \psi(x, 0) &= 0 & &\psi(x,h(x)) = \mathcal{Q},
\end{align}
where, 
\begin{align} \label{n-s_bc_stream_x0}
\int_{0}^y u_\text{Reyn}(x,\hat{y})d\hat{y} 
&= \frac{\mathcal{Q}y^2}{\big[h(x)\big]^3}\bigg(3h(x)-2y\bigg)+\frac{\mathcal{U}y}{\big[h(x)\big]^2}\Big(h(x)-y\Big)^2.
\end{align}
Finally, the outlet pressure is fixed at the lower surface, $p(x_L,0)=0$.
To be consistent with the assumption of a fully developed flow at the inlet and outlet, and to prevent blow up in numerical solutions to the Stokes equation with these boundary conditions, surface variation should be located well within the interior of the domain.

\subsection{Finite difference iterative solution}
We consider a second-order accurate iterative finite difference scheme for the Stokes equation using a nine point stencil. This method is presented in \cite{gupta_new_2005,biswas_hoc_2017} for the full biharmonic Navier-Stokes equation \cref{n-s_psivel}; here we restrict to $\text{Re}=0$ for Stokes flow and use dimensional variables.  

Define the uniform discretisation of the domain $[x_0, x_L] \times [0,h(x)]$, 
\begin{align}
 \{x_i\}_{i=0}^N && x_i = x_0 + i \Delta x && N = \mathcal{N}(x_L - x_0),\\
 \{y_j\}_{i=0}^M && y_j = j \Delta y && M = \mathcal{N}\max h(x),
\end{align}
where $\mathcal{N}$ is the number of grid points per unit length, $\mathcal{N} = 1/\Delta x = 1/\Delta y$. The geometry is chosen so as to ensure an integer number of grid points in either direction, and such that any vertical boundary components lie on the grid. We also assume that the boundary $h(x)$ is not multi-valued, except possibly at discontinuities $x_i$ where $h_{i^+}$ and $h_{i^-}$ are connected by a vertical line segment.

The Stokes equation $\nabla^4 \psi=0$ is discretized as,
\begin{multline}\label{stokes_disc}
 28\psi_{i,j} - 8 \Big(\psi_{i-1,j}+\psi_{i+1,j}+\psi_{i,j-1}+\psi_{i,j+1}\Big) \\ + \Big(\psi_{i-1,j-1}+\psi_{i-1,j+1}+\psi_{i+1,j-1}+\psi_{i+1,j+1}\Big)
 \\ = 
 3\Delta x\big(u_{i,j-1}-u_{i,j+1}+v_{i+1,j}-v_{i-1,j}\big),
\end{multline}
and the stream function \cref{stream} is discretized as,
\begin{align}\label{stream-vel_u_disc}
 u_{i,j} &= \frac{-3}{4\Delta x}\big(\psi_{i,j-1}-\psi_{i,j+1}\big) - \frac{1}{4}\big(u_{i,j-1}+u_{i,j+1}\big), 
 \\ \label{stream-vel_v_disc}
 v_{i,j} &= \frac{3}{4\Delta x}\big(\psi_{i-1,j}-\psi_{i+1,j}\big)-\frac{1}{4}\big(v_{i-1,j}+v_{i+1,j}\big).
\end{align}
The equations \cref{stokes_disc,stream-vel_u_disc,stream-vel_v_disc} are evaluated iteratively. The equation \cref{stokes_disc} is solved for $\psi$, then \cref{stream-vel_u_disc,stream-vel_v_disc} are used to update $u$ and $v$. The initial values for $\psi$, $u$ and $v$ satisfy the boundary conditions and are otherwise zero. In solving \cref{stokes_disc} for $\psi$, an LU factorization of the coefficient matrix corresponding to the left hand side of \cref{stokes_disc} may be precomputed. We implement this method of solution in Python, and the sparse LU factorization is computed and evaluated using scipy.sparse.linalg.splu. The iterative process is terminated when the maximum absolute error in $\psi$ between subsequent iterations falls below $10^{-8}$; this cut-off for convergence is consistent with the literature \cite{gupta_new_2005,biswas_hoc_2017}. 

Once the velocity and stream solutions to the Stokes equation have converged through the iterative method described above, the pressure partial derivatives are determined using a second-order accurate finite difference discretization of the full Navier-Stokes equations \cref{n-s_x,n-s_y}. For \cref{n-s_x},
\begin{multline}\label{n-s-x_disc}
 \frac{\partial p}{\partial x}\Big|_{i,j} = \frac{\eta}{\Delta x^2} \Big(u_{i-1,j} + u_{i+1,j} - 4u_{i,j} + u_{i,j-1} + u_{i,j+1}\Big) \\ -\frac{\rho}{2\Delta x}\Big(u_{i,j}(u_{i+1,j}-u_{i-1,j}) + v_{i,j}(u_{i,j+1}-u_{i,j-1})\Big),
 \end{multline} and \cref{n-s_y} is similar. 
The pressure is then determined by numerical integration using the boundary condition $p(x_L,0) = 0$. We expect path independence for a velocity-stream solution that has sufficiently converged. For convenience, the contour is taken first from the outlet to the inlet along the lower surface, \begin{equation}p(x_i,0)=p(x_{i+1},0) - \Delta x \frac{\partial p}{\partial x}\Big|_{i,0}, \end{equation}
and then from the lower surface to the upper surface at each $x_i$,
\begin{equation}p(x_i,y_j)=p(x_{i},y_{j-1}) + \Delta y \frac{\partial p}{\partial y}\Big|_{i,j}.\end{equation}
This approach assumes the boundary $h(x)$ is not multi-valued, except possibly at jump discontinuities $x_i$ where $h_{i^+}$ and $h_{i^-}$ are connected by a vertical line segment. For example, an S-shaped lower boundary would require a more complicated path with additional components. Path independence is expected for a velocity-stream solution that has sufficiently converged. 

\subsection{Piecewise linear boundary approximation}
Here we consider the treatment of boundary conditions for solving \cref{stokes_disc,stream-vel_u_disc,stream-vel_v_disc} on the grid $\{(x_i,y_j)\}_{i=0,j=0}^{N,M}$ in cases where the boundary, particularly $h(x)$, is non-rectilinear. For non-rectilinear domains, the nine or five point stencil at an interior grid point $(x_i,y_j)$ may include a grid point $(x_s,y_t)$ which is exterior to the domain; in this case it is necessary to approximate the stream and velocity boundary conditions in relation to $(x_i,y_j)$. The approximation is dependent on the geometry at $(x_i,y_j)$, hence the same procedure applies for all of $u$, $v$, and $\psi$. This approximation assumes the boundary is linear between $(x_i,y_j)$ and $(x_s,y_t)$. 

Suppose the stencil centered at an interior grid point $(x_i,y_j)$ includes an exterior grid point $(x_s,y_t)$. We first determine the non-grid-aligned point $(x_p,y_q)$ where the boundary crosses the stencil axis from $(x_i,y_j)$ to $(x_s,y_t)$. Since the lower surface is constant, if $(x_i,y_j)$ is interior, then the neighbor $(x_i,y_{j-1})$ is also interior or on the boundary. For the remaining seven cases, the non-grid-aligned point is given by,
\begin{equation}(x_p,y_q) =\begin{cases}\big(x_i,h_i\big) & (s,t) = (i,j+1)\\
\Big(x_i + \cfrac{y_j-h_{i^\mp}}{G_{s,i}},y_j\Big) & (s,t) = (i\pm1,j)\\
\Big(x_i + \cfrac{y_j-h_{i^\mp}}{G_{s,i}-1}, y_j+ x_p - x_i\Big) & (s,t) = (i\pm1,j\pm1)\\
\Big(x_i + \cfrac{y_j-h_{i^\mp}}{G_{s,i}+1}, y_j - x_p + x_i\Big) & (s,t) = (i\pm1,j\mp1)
\end{cases},\label{n-s_bdry_loc}
\end{equation}
where $G_{s,i}$ approximates the slope of the boundary between $x_i$ and $x_s$,
\begin{equation}
 G_{s,i}=\begin{cases}\cfrac{h_{s^+}-h_{i^-}}{x_s-x_i} & s = i-1\\
 \cfrac{h_{s^-}-h_{i^+}}{x_s-x_i} & s = i+1
 \end{cases},
\end{equation}
and $h_{i^\pm} = \lim_{x\to x_i^\pm}h(x)$ are the boundary values. Note that the cases $G_{s,i}=\{0,1,-1\}$ which appear as potential singularities in  \cref{n-s_bdry_loc} are not possible. For example, consider the case $(s,t)=(i+1,j)$, if $G_{s,i} = 0$, then $(x_s,y_t)$ must be a boundary point and not exterior. Similarly, consider the case $(s,t)=(i+1,j+1)$, if $G_{s,i} = 1$, then $(x_s,y_t)$ must be a boundary point and not exterior. Hence \cref{n-s_bdry_loc} is well defined.

Given the off-grid location $(x_p,y_q)$ where the boundary $h$ crosses the stencil axis between $(x_i,y_j)$ and $(x_s,y_t)$, we determine ghost values for $u_{s,t}$, $v_{s,t}$ and $\psi_{s,t}$ such that linear interpolation along the stencil axis leads to $u$, $v$, and $\psi$ satisfying their respective boundary values at $(x_p,y_q)$. That is, let $\phi$ denote any of $u$, $v$ or $\psi$, the ghost value $\phi_{s,t}$ is determined by linear interpolation,
\begin{equation} \label{n-s_bdry_interp}\phi_{s,t} =\phi_{p,q} +(\phi_{p,q}-\phi_{\overline{s,t}} ) \dfrac{||(x_s,y_t)-(x_p,y_q)||_2}{||(x_{\overline{s}},y_{\overline{t}})-(x_p,y_q)||_2},\end{equation}
where $\phi_{p,q}$ is the boundary value, and $(x_{\overline{s}},y_{\overline{t}})$ is the `opposite' grid point to $(x_s,y_t)$ in the stencil at $(x_i,y_j)$ such that the three points are collinear. The interpolation \cref{n-s_bdry_interp} requires a known value for $\phi_{\overline{s,t}}$. If $(x_i,y_j)$ is not near a change in the surface gradient and $(x_{s},y_{t})$ is exterior, then $(x_{\overline{s}},y_{\overline{t}})$ must be interior and so the value $\phi_{\overline{s,t}}$ is known from a previous iteration. Otherwise, if the two opposite points $(x_{s},y_{t})$ and $(x_{\overline{s}},y_{\overline{t}})$ in the stencil are both exterior, then the slope of the boundary must have changed sign at $x_i$. In particular, $(x_i,y_j)$ must be in an acute corner near the boundary, and few surrounding points in the stencil will be interior, so assume $\phi_{i,j} =\phi_{p,q}$. This approach smooths the inner corner with the boundary value at a constant number of grid points as the grid resolution $\mathcal{N}$ increases. 

\section{Comparison of the Reynolds and Stokes solutions}
In the following sections we discuss the differences in the solutions to the Reynolds and Stokes equations for several variations on the classical backward facing step (BFS), and a lid-driven triangular cavity. 
The flow patterns predicted by the lubrication approximation and the Stokes approximation often vary significantly. Recall, the Reynolds equation relies on the lubrication assumptions $\varepsilon^2 \ll 1$ and $\varepsilon^2 \text{Re}\ll 1$, while the Stokes equation relies only on $\text{Re}=0$ with no assumptions on the length scale ratio. Thus, discrepancies between the Reynolds and Stokes solutions are intrinsically related to the geometry. One should only expect exact agreement between the two models when the height is constant. 

To consistently compare solutions for a particular geometry, we prescribe the same flux $\mathcal{Q}$, boundary velocity $\mathcal{U}$ and ambient pressure $p(x_L,y)=0$. The average pressure drop $\Delta p$ between the solutions are then presumed to vary. 
 
\subsection{The backward facing step}
\begin{figure}[h] 
 \centering 
 \includegraphics[width=.75\textwidth]{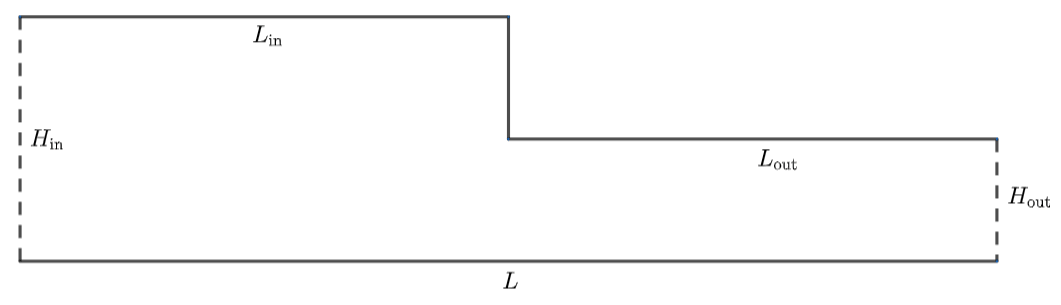}
 \caption{Schematic of the backward facing step with expansion ratio $\mathcal{H}=H_{\text{in}}/H_{\text{out}}$.} \label{schematic_bfs}
\end{figure}

The backward facing step (BFS), shown in \cref{schematic_bfs}, is characterized by the expansion ratio $\mathcal{H}=H_{\text{in}}/H_{\text{out}}$ to quantify the magnitude of the step. The height of the fluid film is, \begin{equation}h(x) = \begin{cases}H_{\text{in}} & \hspace{2em} 0\le x \le L_{\text{in}}\\ H_{\text{out}} &\hspace{2em} L_{\text{in}}\le x \le L \end{cases},\end{equation}
where $L=L_{\text{in}}+L_{\text{out}}$ is the domain length. We compare the Reynolds and Stokes solutions to the BFS for varying expansion ratios $1<\mathcal{H}\ll L$, and for fixed $H_{\text{out}}=1$,  $L_{\text{in}}=L_\text{out} = 8$. The boundary conditions $\mathcal{U}=0$, $\mathcal{Q}=1$, and $p(L,0)=0$ are kept constant.

\begin{figure}[h]
 \centering
 \subfloat[Stokes $\mathcal{H}=1.25$]{\label{stokes_bfs_H1p25_p} \includegraphics[width=.45\textwidth]{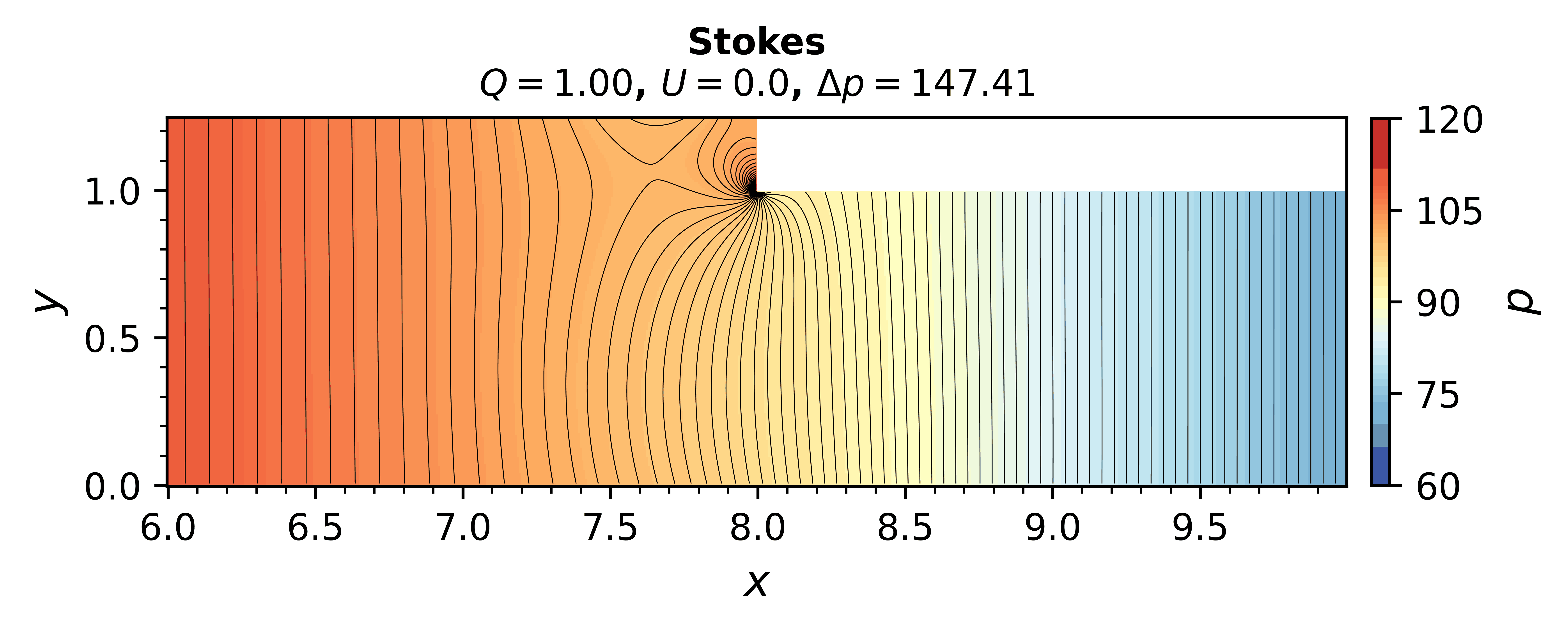}}
 \subfloat[Reynolds $\mathcal{H}=1.25$]{\label{reyn_bfs_H1p25_p}
\includegraphics[width=.45\textwidth]{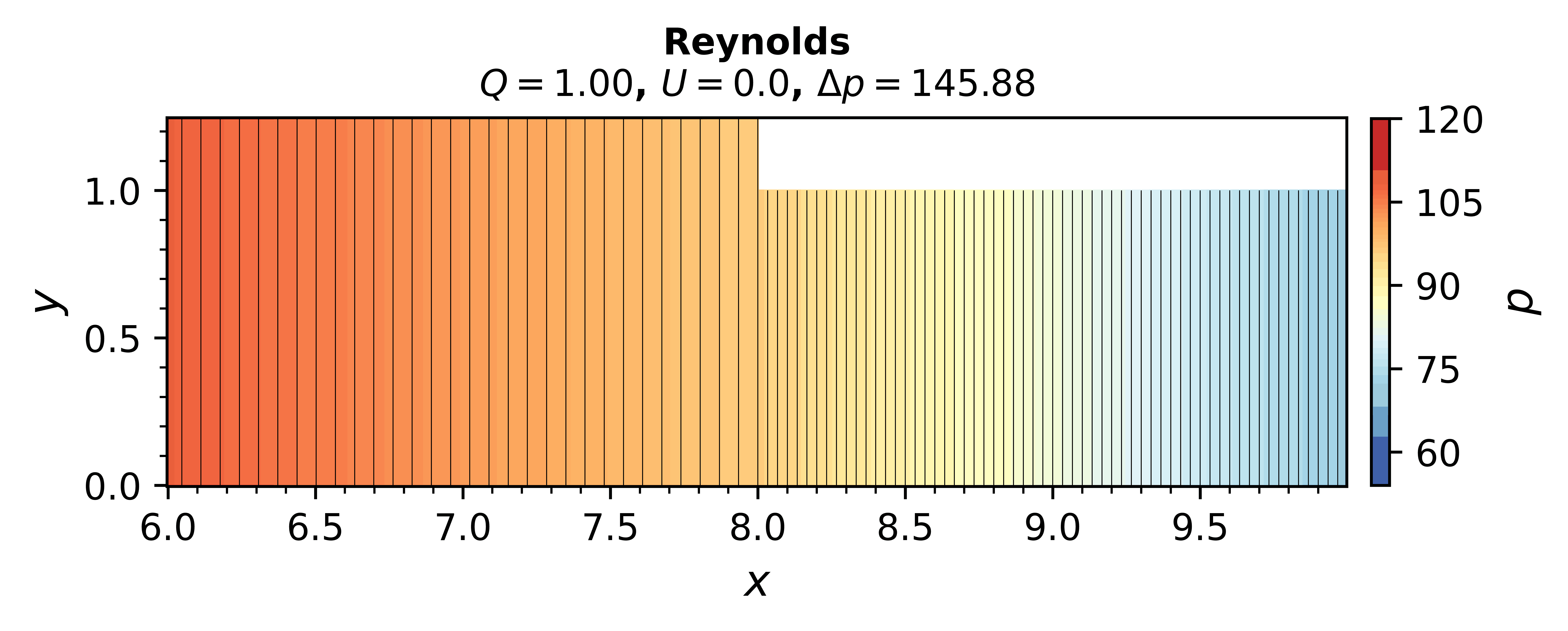}}
 
 \subfloat[Stokes $\mathcal{H}=2.75$]{\label{stokes_bfs_H2p75_p}\includegraphics[width=.45\textwidth]{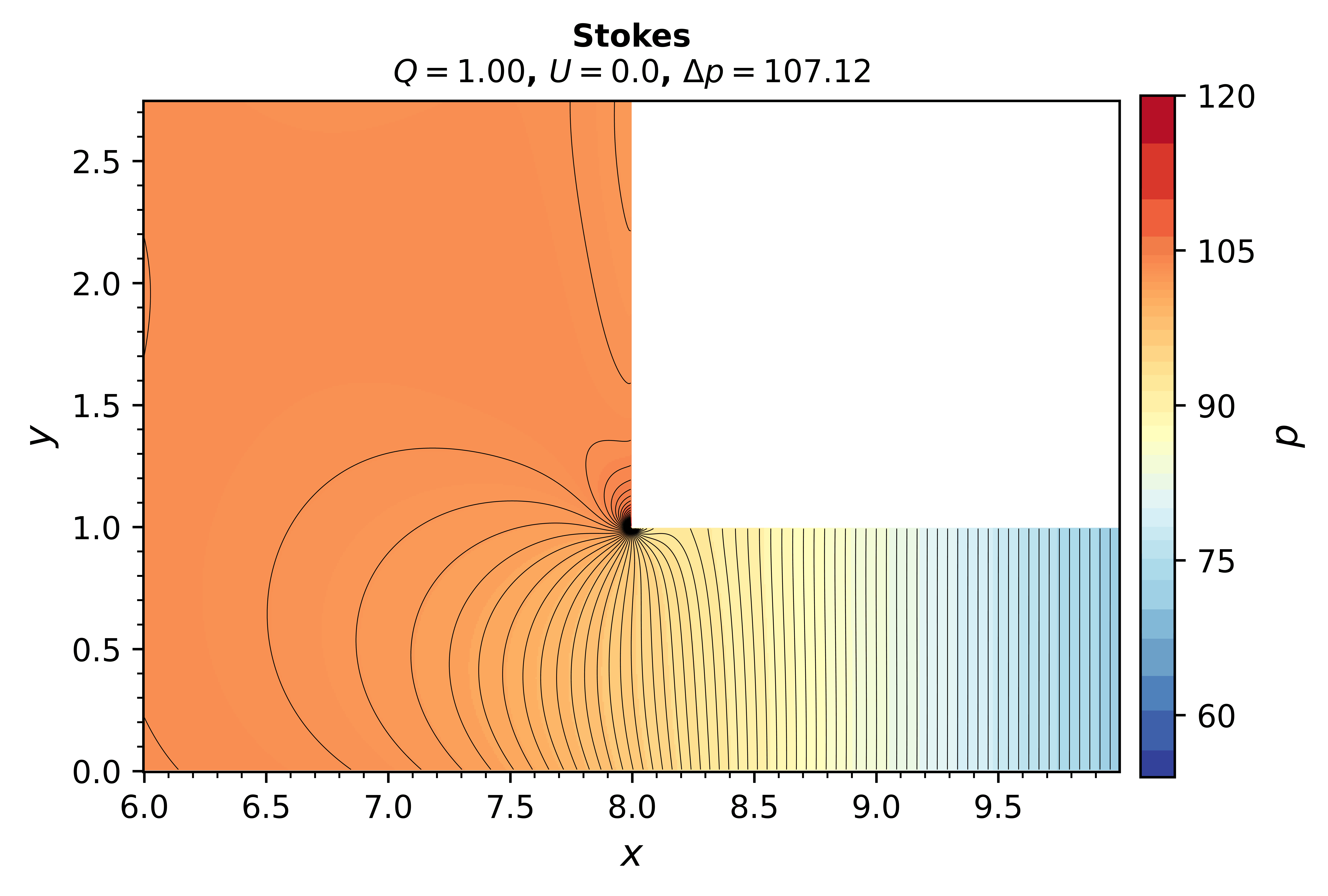}}
 \subfloat[Reynolds $\mathcal{H}=2.75$]{\label{reyn_bfs_H2p75_p}\includegraphics[width=.45\textwidth]{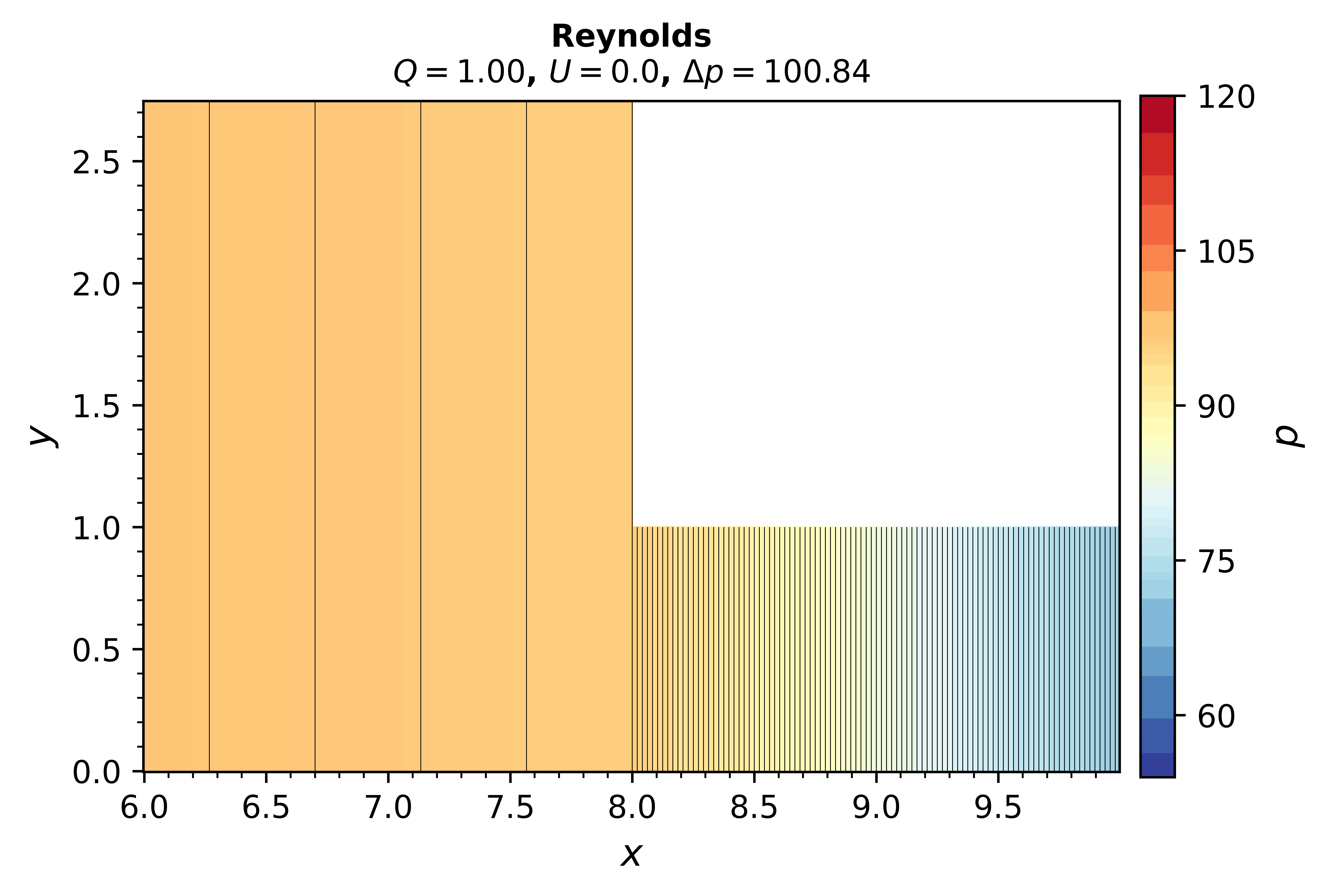}}
 \caption{Pressure contours for the BFS. Curved contours in the Stokes solutions indicate significant pressure gradients at the step. For a particular $\mathcal{H}$, $\Delta p$ is smaller for the Reynolds solution versus the Stokes solution. }\label{bfs_pres}
\end{figure}

\begin{figure}[h]
 \centering 
 \subfloat[$\mathcal{H}=1.25$]{\includegraphics[width=0.3\textwidth]{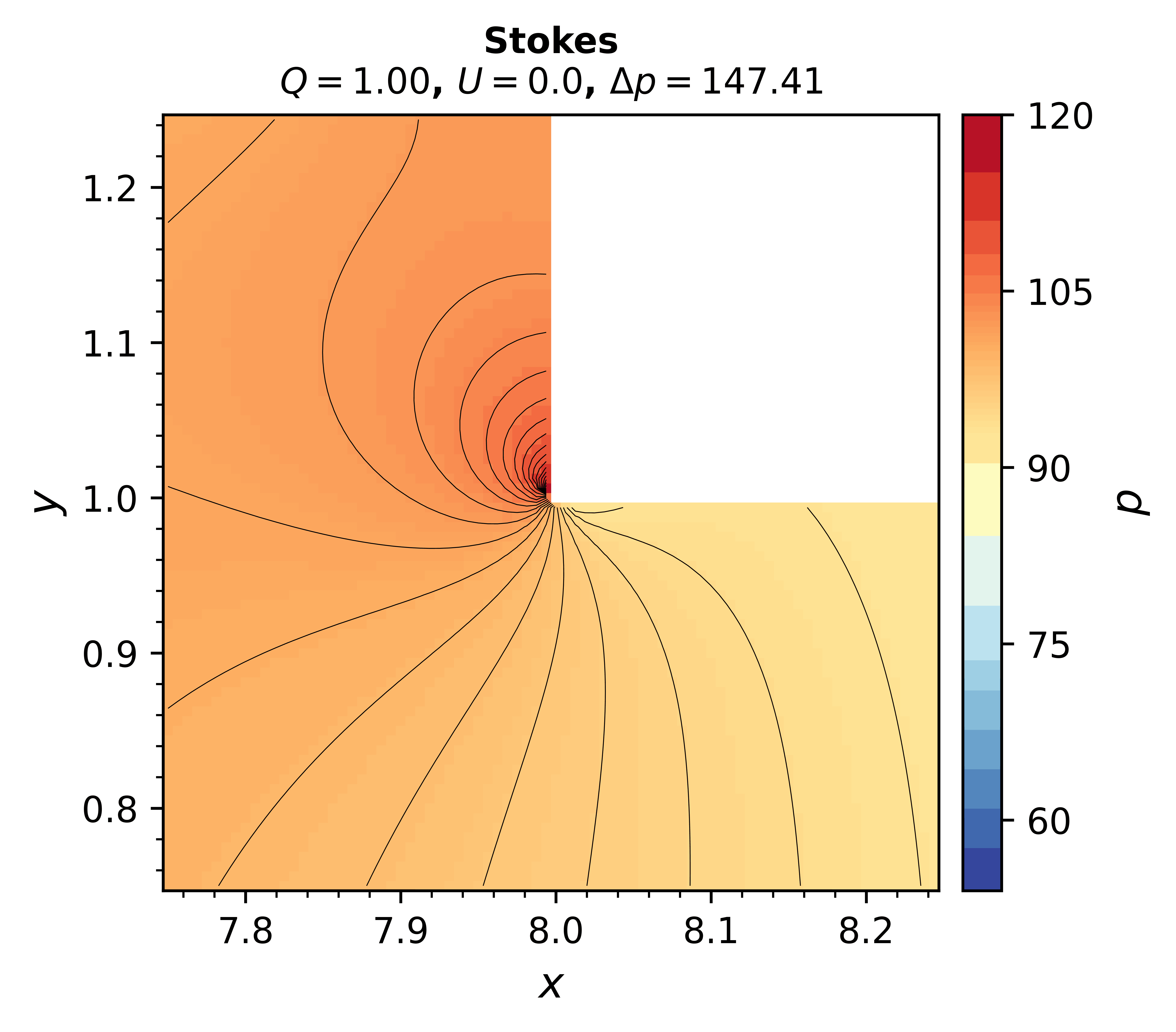}}
 \subfloat[$\mathcal{H}=2$]{\includegraphics[width=0.3\textwidth]{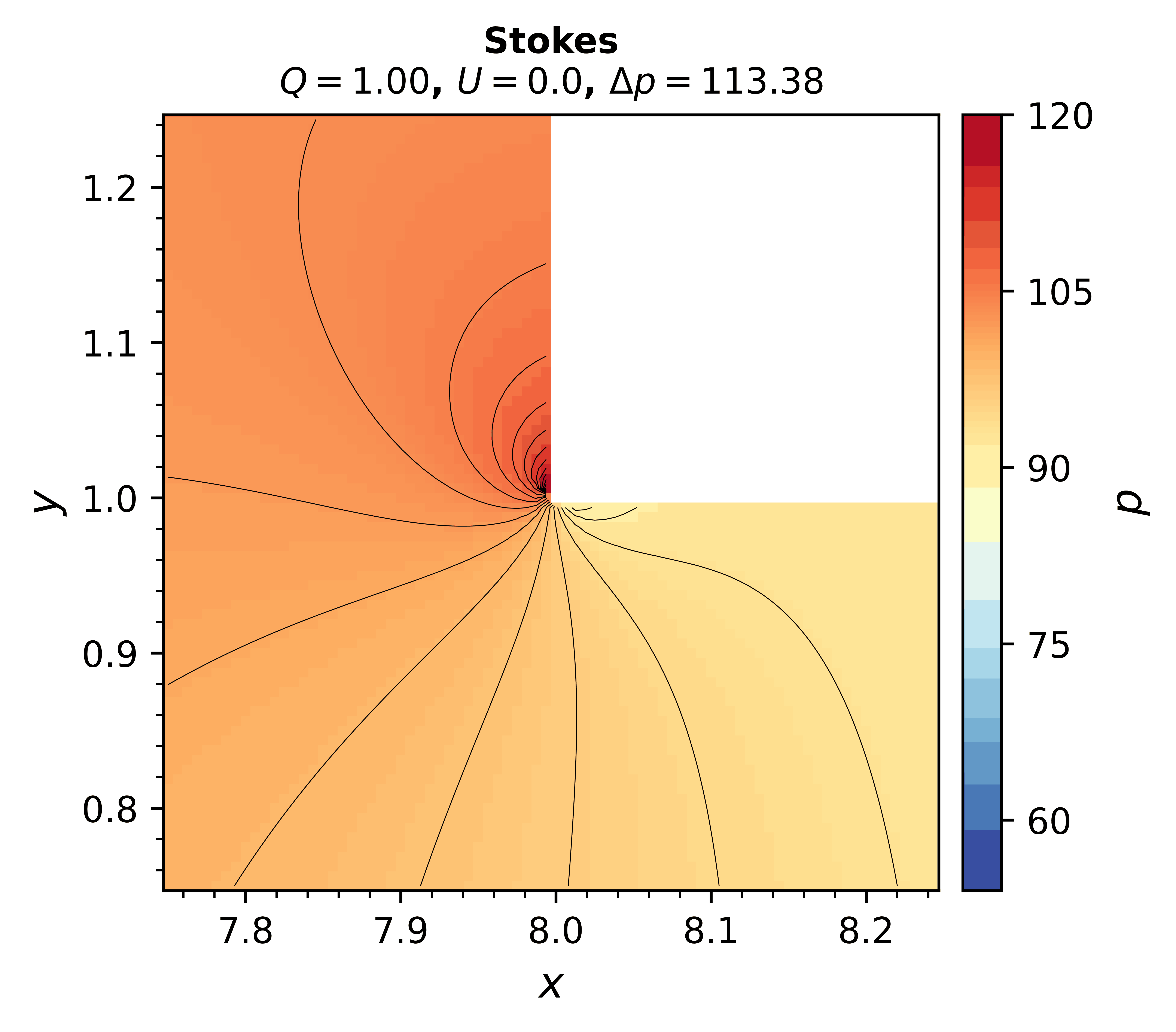}}
 \subfloat[$\mathcal{H}=2.75$]{\includegraphics[width=0.3\textwidth]{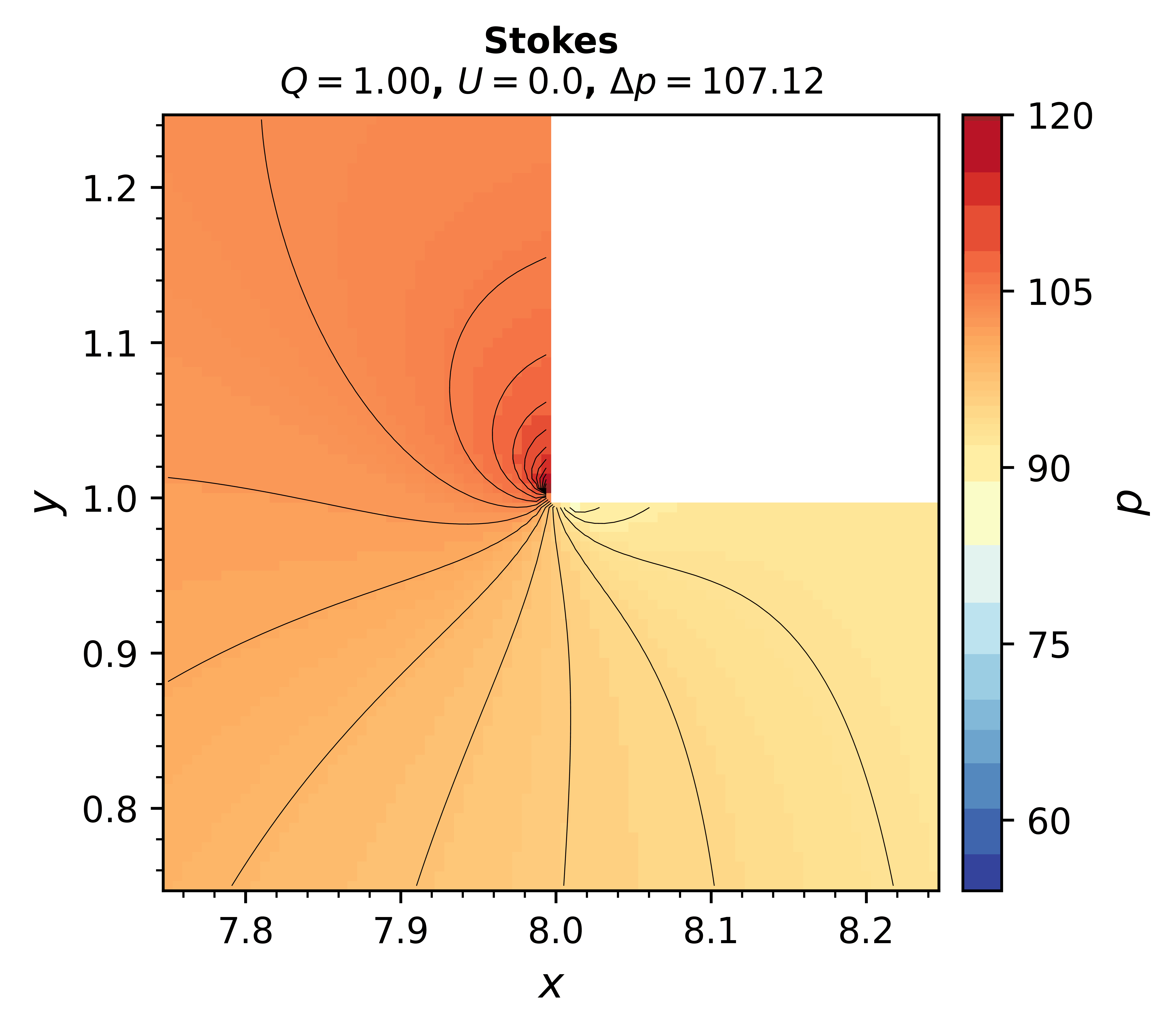}}
 \caption{Pressure contours for the BFS surrounding the step tip, Stokes solutions. The maximum pressure in the domain occurs at the step tip.}\label{bfs_pres_zoom}
\end{figure}

\begin{figure}[h]
 \centering 
 \subfloat[Pressure error]{\includegraphics[width=0.45\textwidth]{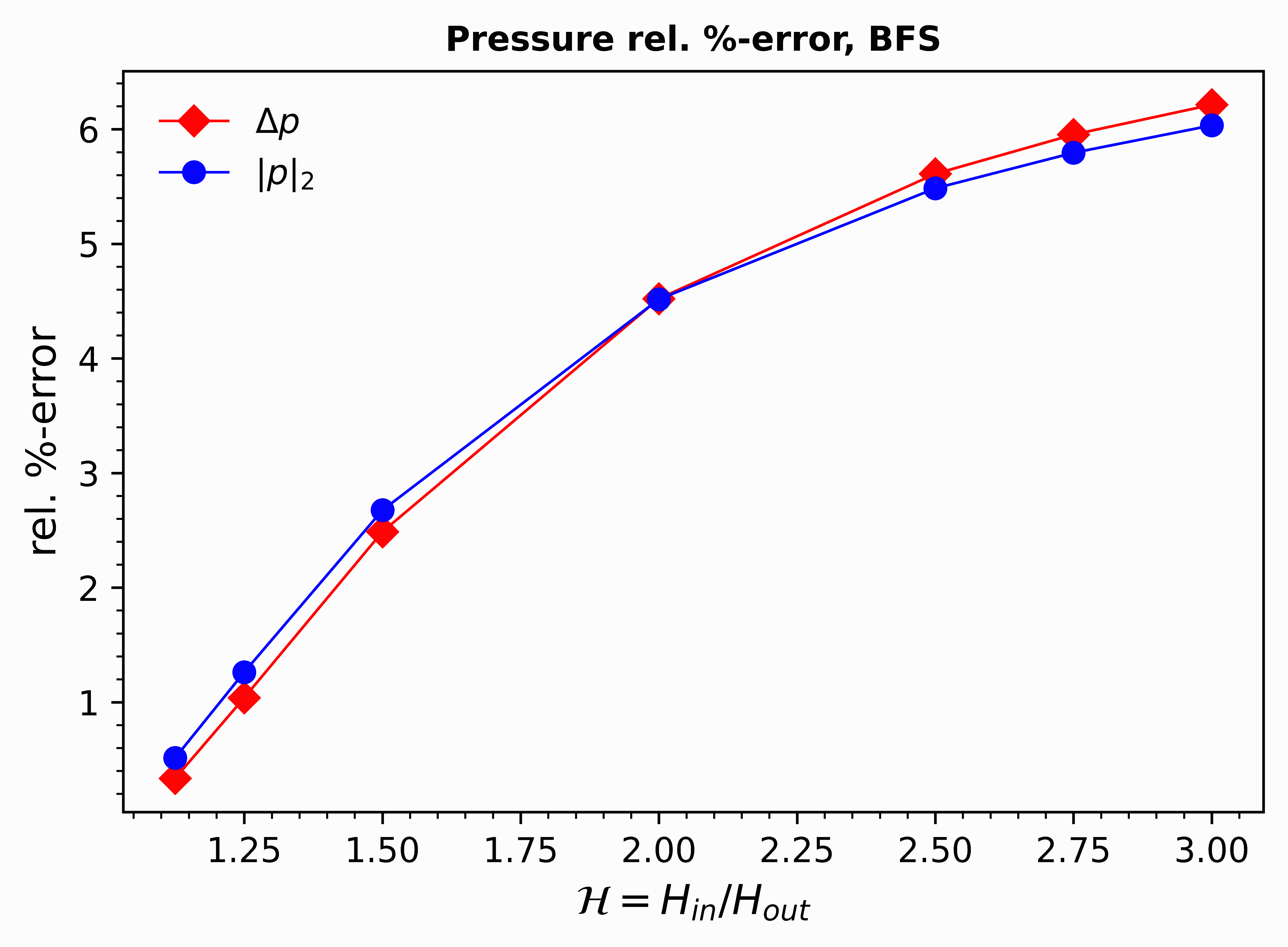}\label{bfs_p_error}}
 \subfloat[Velocity error]{ \includegraphics[width=0.46\linewidth]{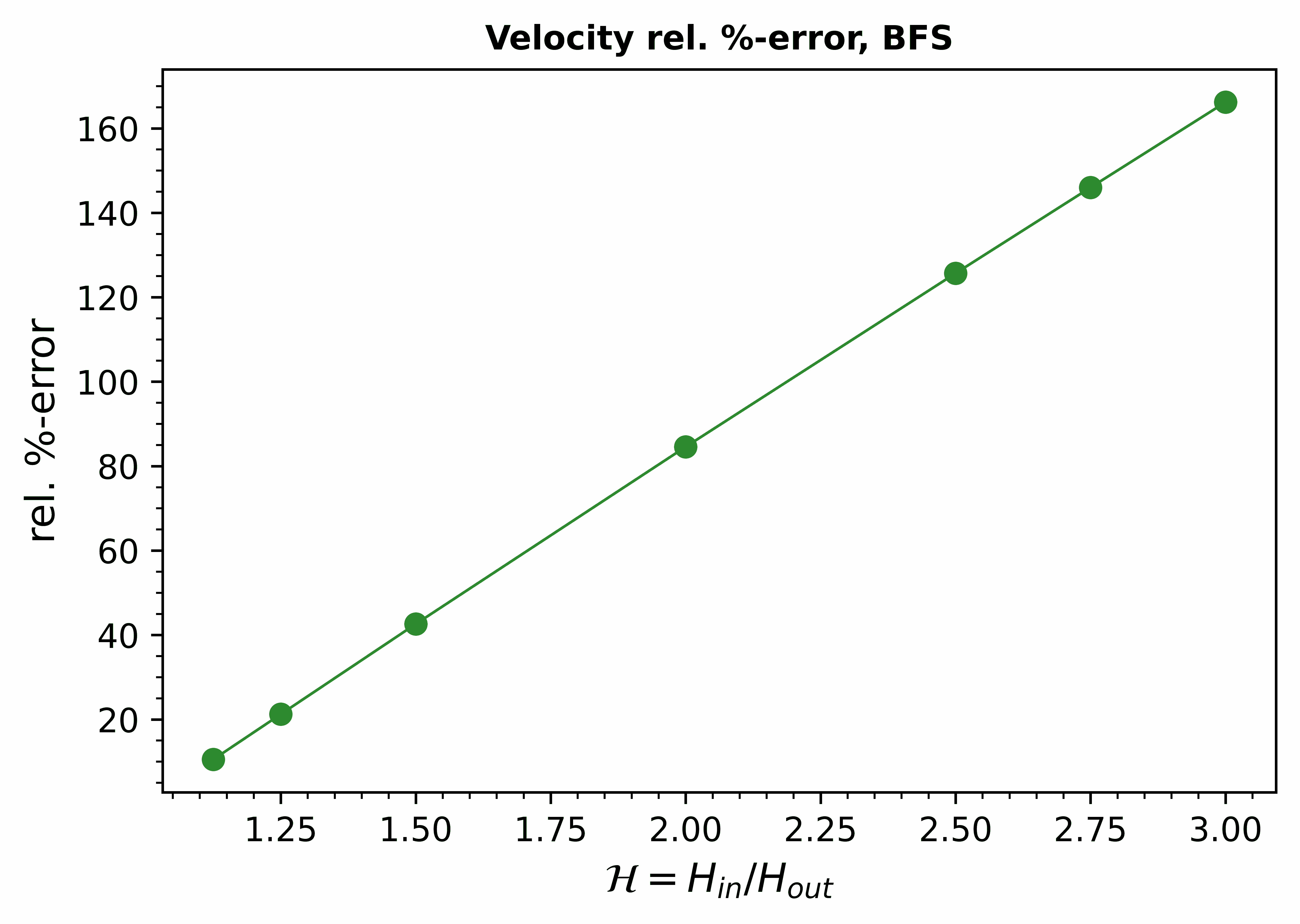}\label{bfs_v_err}}
 
 \caption{Relative percent error between the Stokes and Reynolds solution to the BFS. Error in the average pressure drop $\Delta p$, in the $l_2$ norm of pressure, and in the $l_2$ norm of velocity, increases with increasing expansion ratio $\mathcal{H}$.}
\end{figure}

The pressure contours for the Reynolds and Stokes solutions to the BFS are shown in \cref{bfs_pres} for various $\mathcal{H}$. In the Reynolds solution, the pressure is purely one dimensional, resulting in vertical pressure contours. In the Stokes solution, significant cross-film pressure variation occurs in the vicinity of the step tip. The spiraling pressure contours at the step tip, shown in \cref{bfs_pres_zoom}, indicate that the pressure gradients $\frac{\partial p}{\partial y}$ and $\frac{\partial p}{\partial x}$ are of proportional magnitude. For the Stokes solutions, the maximum pressure in the domain occurs at the step tip, whereas the corresponding Reynolds solutions all achieve their maximum pressure at the inlet. Accordingly, compared with the Stokes solution of the same flux, the Reynolds solution to the BFS consistently underestimates the average pressure drop $\Delta p$. The relative percent error in the average pressure drop $\Delta p$ and in the $l_2$ norm of pressure between the Reynolds and Stokes solutions are shown in \cref{bfs_p_error} for varying expansion ratio $\mathcal{H}$. These results confirm that the error between the Reynolds and Stokes solutions increases with increasing expansion ratio.

\begin{figure}[h]
 \centering 
 \subfloat[Stokes $\mathcal{H}=1.25$]{\label{stokes_bfs_H1p25_v}\includegraphics[width=.45\textwidth]{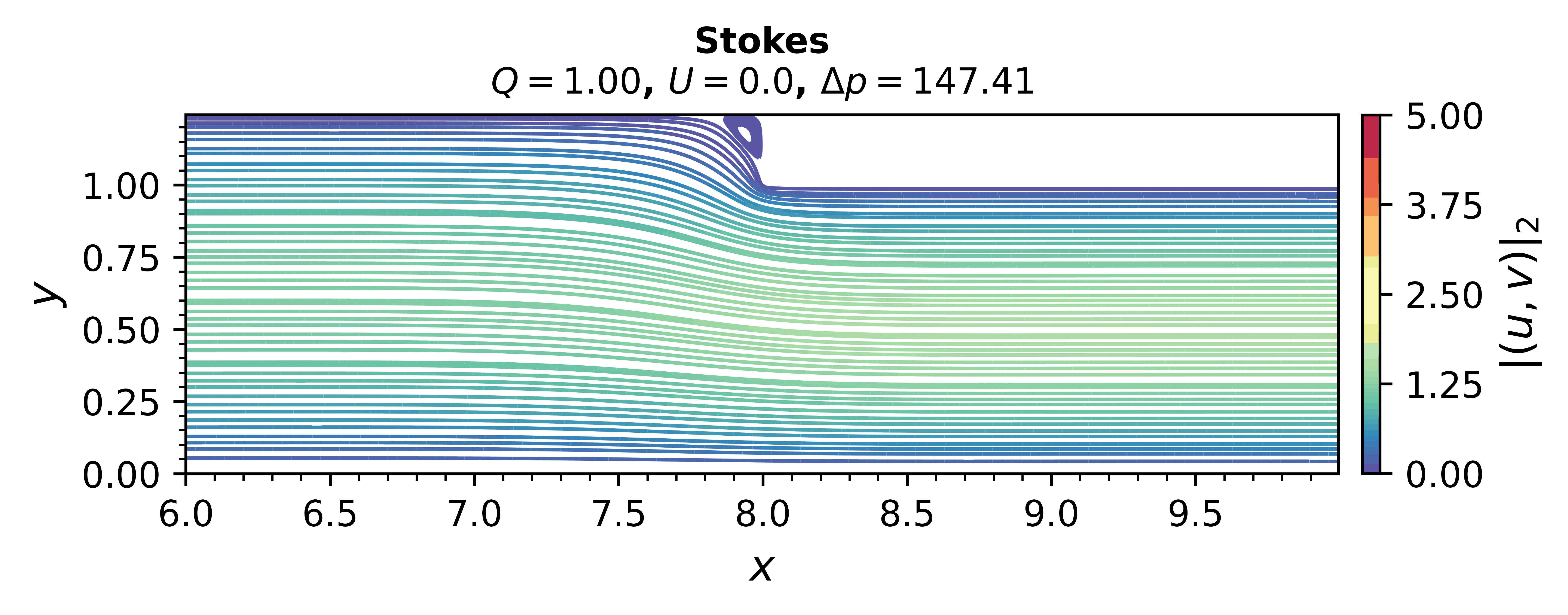}}
 \subfloat[Reynolds $\mathcal{H}=1.25$]{\label{reyn_bfs_H1p25_v}\includegraphics[width=.45\textwidth]{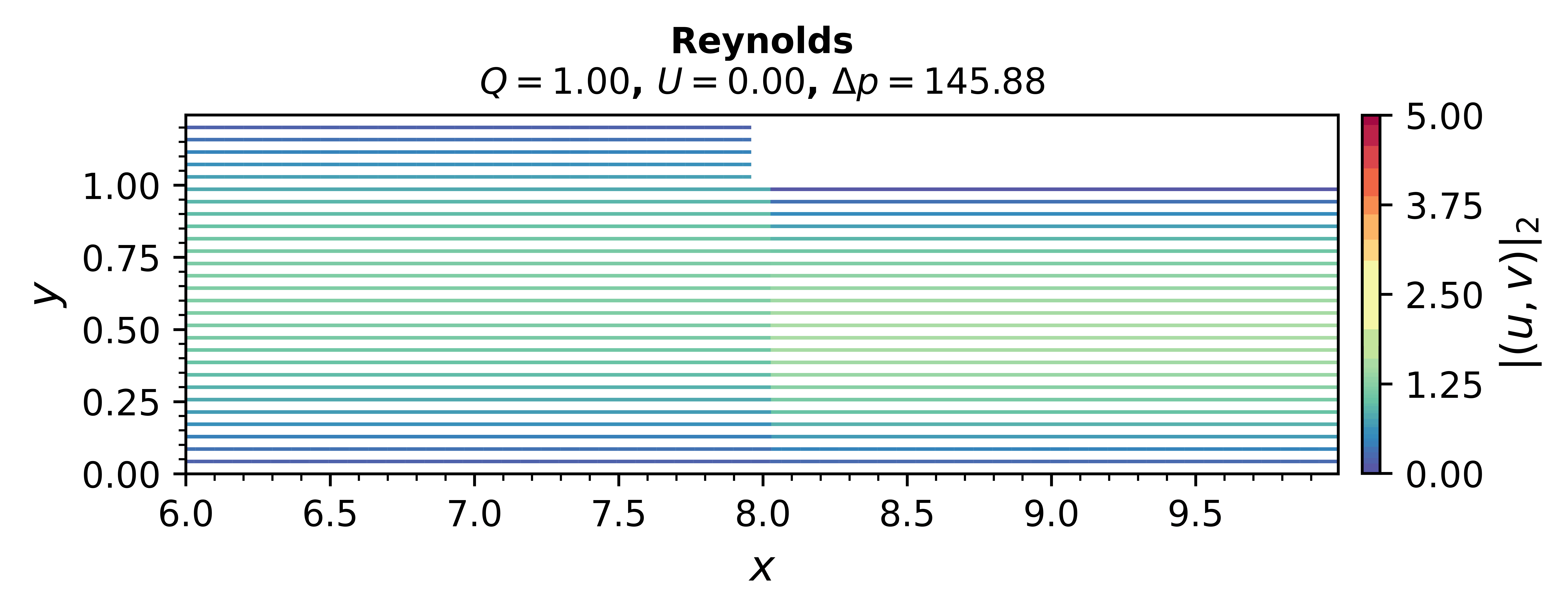}}\\

 \subfloat[Stokes $\mathcal{H}=2.75$]{\label{stokes_bfs_H2p75_v}\includegraphics[width=.45\textwidth]{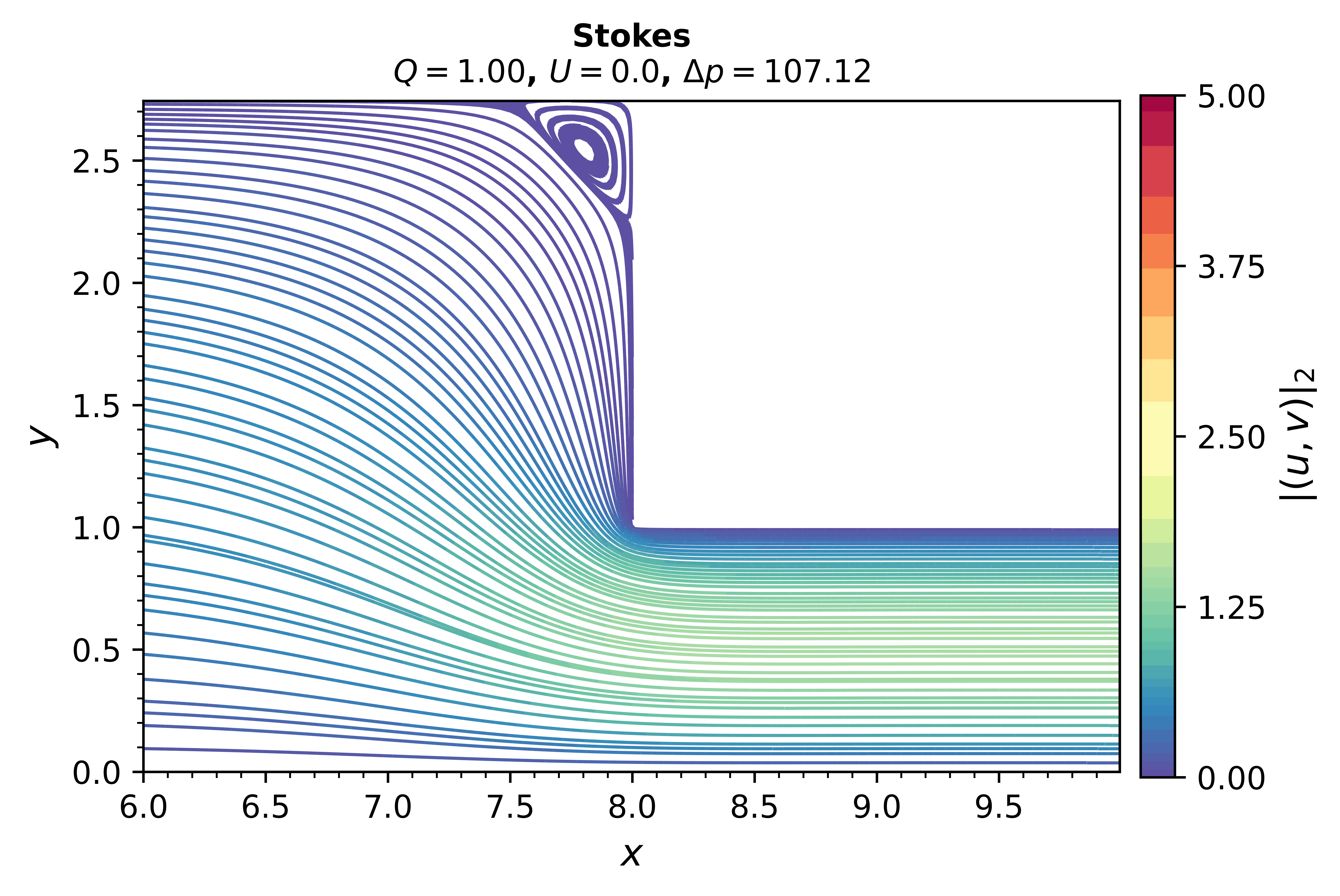}}
 \subfloat[Reynolds $\mathcal{H}=2.75$]{\label{reyn_bfs_H2p75_v}\includegraphics[width=.45\textwidth]{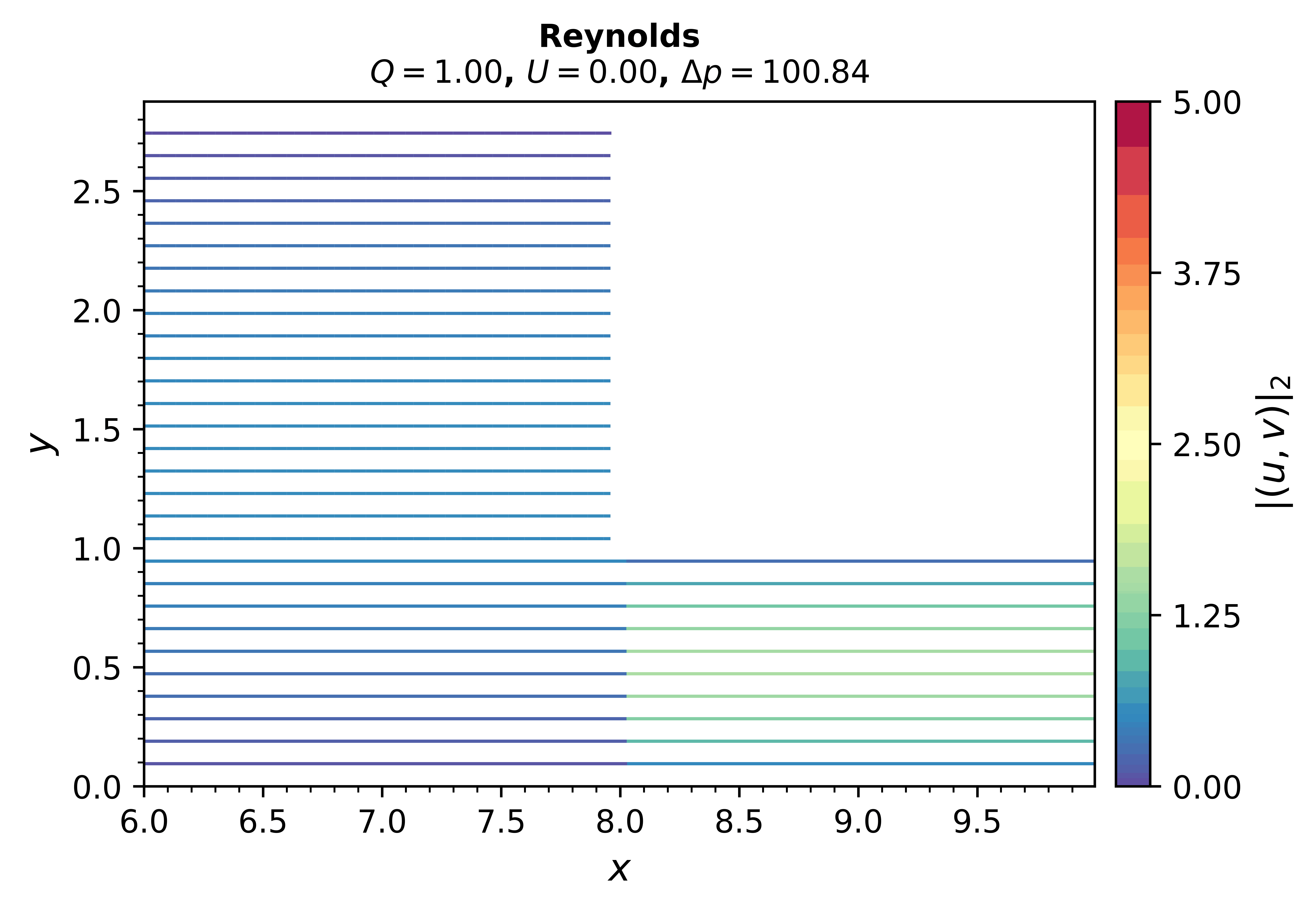}}
 \caption{Velocity streamlines for the BFS. Stokes solutions capture regions of corner recirculation, increasing in size with expansion ratio $\mathcal{H}$.}\label{bfs_vel}
\end{figure}

\begin{figure}[h]
 \centering 
 \subfloat[$\mathcal{H}=2.75$]{\includegraphics[width=0.4\textwidth]{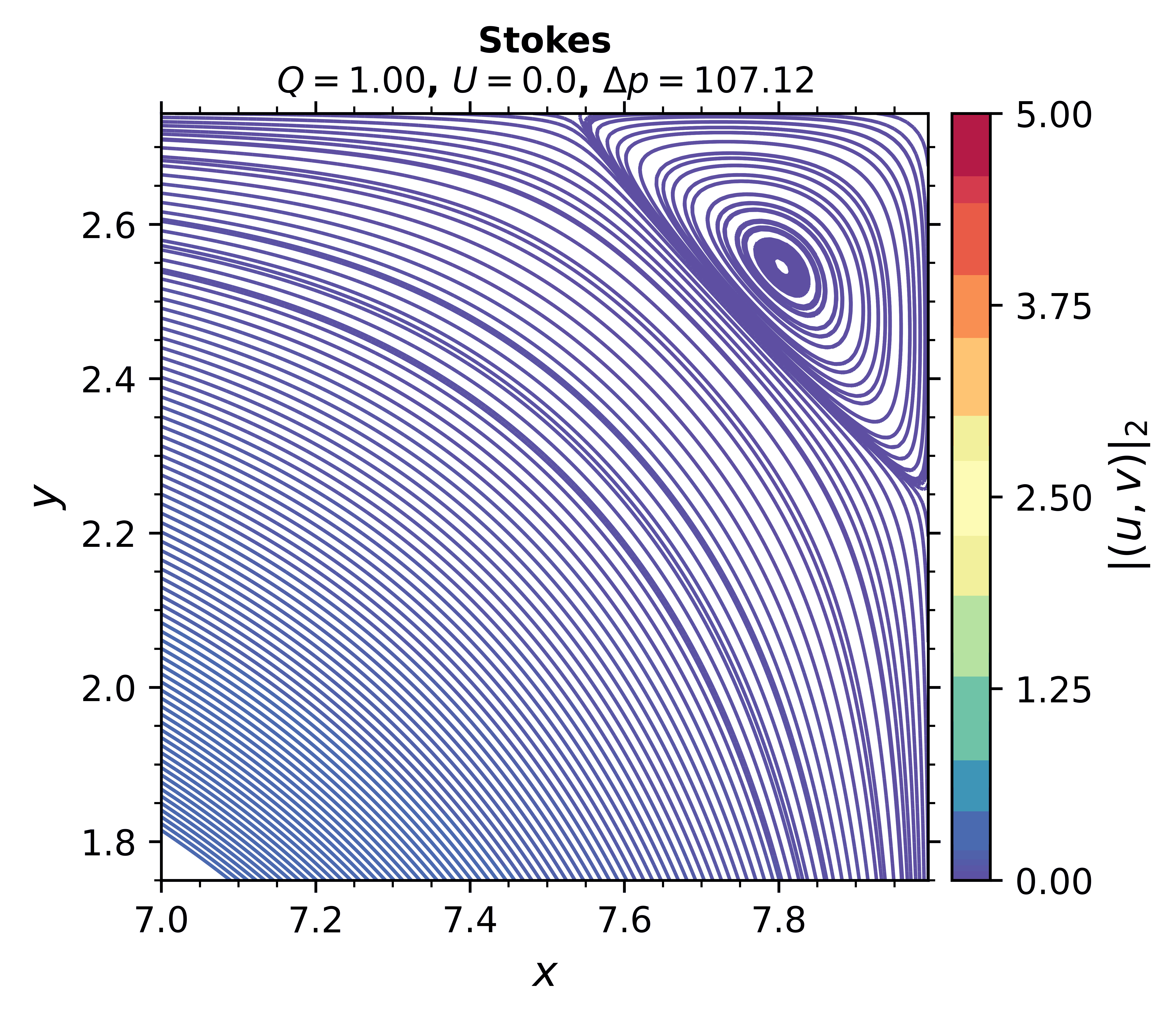}}
 \subfloat[$\mathcal{H}=2.75$]{\includegraphics[width=0.4\textwidth]{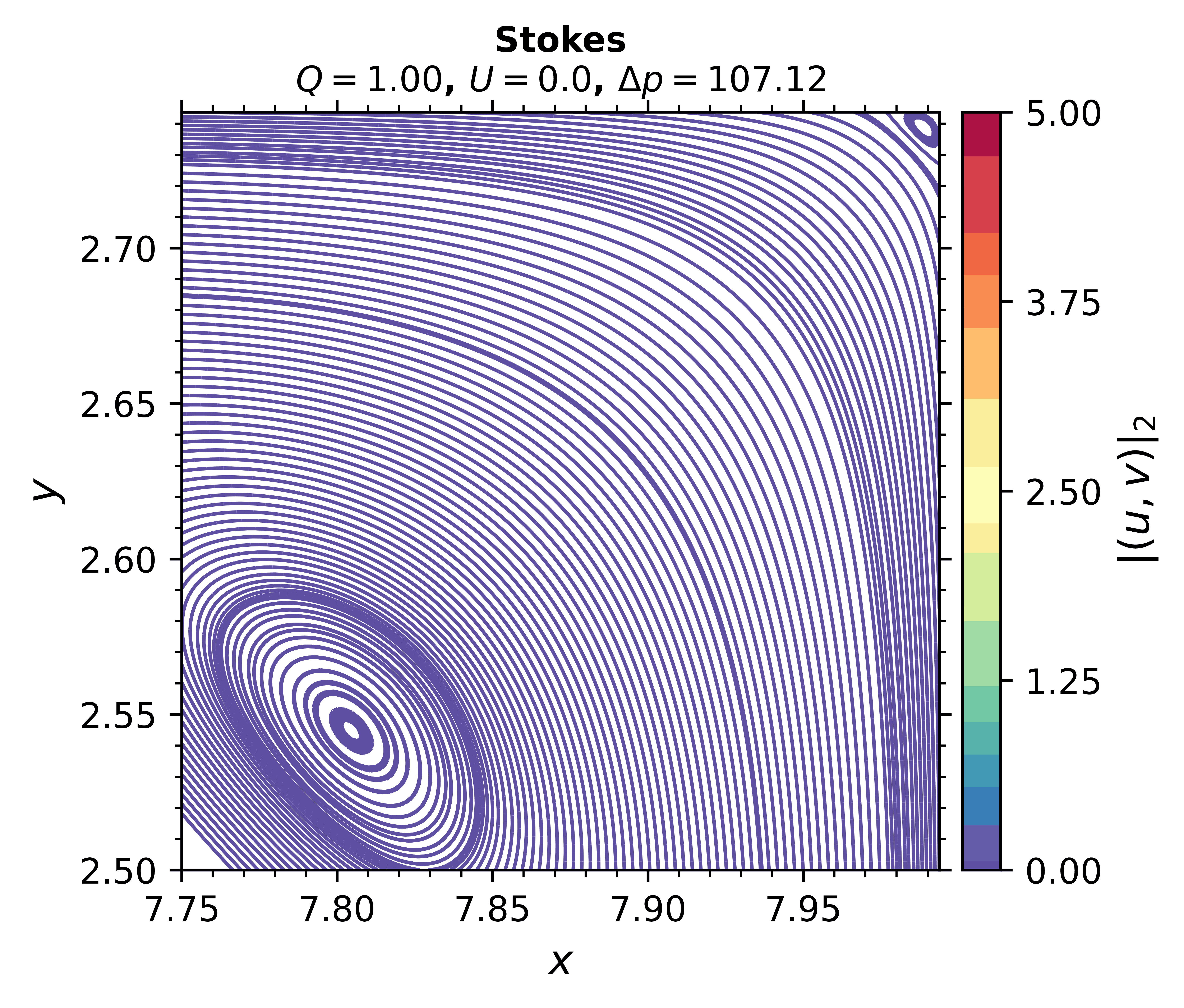}}
 \caption{Velocity streamlines for the BFS with $\mathcal{H}=2.75$, Stokes solutions. The primary recirculation has $x_\text{r}=0.469$ and $y_\text{r} = 0.50$. Secondary recirculation is observed on closer inspection of the inner corner.}\label{bfs_vel_zoom}
\end{figure}

The velocity streamlines for the Reynolds and Stokes solutions to the BFS are shown in \cref{bfs_vel} for various $\mathcal{H}$. The error in velocity between the Reynolds and Stokes solutions is significant. The relative percent error in $l_2$ norm of velocity is show in \cref{bfs_v_err}; as with the error in pressure, the error in velocity increases with increasing expansion ratio. The large error for the Reynolds velocity with the BFS is expected. The Reynolds velocity is discontinuous at the discontinuity in the height, resulting in streamlines which are everywhere flat and are not parallel to the vertical boundary at the step. In the Stokes solutions, flow recirculation occurs in the concave corner induced by the sudden expansion in the film thickness. At large expansion ratios, $\mathcal{H}\ge 2$, a secondary corner recirculation zone is observed in the Stokes solution. Close-up views of the primary and secondary corner recirculation zones are shown in \cref{bfs_vel_zoom} for $\mathcal{H}=2.75$. We expect that with a sufficiently fine grid resolution, secondary recirculation would also be captured at smaller $\mathcal{H}$.

\begin{figure}[h]
 \centering 
 \includegraphics[width=.75\textwidth]{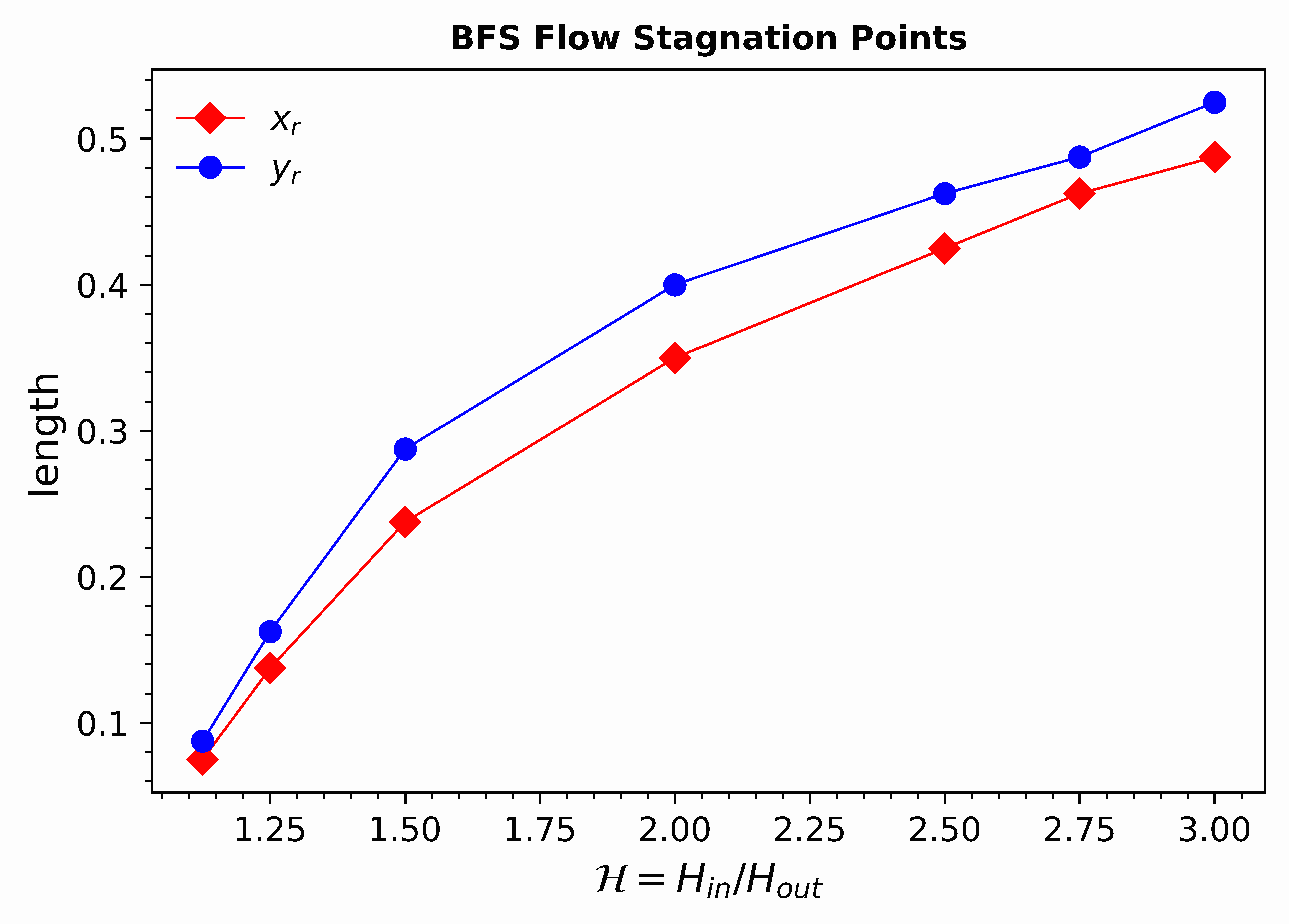}
 \caption{Points of flow separation $(L_\text{in} -x_\text{r},H_\text{in})$ and $(L_\text{in},H_\text{in} -y_\text{r})$ in the BFS, Stokes solutions. The corner recirculation region increases with expansion ratio $\mathcal{H}$.} \label{bfs_attachments}
\end{figure}

We further examine the flow separation for the Stokes solution to the BFS with various expansion ratios $\mathcal{H}$. The corner separated region is characterized by the points of primary flow separation: the half-saddle points of the stream function, $(L_\text{in} -x_\text{r},H_\text{in})$ and $(L_\text{in},H_\text{in} -y_\text{r})$ furthest from the corner $(L_\text{in}, H_\text{in})$. As shown in \cref{bfs_attachments}, the lengths $x_\text{r}$ and $y_\text{r}$ increase with increasing expansion ratio. Validation of our result $x_r=0.356$ at $\mathcal{H} =2$ and $\text{Re} = 0$ is considered in \cref{bfs_xr_H2}. Previously reported results of $x_r$ for the similar example $\mathcal{H}=1.9423$ and $\text{Re} = 10^{-4}$ are $x_\text{r} = 0.350$ from Biswas et al. \cite{biswas_backward-facing_2004} and $x_\text{r} = 0.3491$ from Saleel et al. \cite{saleel_simulation_2013}. Our results for $x_\text{r}$ is on par with the literature results.

\begin{table}[h]
\centering
\begin{tabular}{|c|c|c|c|}
\hline
\multirow{2}{*}{$\mathcal{H}$}  & ($\text{Re}=0$)  & ($\text{Re}=10^{-4}$, \cite{biswas_backward-facing_2004}) & ($\text{Re}=10^{-4}$, \cite{saleel_simulation_2013}) \\
            &    $2$     & $1.9423$ & $1.9423$ \\ \hline\hline
$x_\text{r}$& $0.356$  & $0.350$ & $0.349$  \\ \hline
\end{tabular}
\vspace{0.5em}
\caption{Lengths $x_r$ corresponding to the point of flow separation $(L_\text{in} -x_\text{r},H_\text{in})$ in the BFS. Our results for $\mathcal{H}=2$ are consistent with the literature results for $\mathcal{H}=1.9423$ and $\text{Re}=10^{-4}$.} \label{bfs_xr_H2}
\end{table}

\subsection{The wedged BFS}
\begin{figure}[h]
 \centering 
 \includegraphics[width=0.75\textwidth]{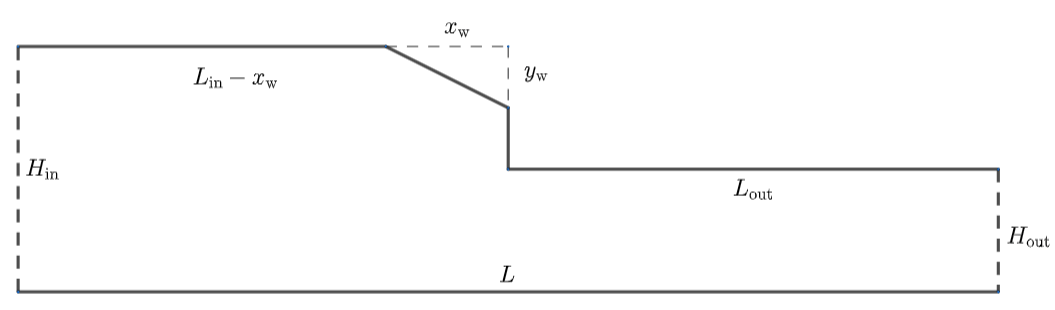}
 \caption{Schematic of the wedged BFS with expansion ratio $\mathcal{H}=H_{\text{out}}/H_{\text{in}}$, wedge height $y_\text{w}$ and length $x_\text{w}$.} \label{schematic_wedge}
\end{figure}

This section presents a variation of the BFS, inspired by our investigation of corner recirculation. As with the BFS, the wedged BFS is characterized by the expansion ratio $\mathcal{H}=H_{\text{in}}/H_{\text{out}}$, but now the concave corner of the step is smoothed by a wedge to fully or partially occlude the corner recirculation zone that occurs in the standard BFS. 
The film height for the wedged BFS is given by,
\begin{equation}
 h(x)=\begin{cases}
 H_{\text{in}} & 0\le x \le L_{\text{in}}-x_\text{w}\\
 H_{\text{in}}-y_\text{w}+\frac{y_\text{w}}{x_\text{w}}(L_{\text{in}}-x) & L_{\text{in}}-x_\text{w} \le x \le L_{\text{in}}\\
 H_{\text{out}} &L_{\text{in}}\le x \le L
 \end{cases},
\end{equation}
and a schematic is shown in \cref{schematic_wedge}.
We compare the Stokes solutions to the wedged BFS with the standard BFS for varying expansion ratios $1 < \mathcal{H}\ll L$ and with various corner wedges $0<x_\text{w}\le x_\text{r}$, $0<y_\text{w}\le y_\text{r}$. The parameters $H_{\text{out}}=1$ and $L_{\text{in}}=L_\text{out}=8$ are kept constant, along with the boundary conditions $\mathcal{U}=0$, $\mathcal{Q}=1$ and $p(L,0)=0$.

For a first example of the wedged BFS, we consider $x_\text{w}=x_\text{r}$ and $y_\text{w}=y_\text{r}$, where $x_\text{r}$ and $y_\text{r}$ are determined through our analysis of flow separation in the BFS. This configuration of the wedged BFS aims to exactly occlude the corner recirculation region. Note that because the velocity along the streamline partitioning the bulk flow from a corner recirculation region is zero, adjusting the domain boundary to lie along this null streamline will preserve the flow structure and the overall pressure drop for the domain. In this case, we approximate the partition between the bulk flow and the recirculation region as linear between the points of flow separation, and examine the extent to which the bulk flow and pressure drop are preserved. 

\begin{figure}[h]
 \centering
 \subfloat[BFS velocity]{\includegraphics[width=.45\textwidth]{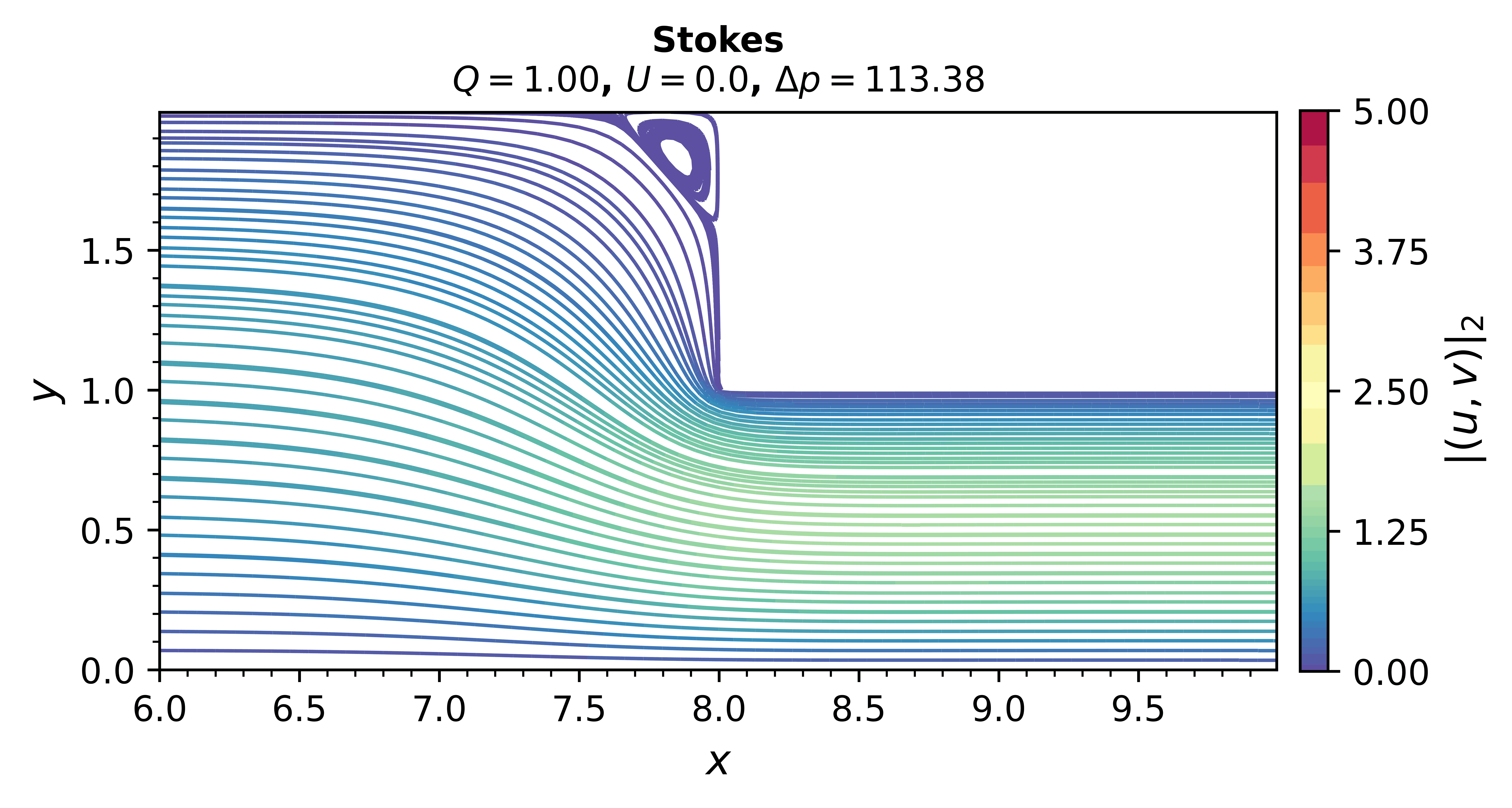}}
\subfloat[BFS pressure]{\includegraphics[width=.45\textwidth]{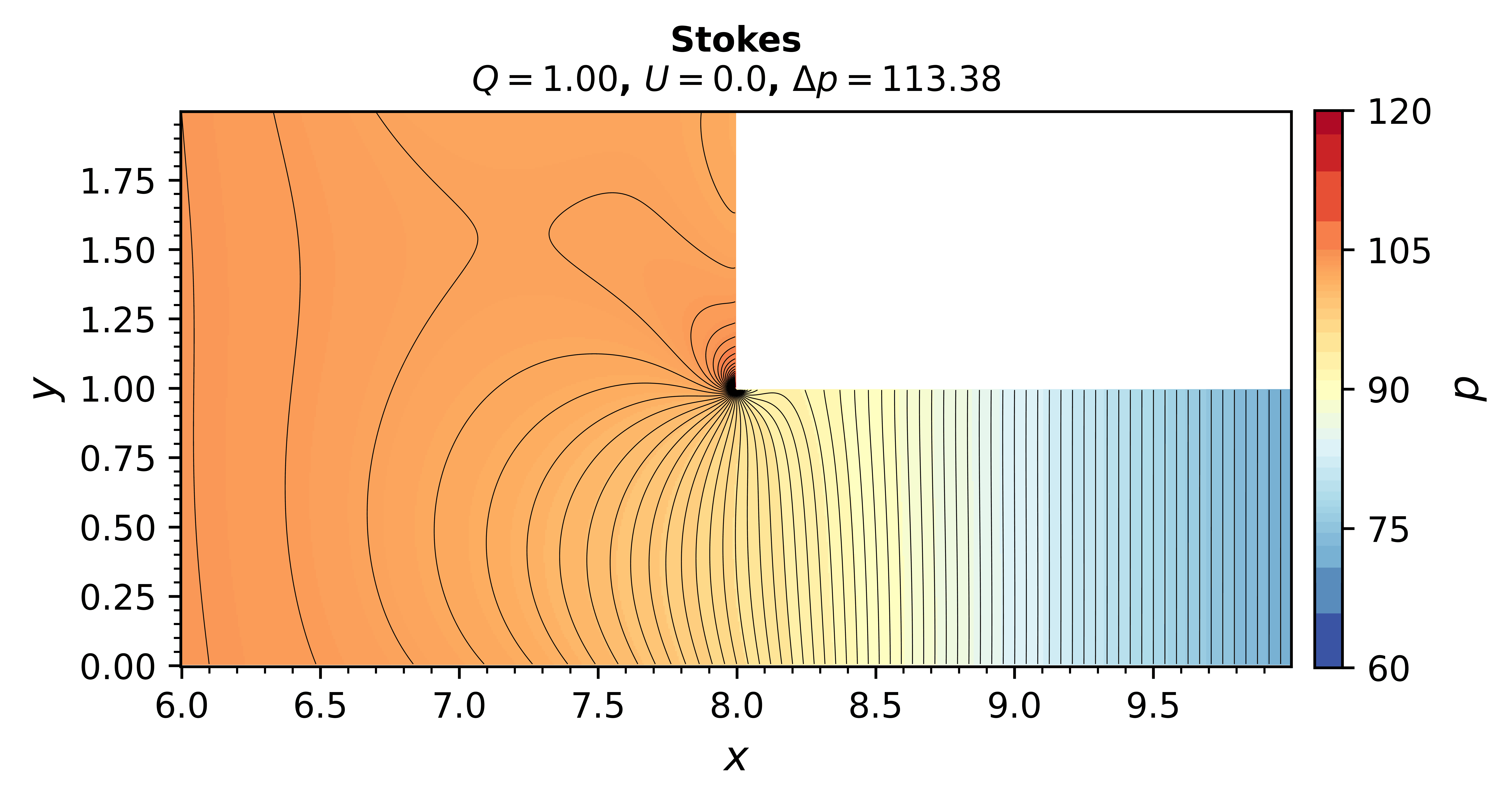}}\\
 \subfloat[Wedged BFS velocity]{\includegraphics[width=.45\textwidth]{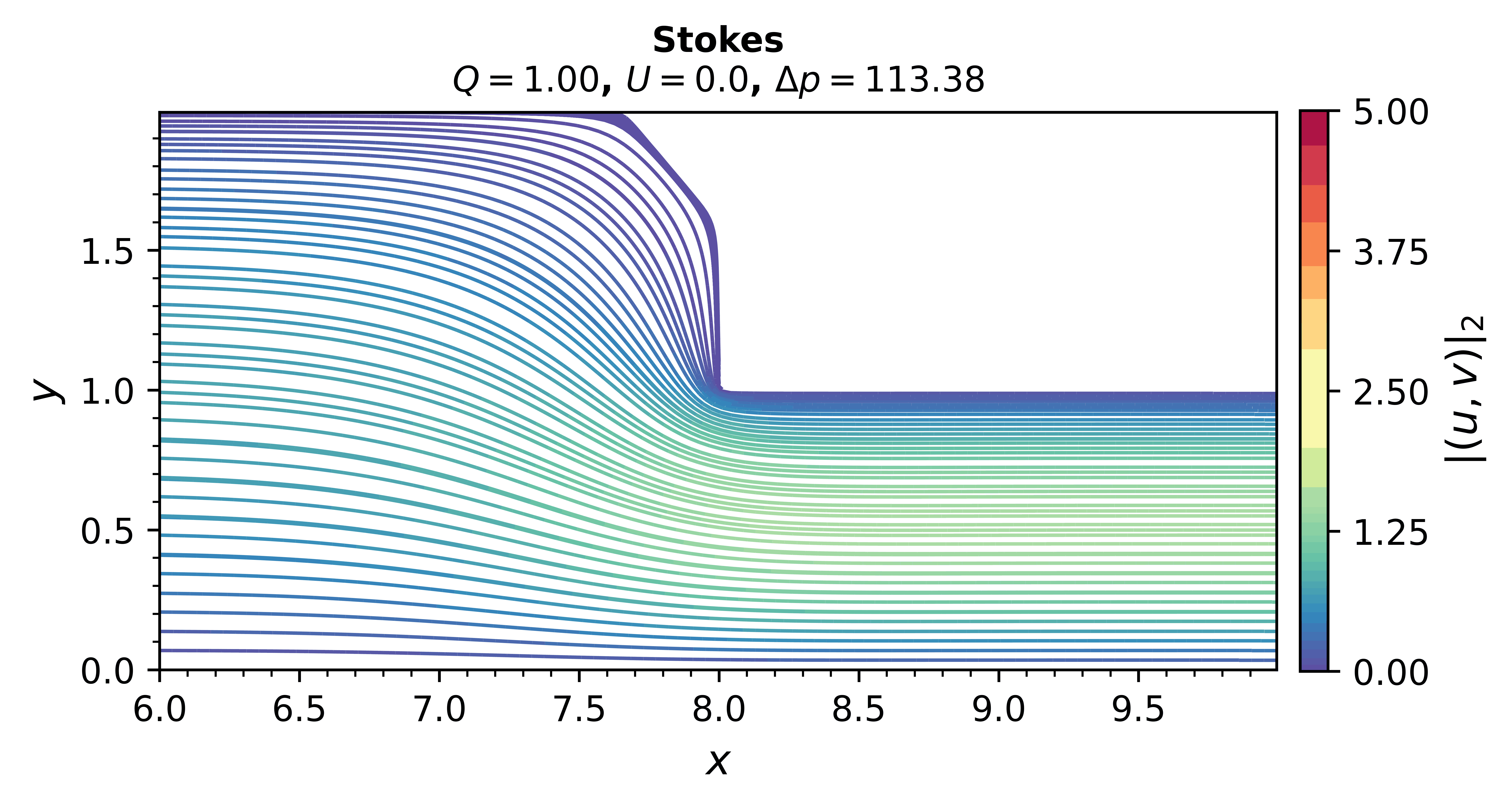}}
 \subfloat[Wedged BFS pressure]{\includegraphics[width=.45\textwidth]{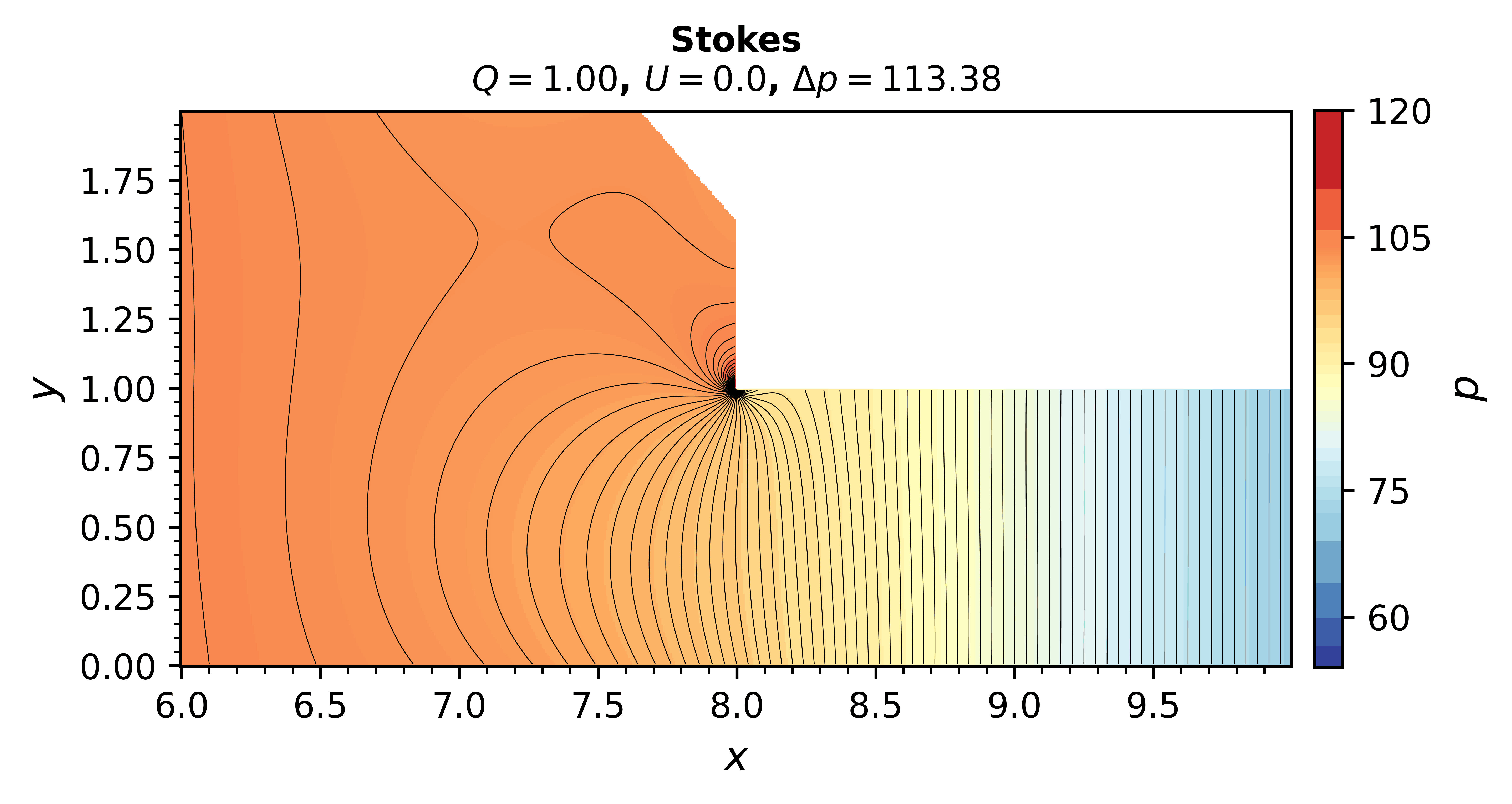}}
 \caption{Stokes solutions for the BFS and wedged BFS with $\mathcal{H} =2$, $x_\text{w} = x_\text{r}= 0.356$, $y_\text{w}=y_\text{r}=0.41$. The velocity and pressure profiles are very similar.}\label{wedge}
\end{figure}

The Stokes solutions of velocity and pressure to the BFS and the wedged BFS with $\mathcal{H} = 2$, $x_\text{r}=x_\text{w}=0.356$ and $y_\text{r}=y_\text{w}=0.406$ are presented in \cref{wedge}. As anticipated, the solutions of velocity and pressure for the BFS and wedged BFS are very similar. The average pressure drops $\Delta p$ are equivalent, $\Delta p = 113.38$ for $\mathcal{H}=2$, although corner recirculation is no longer apparent in the wedged BFS. This behavior is similarly observed at the various expansion ratios $1.125 \le \mathcal{H}\le 3$ for which we determined $x_\text{r}$ and $y_\text{r}$.

\begin{figure}[h]
 \centering 
 \subfloat[$x_\text{w} =\tfrac{1}{2}x_
 \text{r}$]{\includegraphics[width=.3\textwidth]{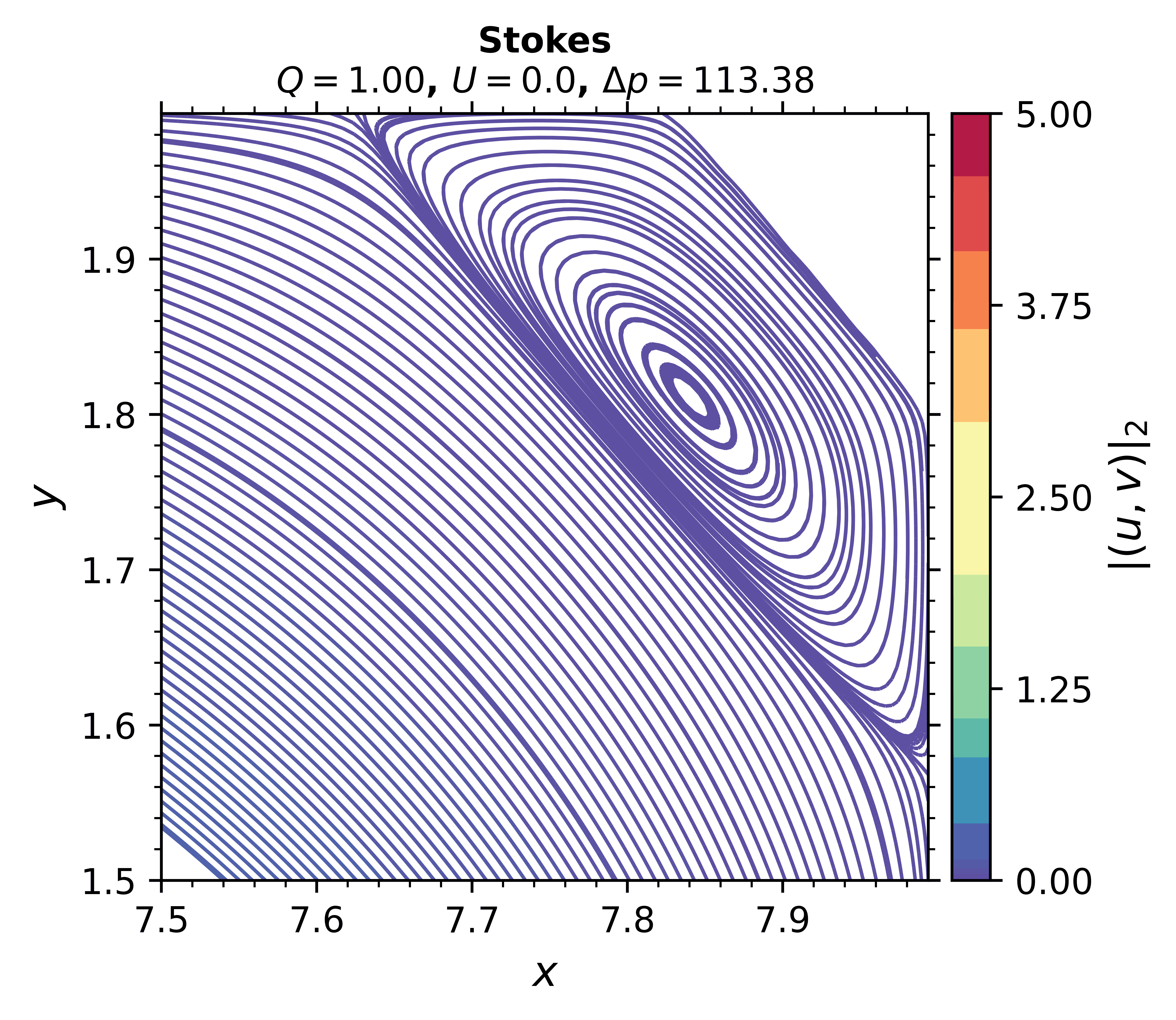}}
 \subfloat[$x_\text{w} =\tfrac{3}{4}x_
 \text{r}$]{\includegraphics[width=.3\textwidth]{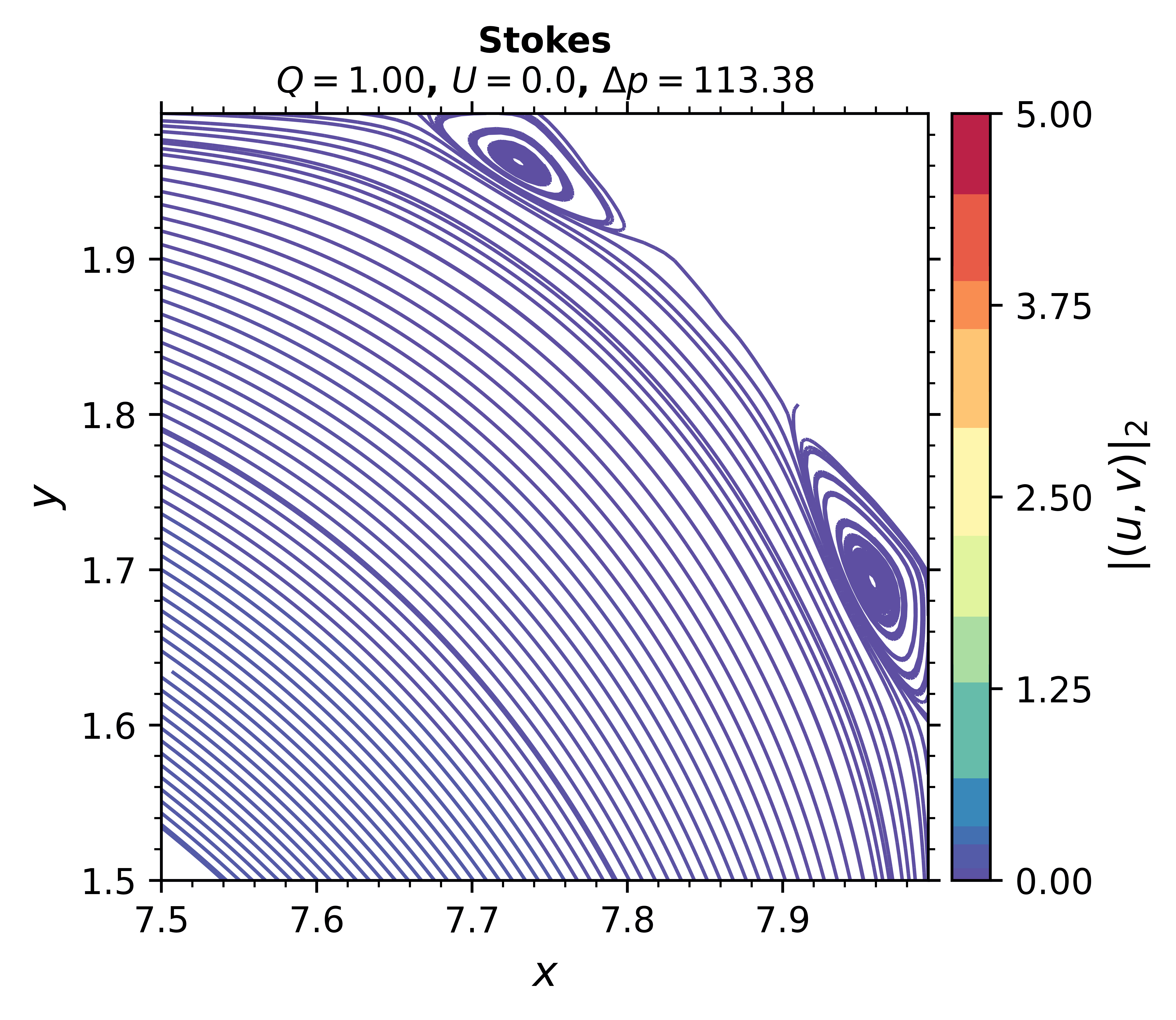}}
 \subfloat[$x_\text{w}=x_\text{r}=0.356$]{\includegraphics[width=.3\textwidth]{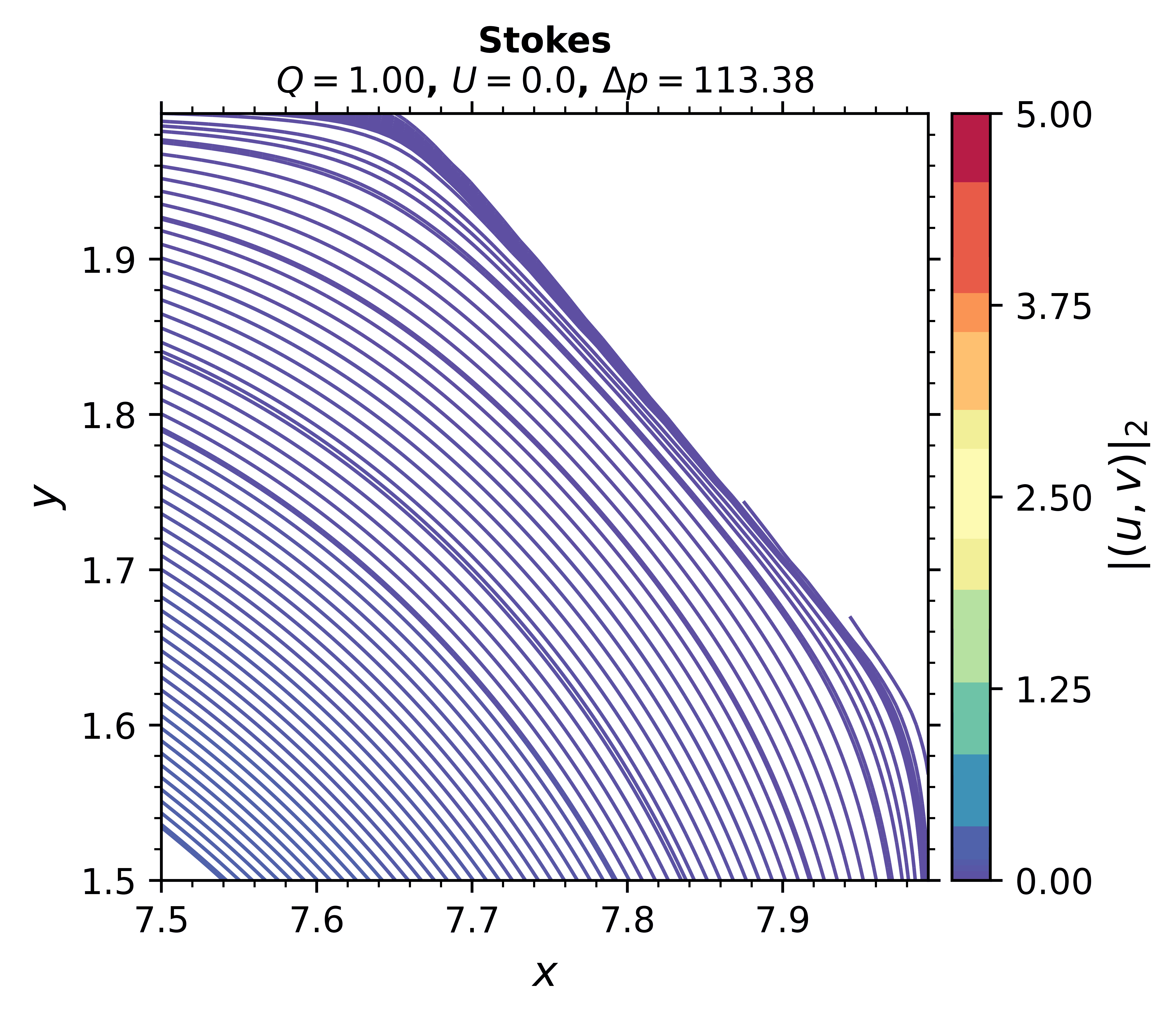}}
 
 \caption{Velocity streamlines in the vicinity of the wedged corner for $\mathcal{H}=2$, $y_\text{w}/x_\text{w} = 0.41/0.356$. The average pressure drop and points of flow separation are approximately stable. As $x_\text{w}\to x_\text{r}$ and $y_\text{w}\to y_\text{r}$, the primary recirculation splits in two, eventually vanishing.}\label{wedge_vary}
\end{figure}

Now we analyze the effect of resizing the corner wedge. For each expansion ratio $\mathcal{H}$, we consider the class of examples with fixed wedge slope $y_\text{w}/x_\text{w}=y_\text{r}/x_\text{r}$ and varying wedge lengths $x_\text{w} \le x_\text{r}$. \cref{wedge_vary} depicts the Stokes solutions of velocity in the vicinity of the wedge corner at $\mathcal{H}=2$ and various $x_\text{w}$, where $y_\text{w}/x_\text{w} = y_\text{r}/x_\text{r}= 0.41/0.356$. We observe that for various wedge sizes, the flow separation points remain stable. For large wedges, where $x_\text{w}\to x_\text{r}$ and $y_\text{w}\to y_\text{r}$, the primary recirculation splits, leaving distinct recirculation zones of similar size in each of the corners $(x_\text{w},h)$ and $(L_\text{in}, y_\text{w})$. At $x_\text{w}= x_\text{r}$ and $y_\text{w}=y_\text{r}$, the corner recirculation zones are no longer visible with grid resolutions at our disposal.

\begin{figure}
    \centering
    \includegraphics[width=0.75\linewidth]{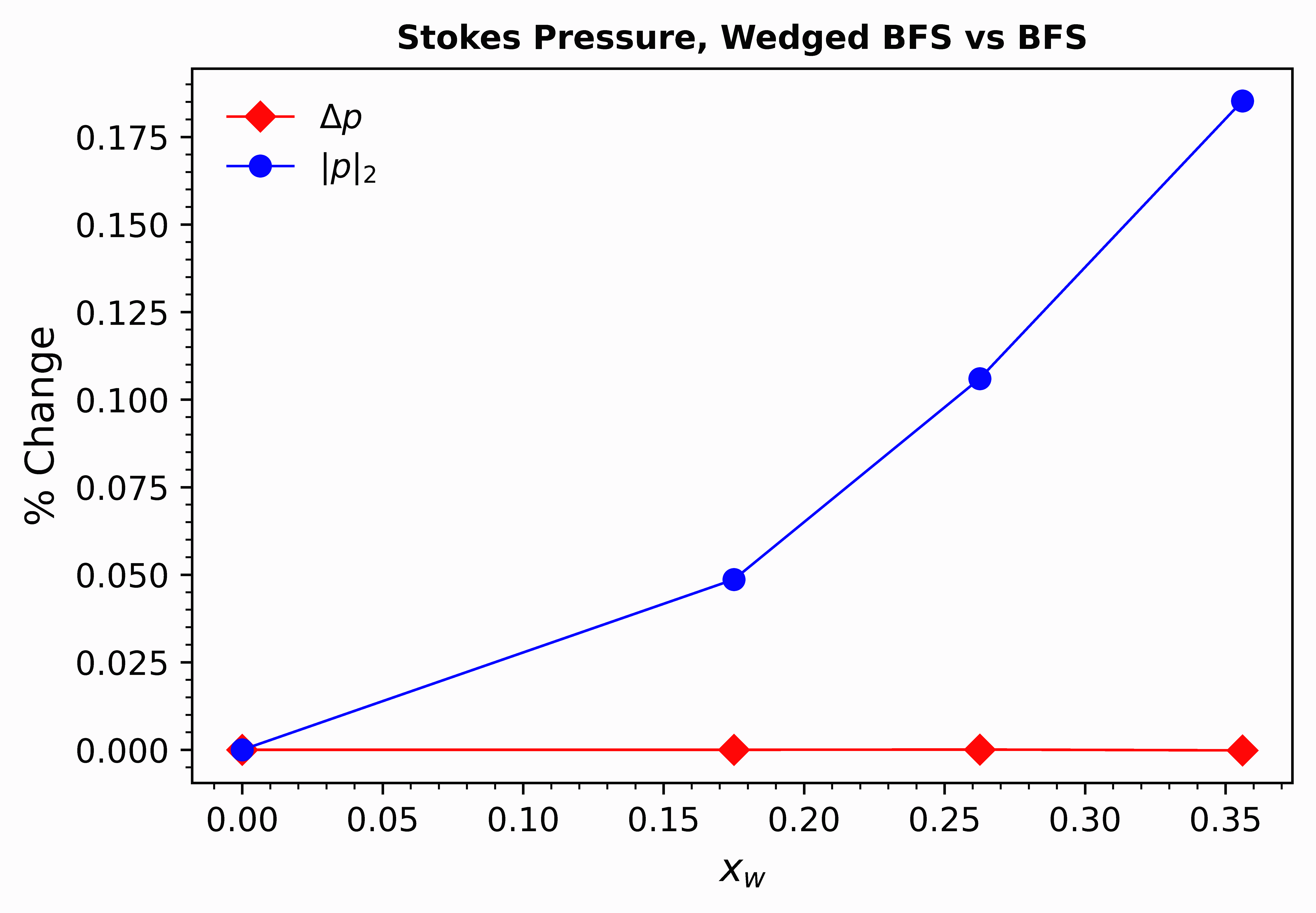}
    \caption{The $l_2$ norm of pressure and the average pressure drop $\Delta p$ for the wedged BFS at various $x_\text{w}$ relative to the standard BFS $(x_\text{w} = 0)$, Stokes solutions. The average pressure drop is stable for various wedge sizes.}
    \label{wedge_vary_pres}
\end{figure}

In the wedged BFS with $y_\text{w}/x_\text{w}=y_\text{r}/x_\text{r}$, we also observe that as the wedge size varies, the average pressure drop $\Delta p$ is constant. \cref{wedge_vary_pres} shows the pressure drop $\Delta p$ and the $l_2$ norm of pressure for varying $x_\text{w}$ relative to the standard BFS $(x_\text{w} = 0)$. The $l_2$ norm of pressure changes slightly (less than 0.2\%) for increasing wedge size, while the average pressure drop remains fixed. This result that the average pressure drop remains fixed while the wedge size varies is consistent with our observation that the points of flow separation also remain stable. All together, we find that the wedged BFS gives a family of geometries all with the same bulk flow characteristics, and for which flow recirculation can be minimized. 

\subsection{The regularized BFS}
\begin{figure}[h]
 \centering 
 \includegraphics[width=.75\textwidth]{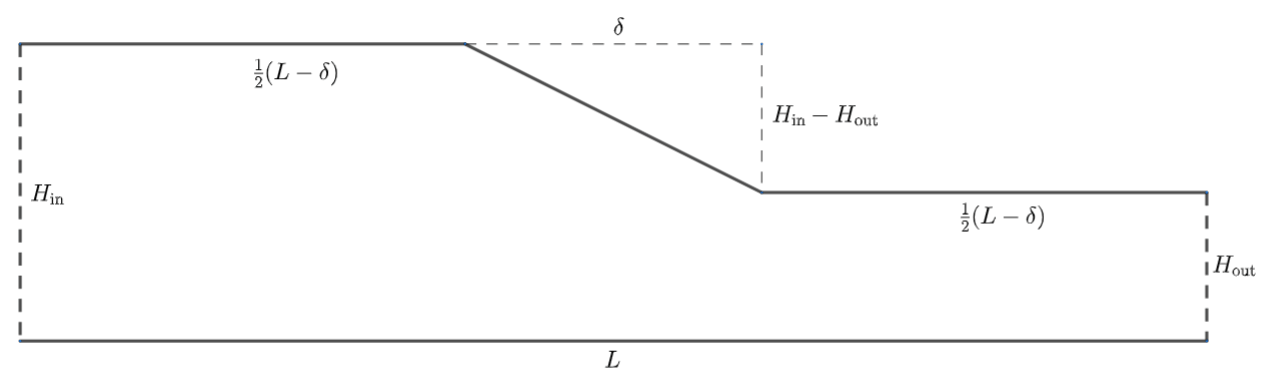}
 \caption{Schematic of the regularized BFS with expansion ratio $\mathcal{H}=H_{\text{in}}/H_{\text{out}}$ and slope $(H_\text{in}-H_\text{out})/\delta$.} \label{schematic_slope}
\end{figure}

The following section considers a further variation on the classical BFS. The regularized BFS is characterized by expansion ratio $\mathcal{H}=H_{\text{in}}/H_{\text{out}}$ in a channel of length $L$. Now, the jump discontinuity in film height is approximated with a line segment of various slopes. A schematic of the regularized BFS is shown in \cref{schematic_slope}. The film height is given by,
\begin{equation}
 h(x) = \begin{cases}
 H_{\text{in}} &\hspace{2em} 0\le x \le \tfrac{1}{2}(L-\delta)\\
 H_{\text{in}}+\frac{H_{\text{in}}-H_{\text{out}}}{\delta} \big(\tfrac{1}{2}(L-\delta)-x\big)&\hspace{2em} \tfrac{1}{2}(L-\delta)\le x \le \tfrac{1}{2}(L+\delta)\\
 H_{\text{out}} &\hspace{2em} \tfrac{1}{2}(L+\delta) \le x \le L
 \end{cases},
\end{equation} for $H_\text{in} \ge H_\text{out}$ and $\delta > 0$. The slope of the wedge is $\frac{H_{\text{in}}-H_{\text{out}}}{\delta}$, and as $\delta \to 0^+$, the regularized BFS gives exactly the BFS.
We compare the Reynolds and Stokes solutions to the regularized BFS for varying $0 < \delta\ll L$ and $1 < \mathcal{H}\ll L$, and for fixed $H_{\text{out}}=1$ and $L=16$. The boundary conditions $\mathcal{U}=0$, $\mathcal{Q}=1$, and $p(L,0)=0$ are kept constant. 

\begin{figure}[h]
 \centering 
 \subfloat[Stokes $\delta=\tfrac{1}{4}$]{\includegraphics[width=0.45\textwidth]{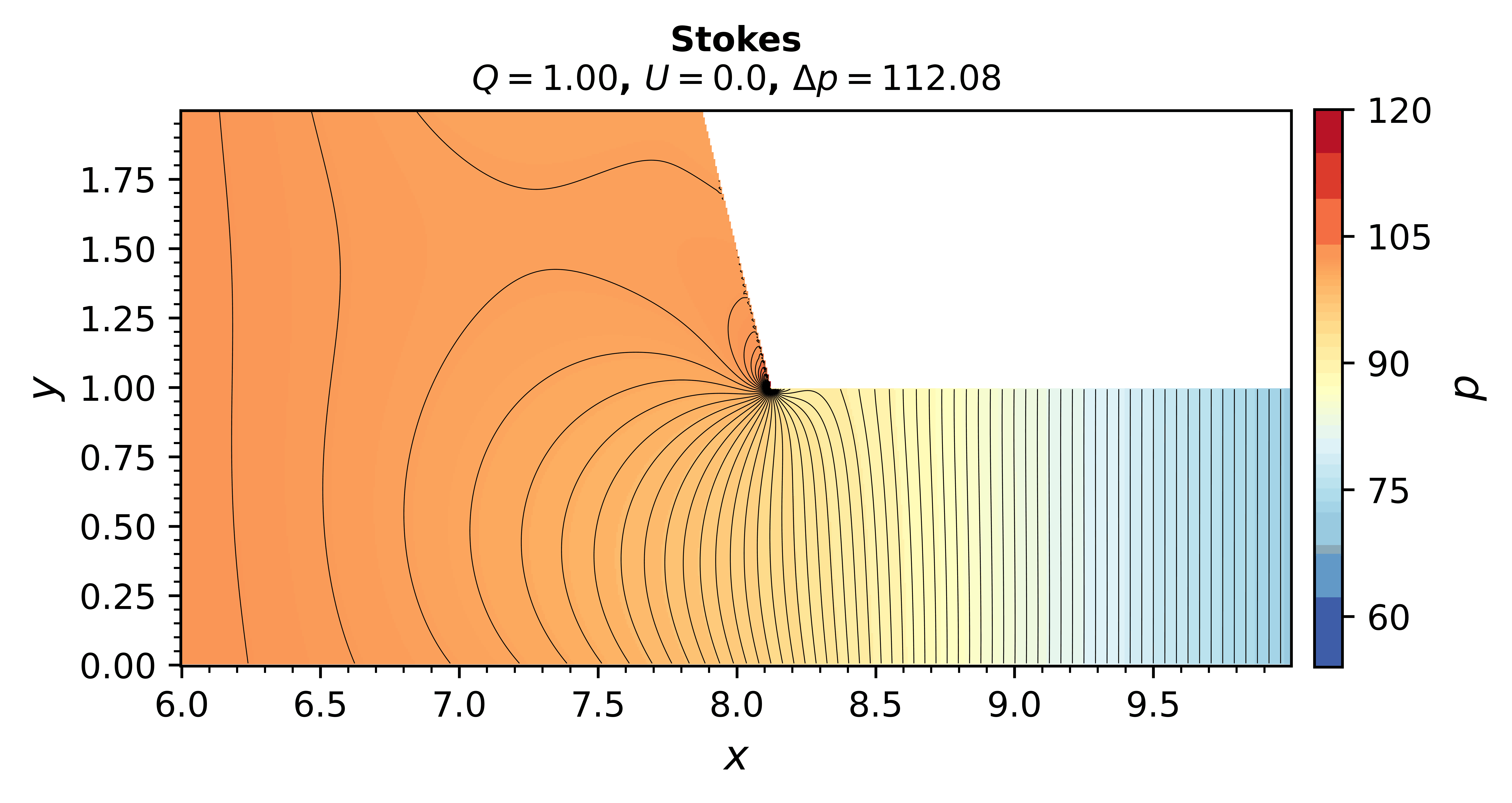}}
 \subfloat[Reynolds $\delta=\tfrac{1}{4}$]{\includegraphics[width=0.45\textwidth]{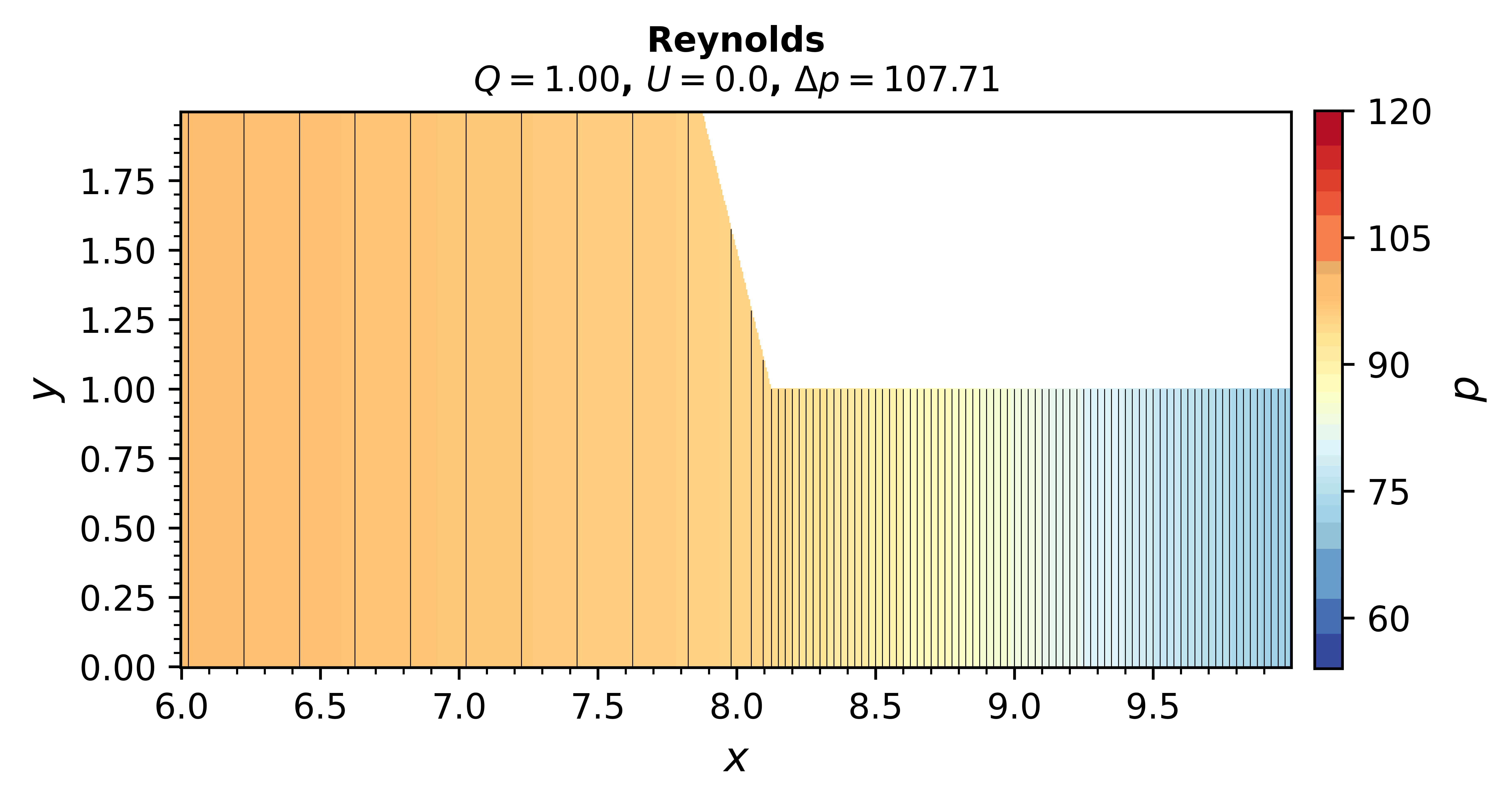}}
 
 \subfloat[Stokes $\delta=\tfrac{1}{8}$]{\includegraphics[width=0.45\textwidth]{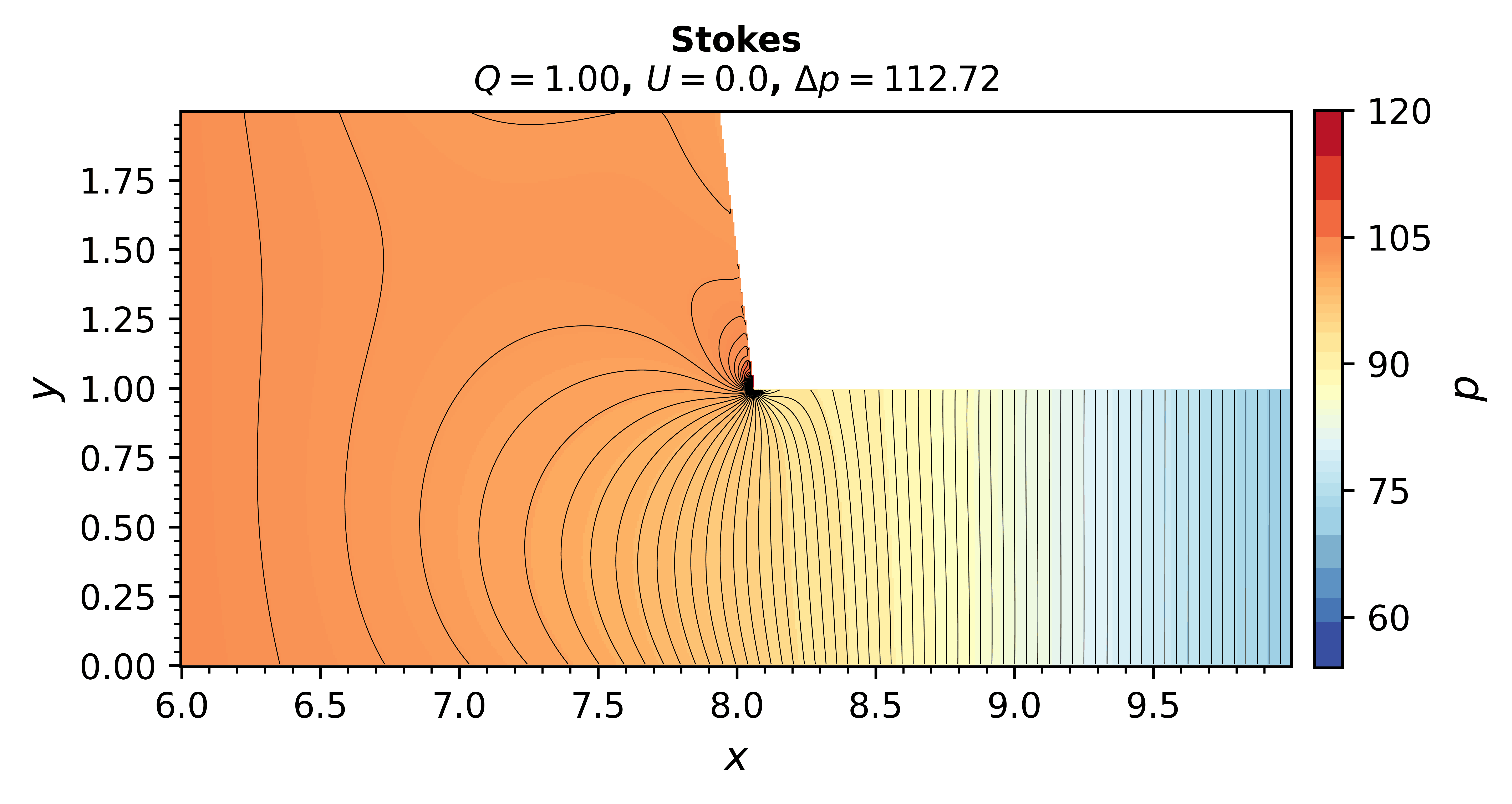}}
 \subfloat[Reynolds $\delta=\tfrac{1}{8}$]{\includegraphics[width=0.45\textwidth]{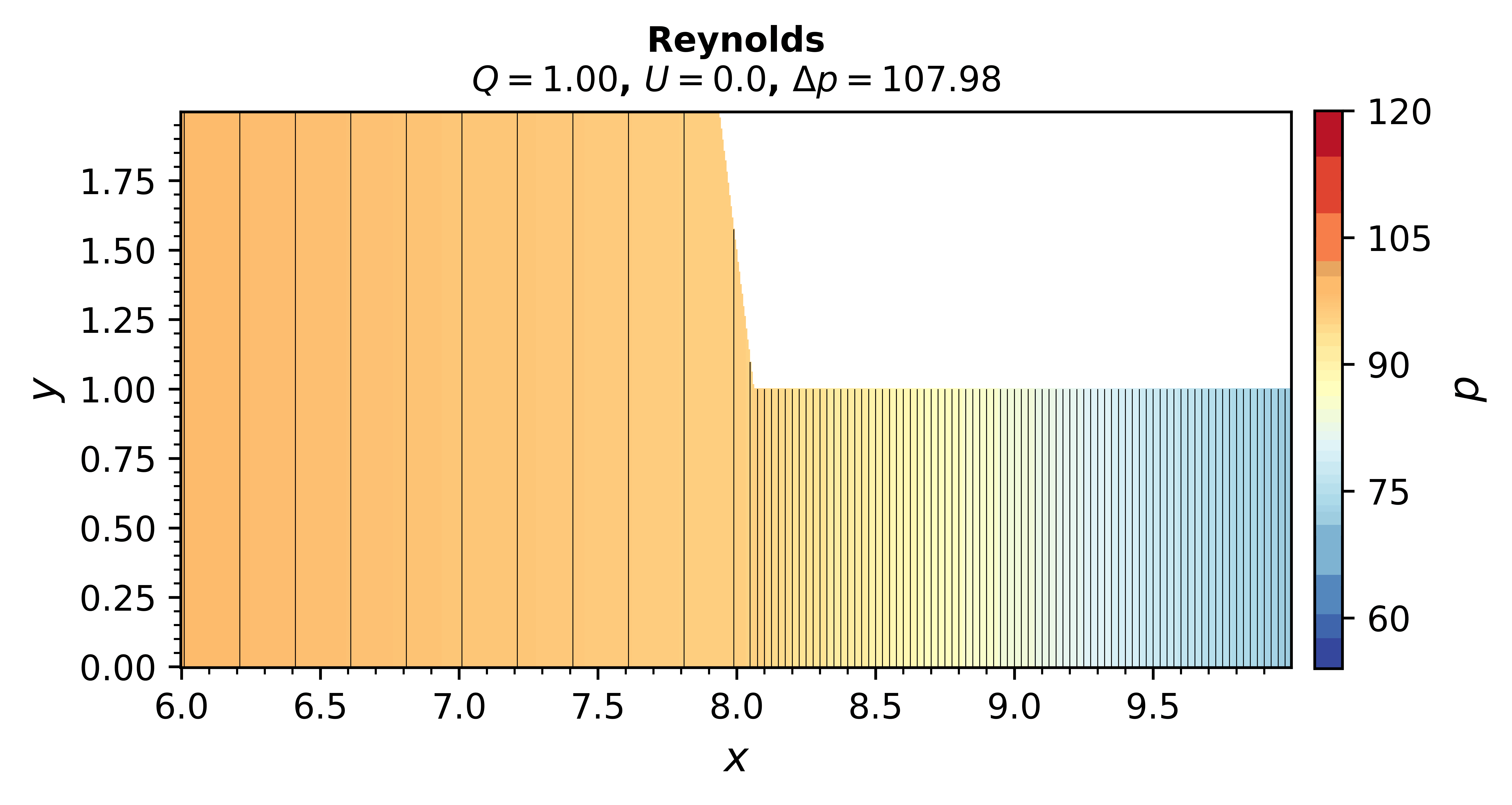}}
 
 \caption{Pressure contours for the regularized BFS, $\mathcal{H}= 2$. Contours from the Stokes solutions show significant pressure variation at the point of expansion in the film height.}\label{slope_pres}
\end{figure}

The pressure contours for the Reynolds and Stokes solutions are shown in \cref{slope_pres} for $\mathcal{H} = 2$ and various $\delta$. In the step limit of $\delta \to 0^+$, the pressure drop $\Delta p$ is more extreme. Moreover, for smaller $\delta$ (steeper slopes), the Stokes solutions obtain their maximum pressure at the step. Whereas for larger $\delta$ (more moderate slopes), the maximum pressure for the Reynolds and Stokes solutions occur at the inlet. These observations are consistent with results for the BFS in that both $\frac{\partial p}{\partial y}$ and $\frac{\partial p}{\partial x}$ are significant in the vicinity of the large surface gradient. 

\begin{figure}[h]
 \centering 
 \subfloat[Pressure error]{\includegraphics[width=0.45\textwidth]{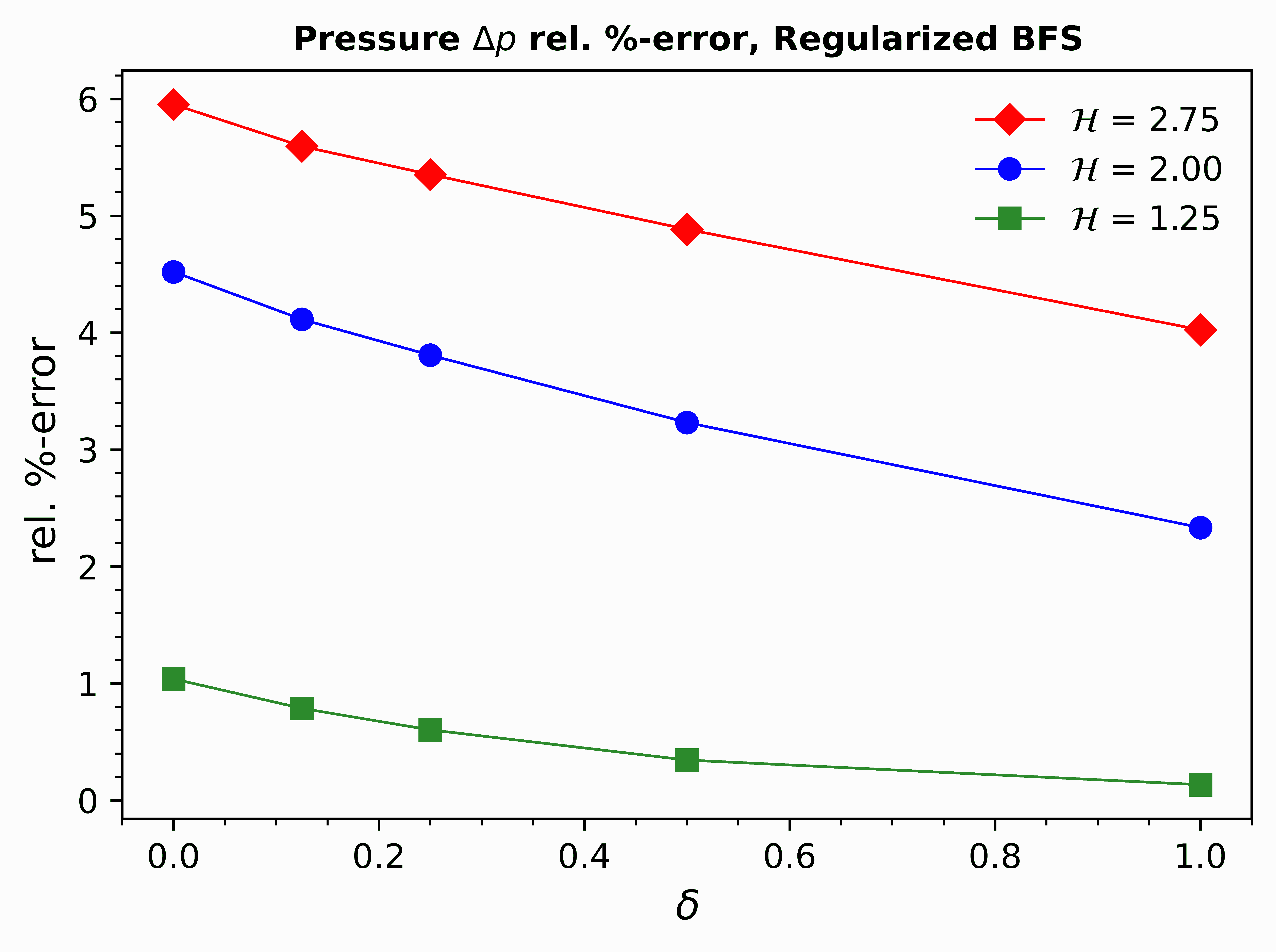}\label{slope_p_err}}
 \subfloat[Velocity error]{\includegraphics[width=0.46\linewidth]{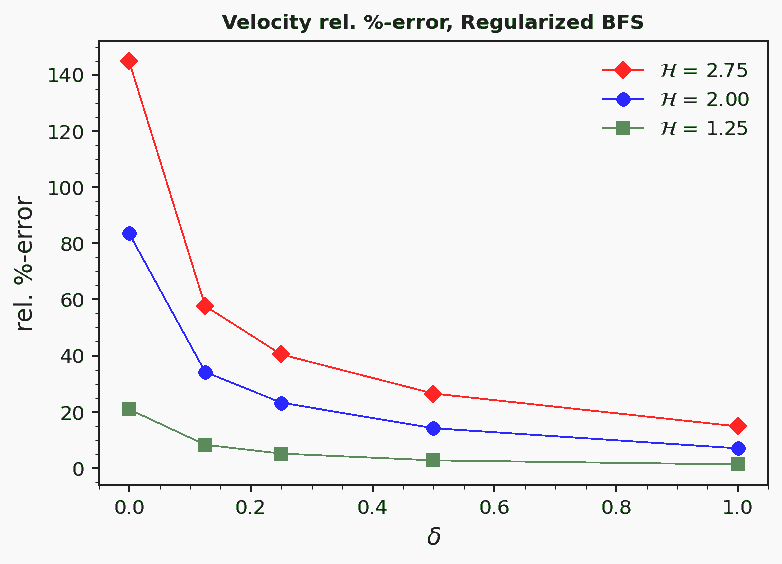}\label{slope_v_err}}
 
 \caption{Relative percent error between the Stokes and Reynolds solution to the regularized BFS. Error in the average pressure drop $\Delta p$ and in the $l_2$ norm of velocity increases with increasing expansion ratio $\mathcal{H}=H_\text{in}/ H_\text{out}$, and with increasing magnitude of the surface gradient $(H_\text{in}-H_\text{out})/\delta$.}
\end{figure}

The relative percent error in average pressure drop $\Delta p$ as a function of $\delta$ are shown in \cref{slope_p_err} for various expansion ratios $\mathcal{H}$; we include the results for the BFS at $\delta = 0$. For all expansion ratios, relative percent error in $\Delta p$ increases as $\delta \to 0^+$, and the BFS exhibits the largest error between the models. Moreover, relative percent error in $\Delta p$ increases with increasing expansion ratio.

\begin{figure}[h]
 \centering 
 \subfloat[Stokes $\delta=\tfrac{1}{4}$]{\includegraphics[width=.45\textwidth]{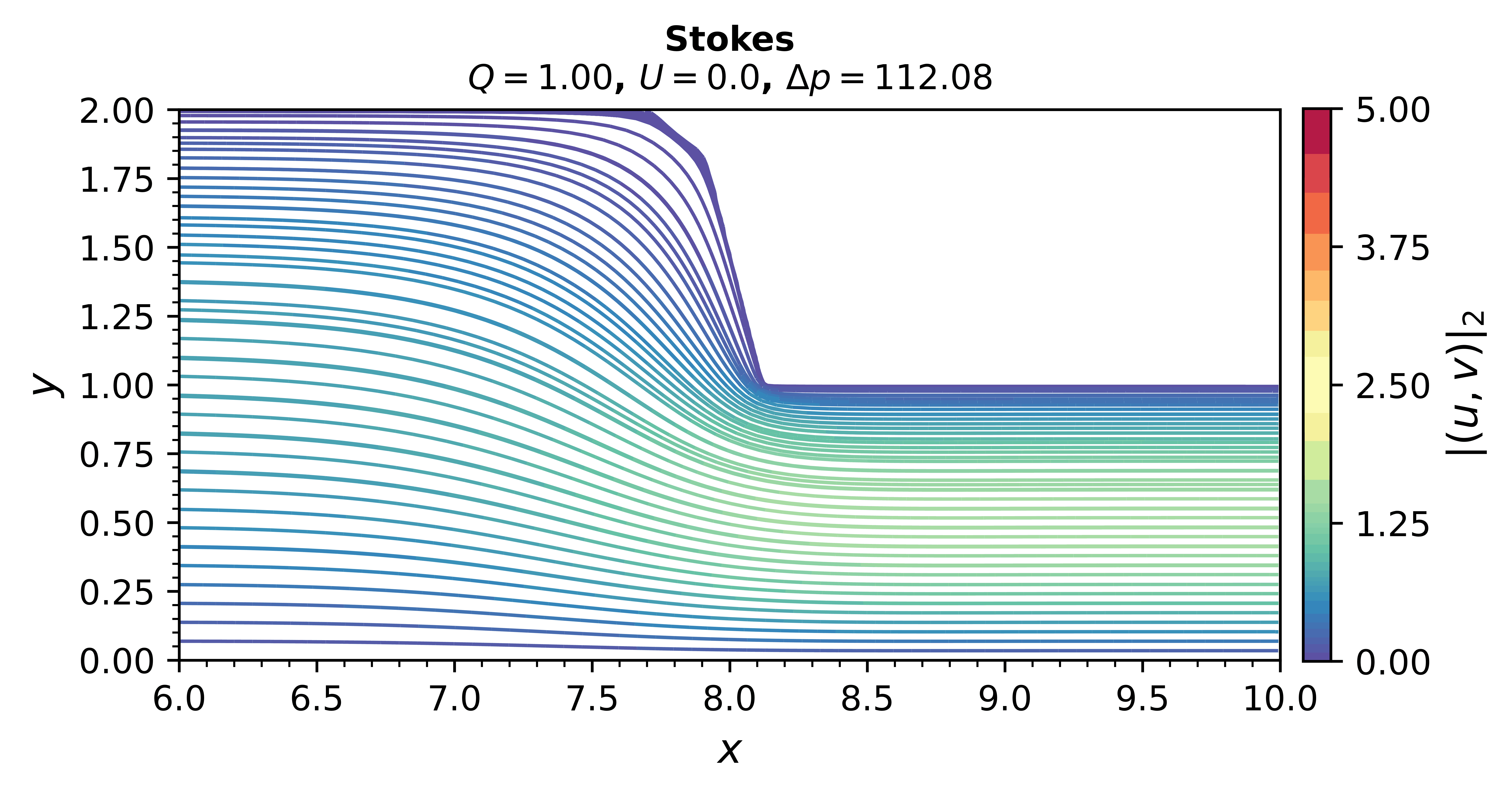}}
 \subfloat[Reynolds $\delta=\tfrac{1}{4}$]{\includegraphics[width=.45\textwidth]{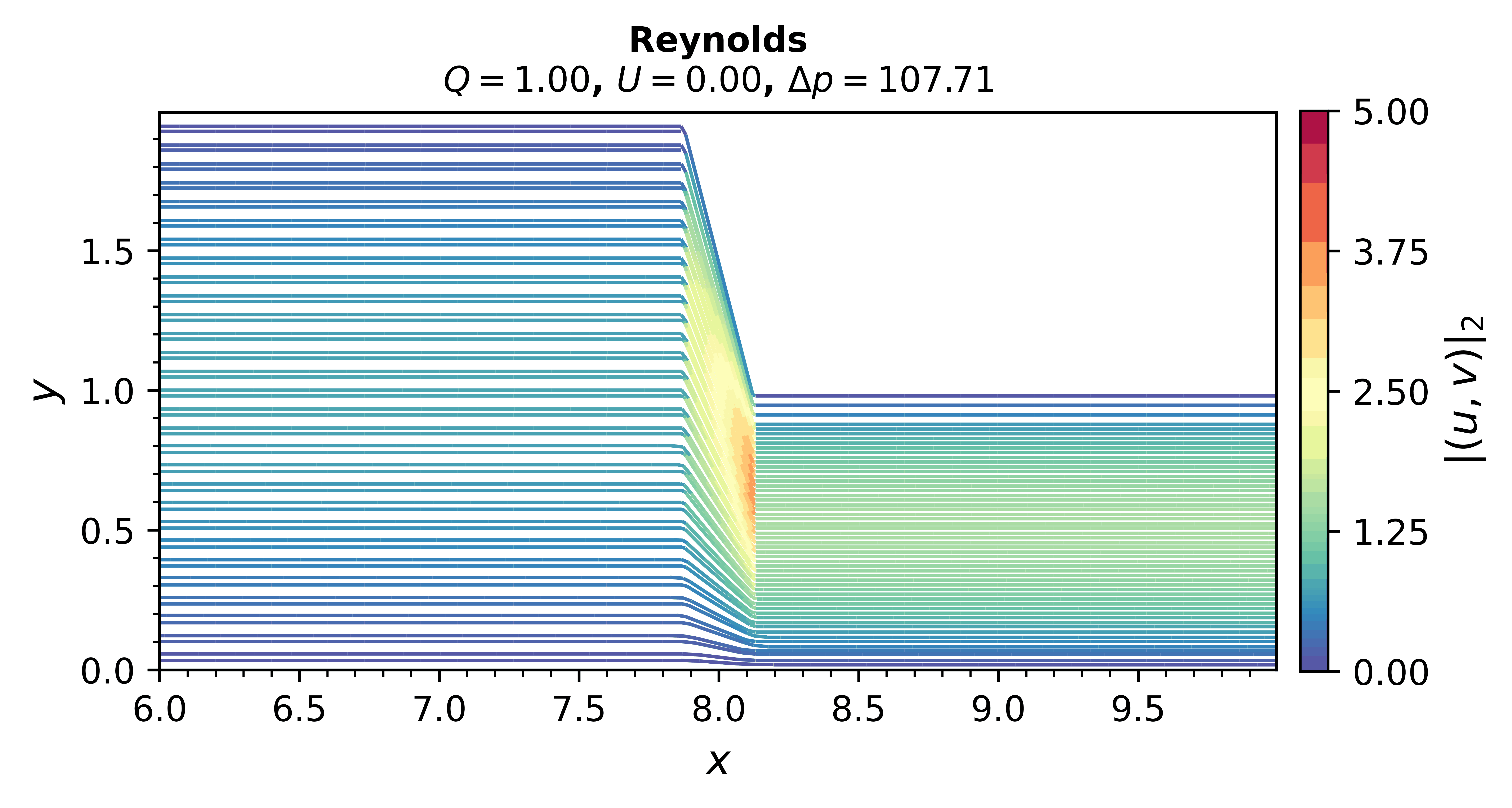}}
 
 \subfloat[Stokes $\delta=\tfrac{1}{8}$]{\includegraphics[width=.45\textwidth]{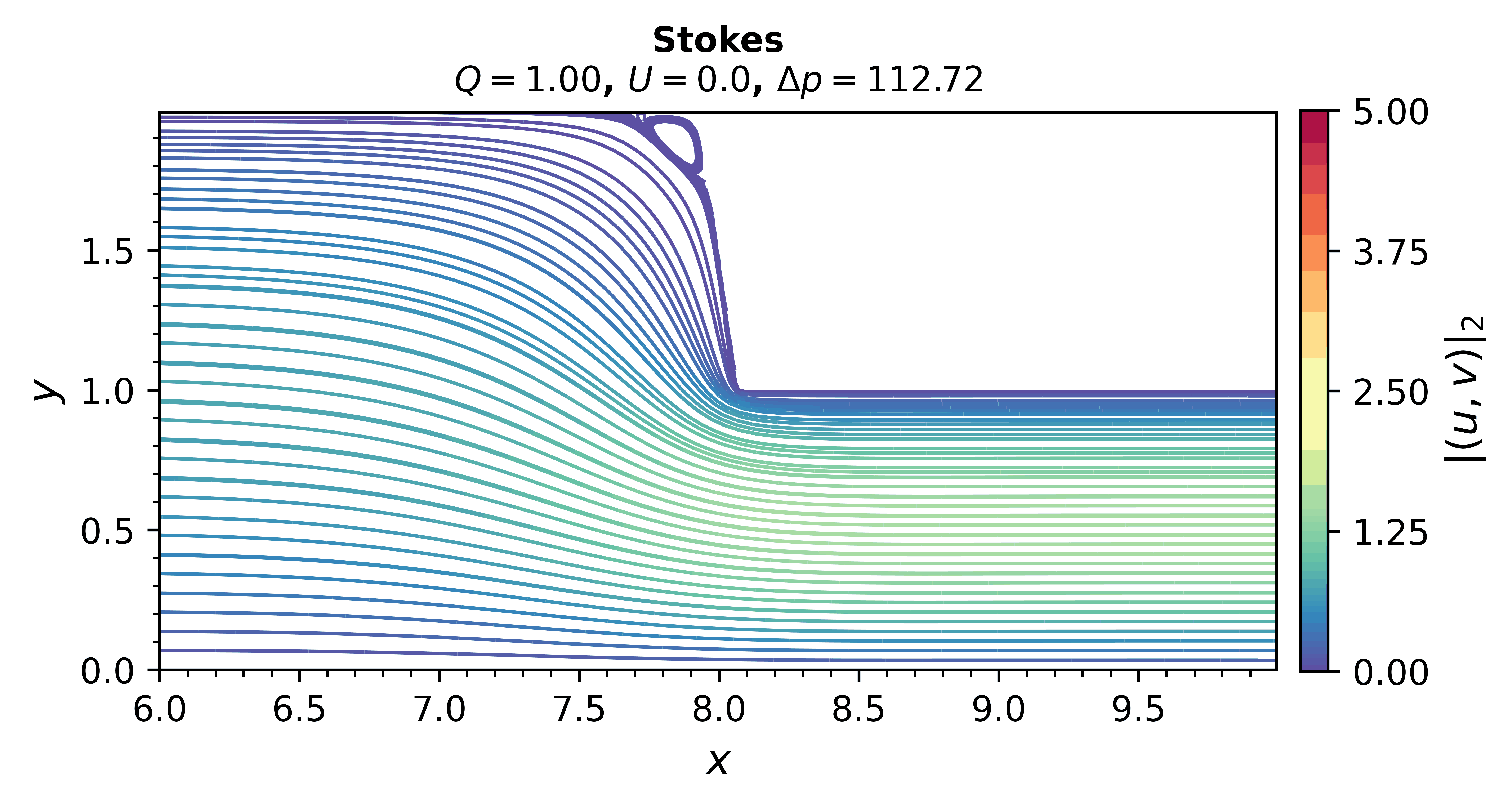}}
 \subfloat[Reynolds $\delta=\tfrac{1}{8}$]{\includegraphics[width=.45\textwidth]{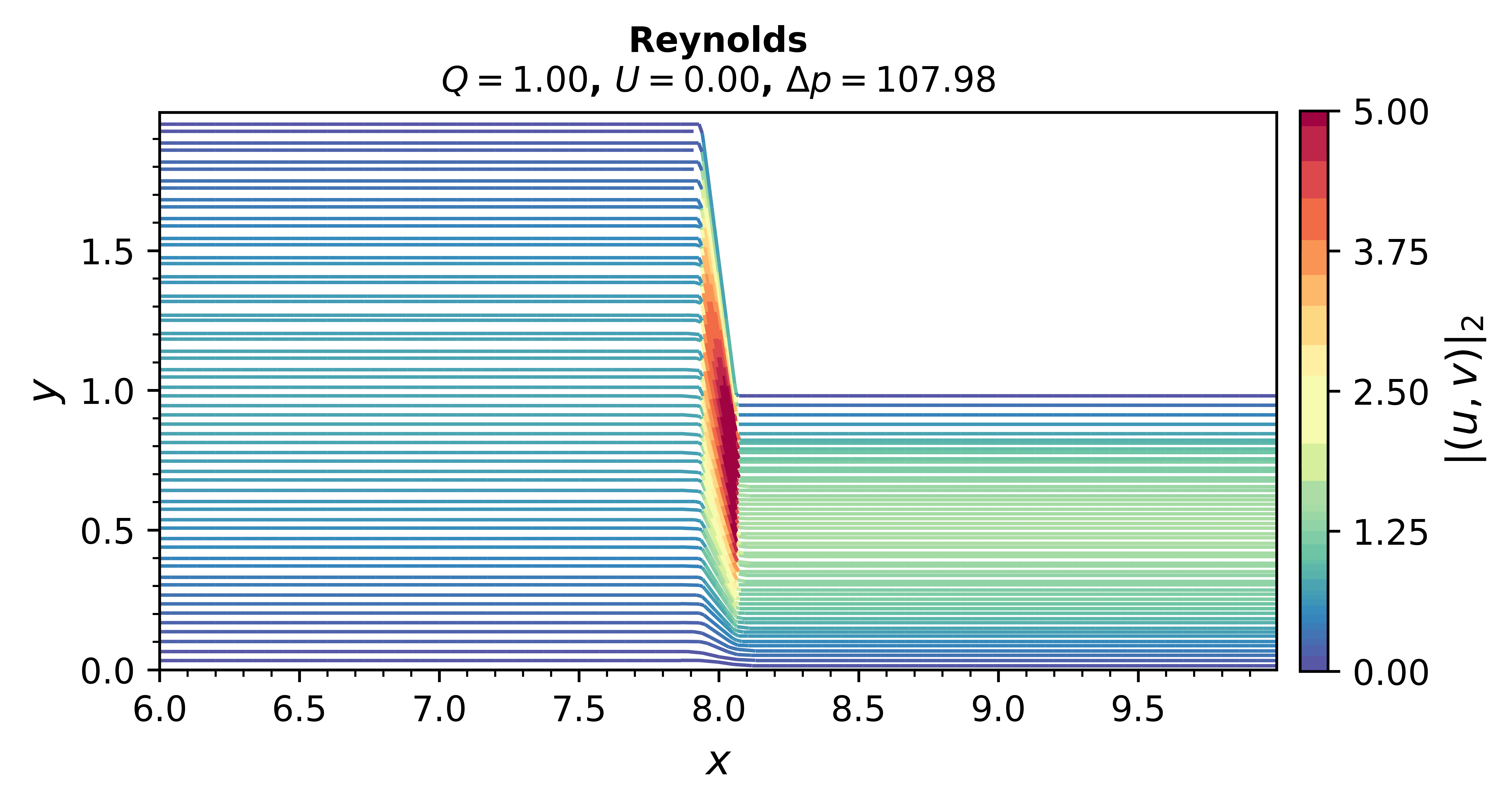}}
 
 \caption{Velocity streamlines for regularized BFS, $\mathcal{H}=2$. In the Stokes solutions, recirculation increases as $\delta \to 0^+$. In the Reynolds solutions, the magnitude of velocity down the slope is larger than in the corresponding Stokes solution}\label{slope_vel}
\end{figure}

The velocity streamlines for the Reynolds and Stokes solutions to the regularized BFS are presented in \cref{slope_vel} for $\mathcal{H} =2$ and various $\delta$. As with the BFS example, the velocity solutions from the Stokes and Reynolds equations vary significantly. Due to the discontinuity in the surface gradient, the Reynolds solution for velocity has continuous $u$ but discontinuous $v$. The Reynolds solution overestimates the magnitude of velocity in the vicinity of large surface gradients, particularly when the large surface gradient occurs in a region with minimal height. In contrast, for sufficiently large slopes, the Stokes solution yields flow separation and corner flow recirculation with similar structure to that of the BFS. \cref{slope_v_err} depicts the relative percent error in the $l_2$ norm of velocity from the Stokes and Reynolds equations. As with the error in pressure, we find that the error in velocity increases with increasing expansion ratio $\mathcal{H}$ and with increasing slope of expansion $(H_\text{in}-H_\text{out})/\delta$. These results highlight how both the magnitude of surface gradient and the expansion ratio are important factors in the validity of the Reynolds equation.

\begin{figure}[h]
 \centering 
 \subfloat[$\delta=\tfrac14$]{\includegraphics[width=0.3\textwidth]{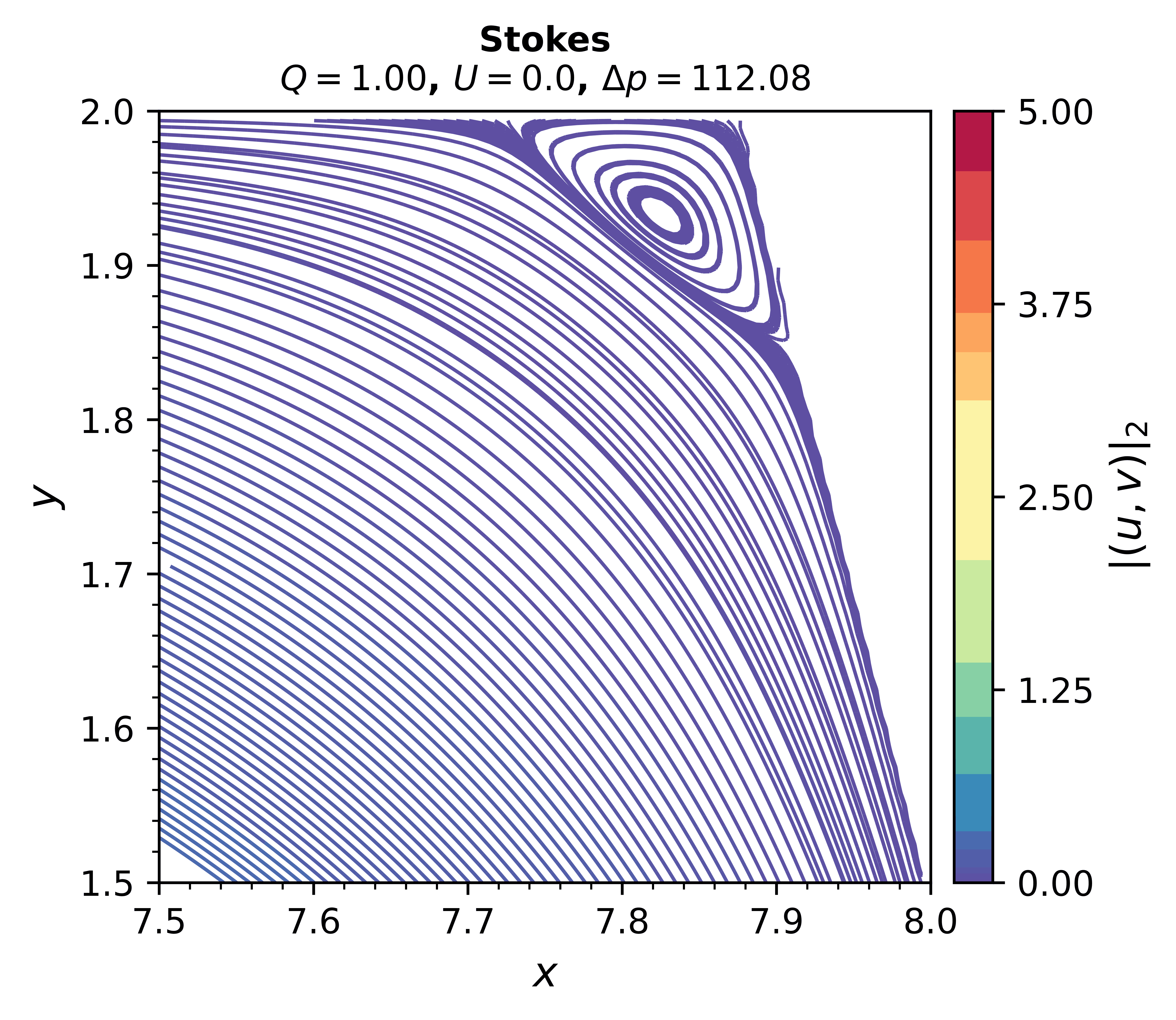}}
 \subfloat[$\delta=\tfrac18$]{\includegraphics[width=0.3\textwidth]{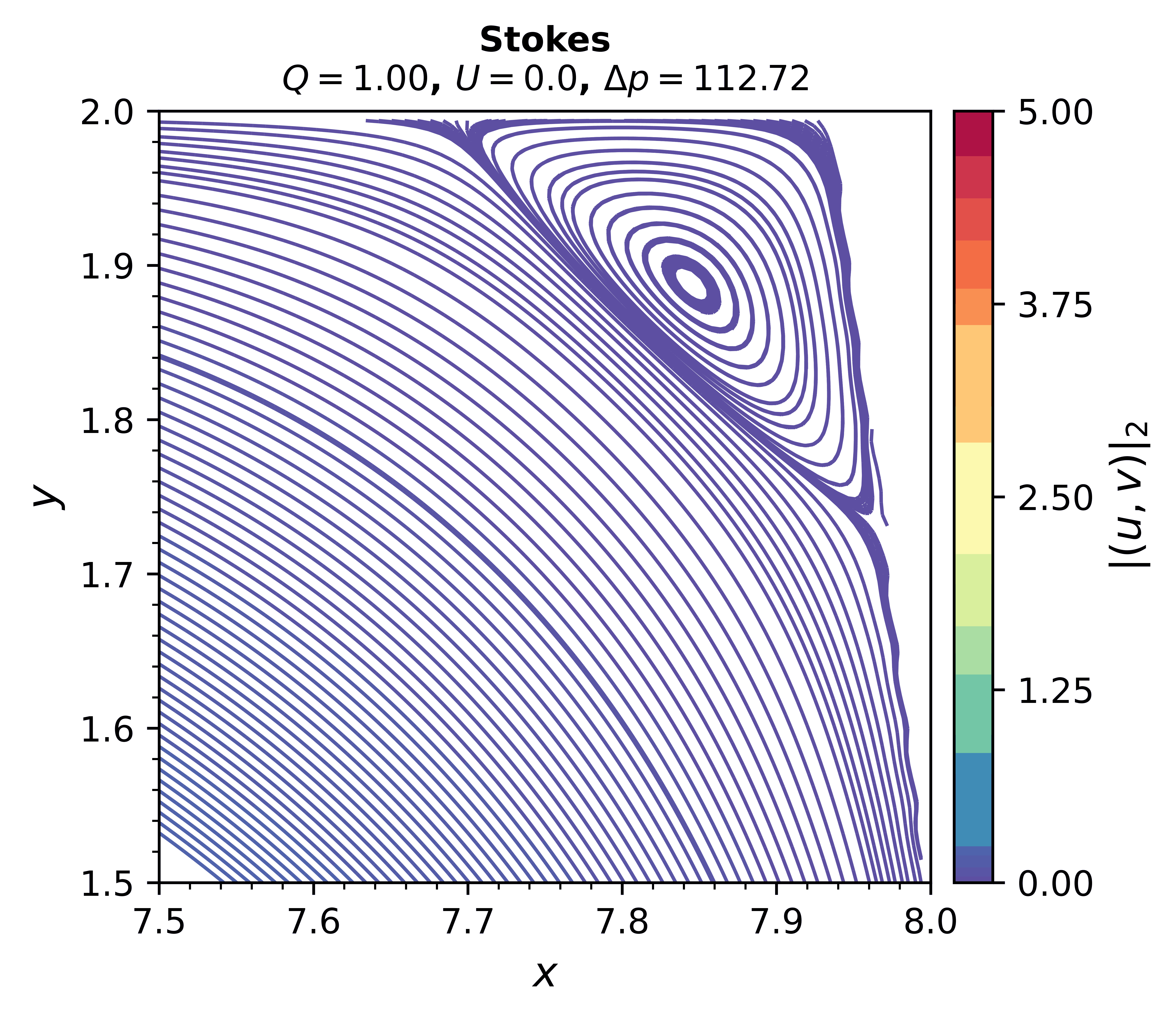}}
 \subfloat[BFS ($\delta = 0$)]{\includegraphics[width=0.3\textwidth]{imgs/slope/stokes_bfs_h2_v_corner.png}}
 
 \caption{Velocity streamlines for the regularized BFS observing the inner corner, Stokes solutions for $\mathcal{H}=2$ and slope of expansion $1/\delta$. Recirculation increases with increasing slope.}\label{slope_vel_zoom}
\end{figure}

\begin{figure}[h]
 \centering 
 \includegraphics[width=.75\textwidth]{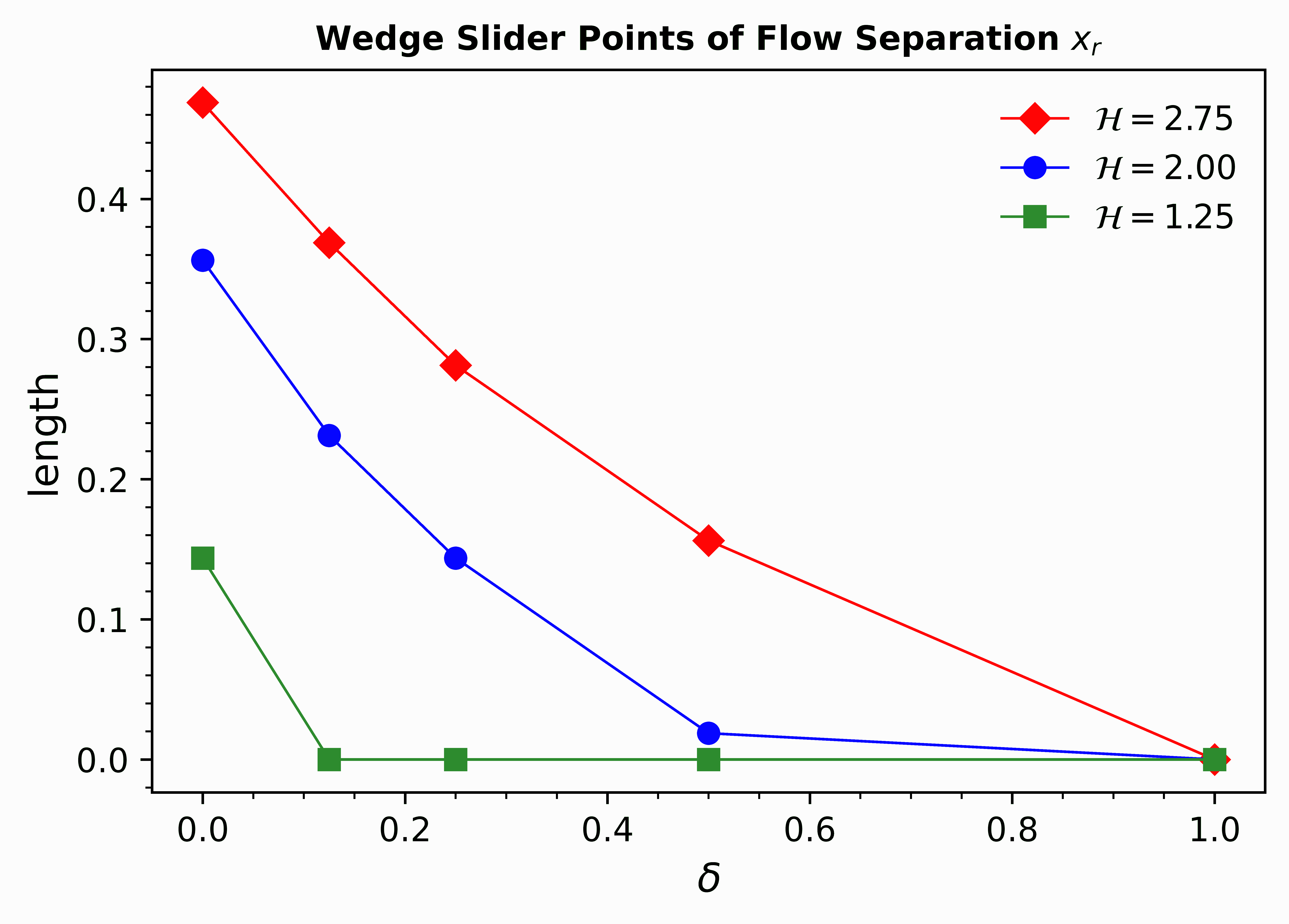}
 \caption{Points of flow separation $(\tfrac{1}{2}(L-\delta)-x_\text{r},H_\text{in})$ in the regularized BFS, Stokes solutions. The corner recirculation region increases in size with expansion ratio $\mathcal{H} = H_\text{in}/H_\text{out}$ and with increasing slope of expansion $(H_\text{in}-H_\text{out})/\delta$.} \label{slope_attachments}
\end{figure}

We now further assess corner flow recirculation in the Stokes solution to the regularized BFS. Close-up views of the concave are shown in \cref{slope_vel_zoom} for $\mathcal{H}=2$ and various $\delta$, including the BFS case of $\delta=0$. The change in the point of flow separation $(\tfrac{1}{2}(L-\delta)- x_r, H_\text{in})$ is charted in \cref{slope_attachments} for various $\mathcal{H}$ as a function of $\delta$. The size of the recirculation zone increases with increasing expansion ratio $\mathcal{H}= H_\text{in}/H_\text{out}$ and with increasing slope of expansion $\big(H_\text{in}-H_\text{out}\big)/\delta$. 

We observe corner flow recirculation in the Stokes solution to the regularized BFS for slopes approximately $\big(H_\text{in}-H_\text{out}\big)/\delta > 2$. This gives the critical angle $\theta < 117^\circ$ at the corner $(\tfrac{1}{2}(L-\delta)- x_r, H_\text{in})$. Note that this is smaller than the critical angle $\theta < 146^\circ$ determined analytically for flow recirculation in the triangular cavity \cite{dean_steady_1949,moffatt_viscous_1964}; the angle $\theta < 146^\circ$ corresponds to wedge slopes $\big(H_\text{in}-H_\text{out}\big)/\delta > 2/3$. This discrepancy in the critical angle is likely due to the small size of the flow recirculation at large angles, being too small to capture with grid resolutions at our disposal. 

\subsection{The triangular cavity}
\begin{figure}[h]
 \centering 
 \includegraphics[width=.75\textwidth]{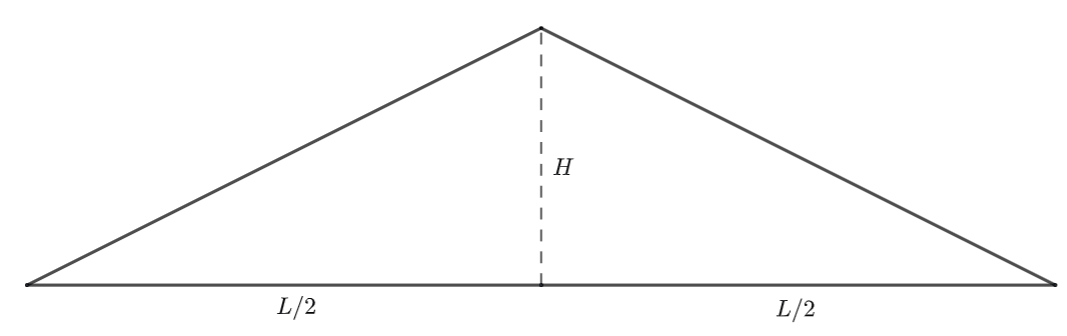}
 \caption{Schematic of the triangular cavity with height $H$ and length $L$.} \label{schematic_tri}
\end{figure}

The following section considers the lid-driven triangular cavity. The triangular cavity is a classical example for viscous flows and studies on corner flow recirculation \cite{moffatt_viscous_1964,ribbens_steady_1994,biswas_hoc_2017}. The film height for the equilateral triangle is given by,
\begin{equation}
 h(x) = \begin{cases}
 \frac{H}{L/2}x &\hspace{2em} 0\le x \le L/2\\
 \frac{-H}{L/2}(x-L) &\hspace{2em} L/2 \le x \le L
 \end{cases},
\end{equation}
where $H$ gives the triangle height and $L$ gives the triangle length; a schematic for the triangular cavity is shown in \cref{schematic_tri}. We compare the Reynolds and Stokes solutions to the triangular cavity for $0.35\le H \le 4$  and $L=2$. The boundary conditions $\mathcal{U}=1$ and $\mathcal{Q}=0$ are kept constant. 

Note that due to the discontinuity in the velocity boundary condition, the pressure gradients in the triangle corners $(0,0)$ and $(L,0)$ are very large and are not numerically stable, for both the Reynolds and Stokes solutions. Hence we consider only the solutions of velocity for the triangular cavity, which we compute independently of the pressure. 

\begin{figure}[h]
 \centering 
 \subfloat[Stokes $H=4$]{\includegraphics[width=.45\textwidth]{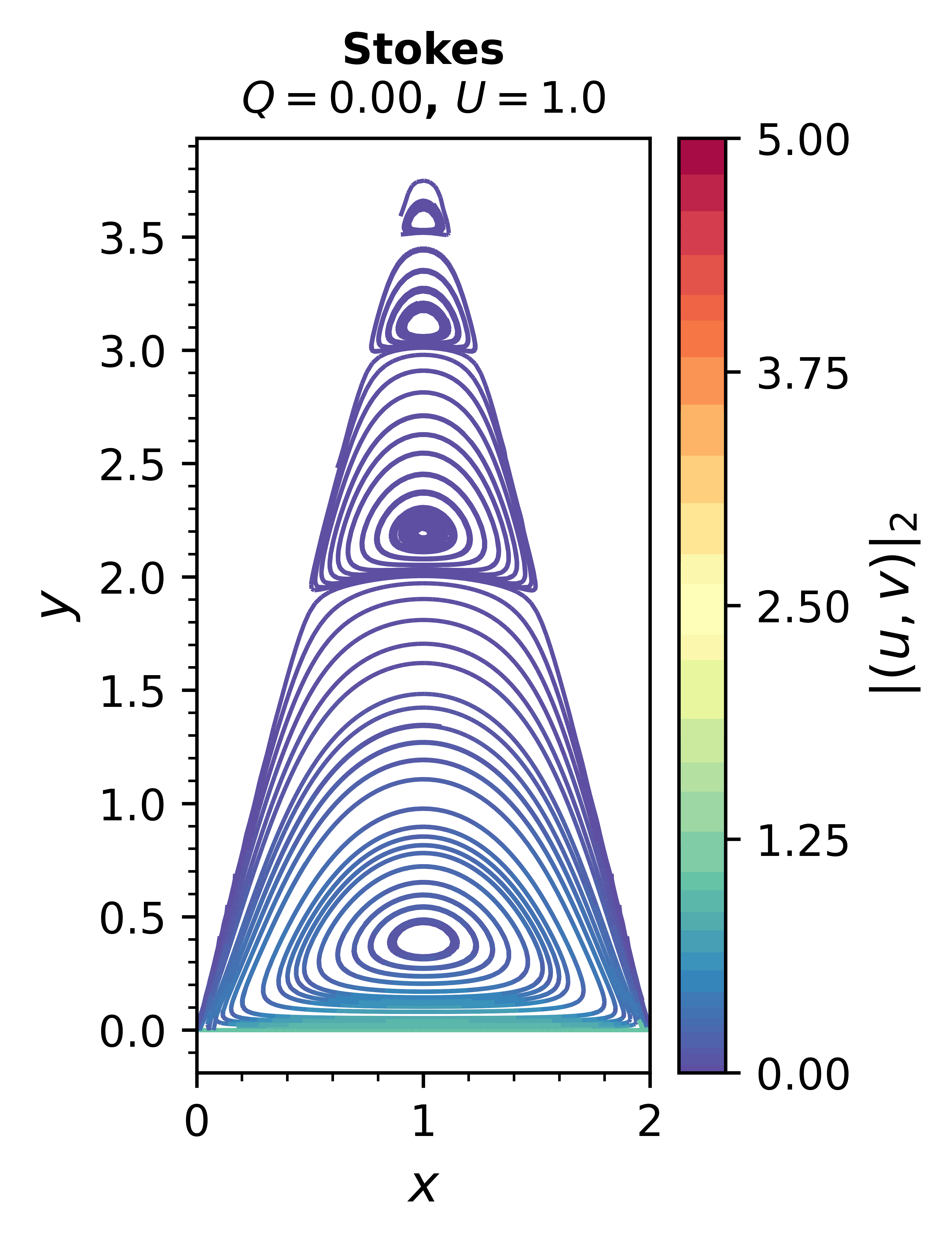}}
 \subfloat[Reynolds $H=4$]{\includegraphics[width=.45\textwidth]{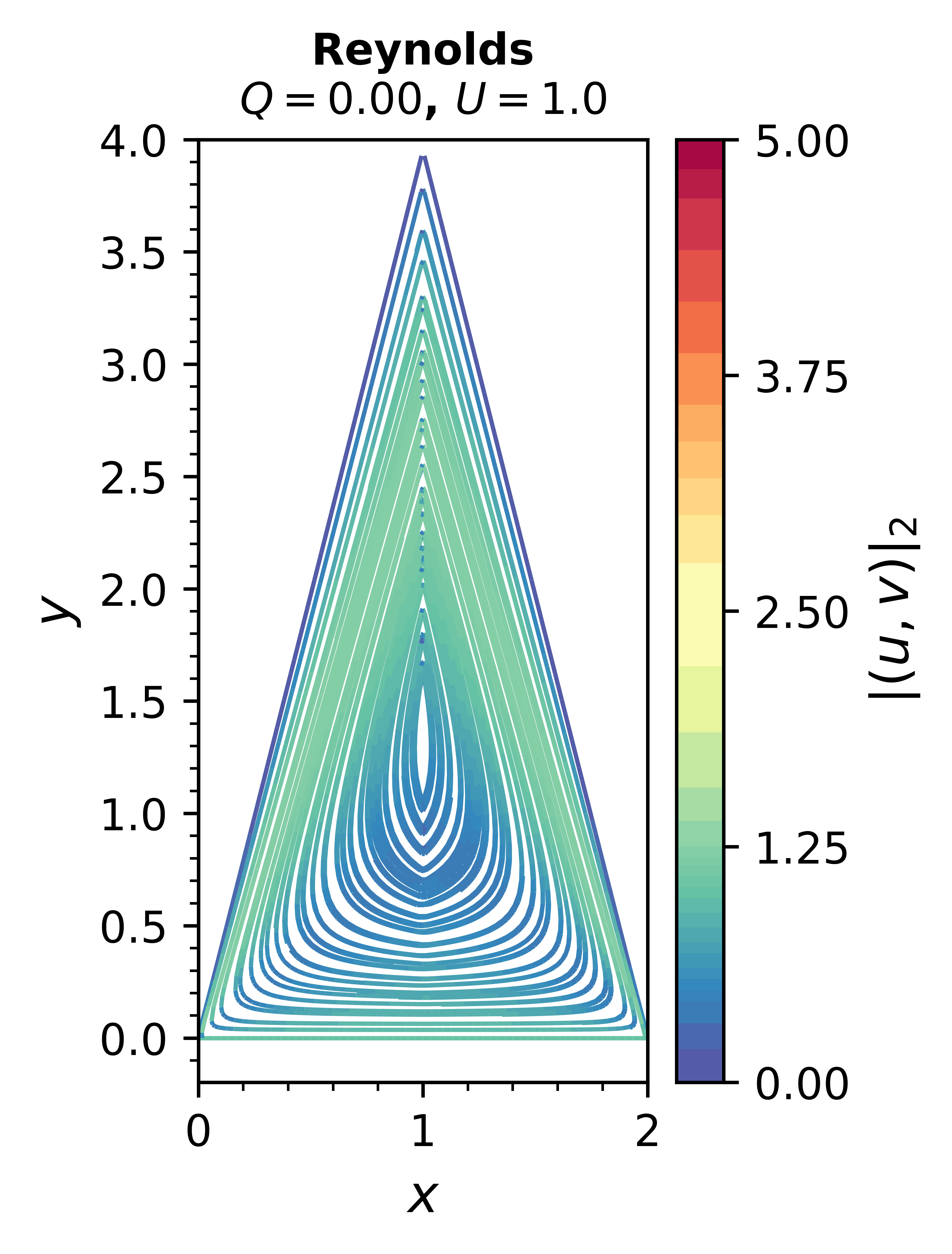}}
 
 \subfloat[Stokes $H=2$]{\includegraphics[width=.45\textwidth]{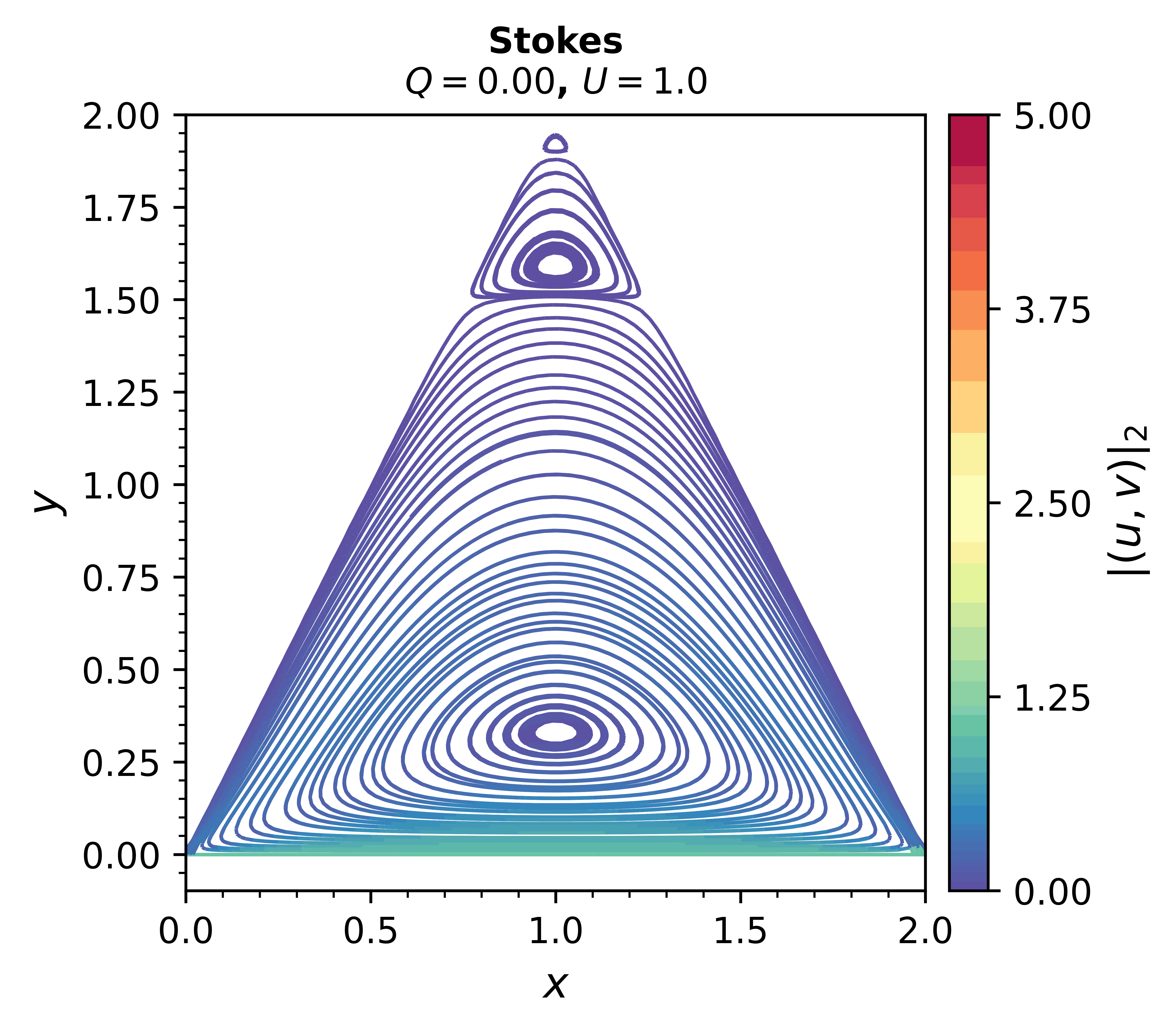}}
 \subfloat[Reynolds $H=2$]{\includegraphics[width=.45\textwidth]{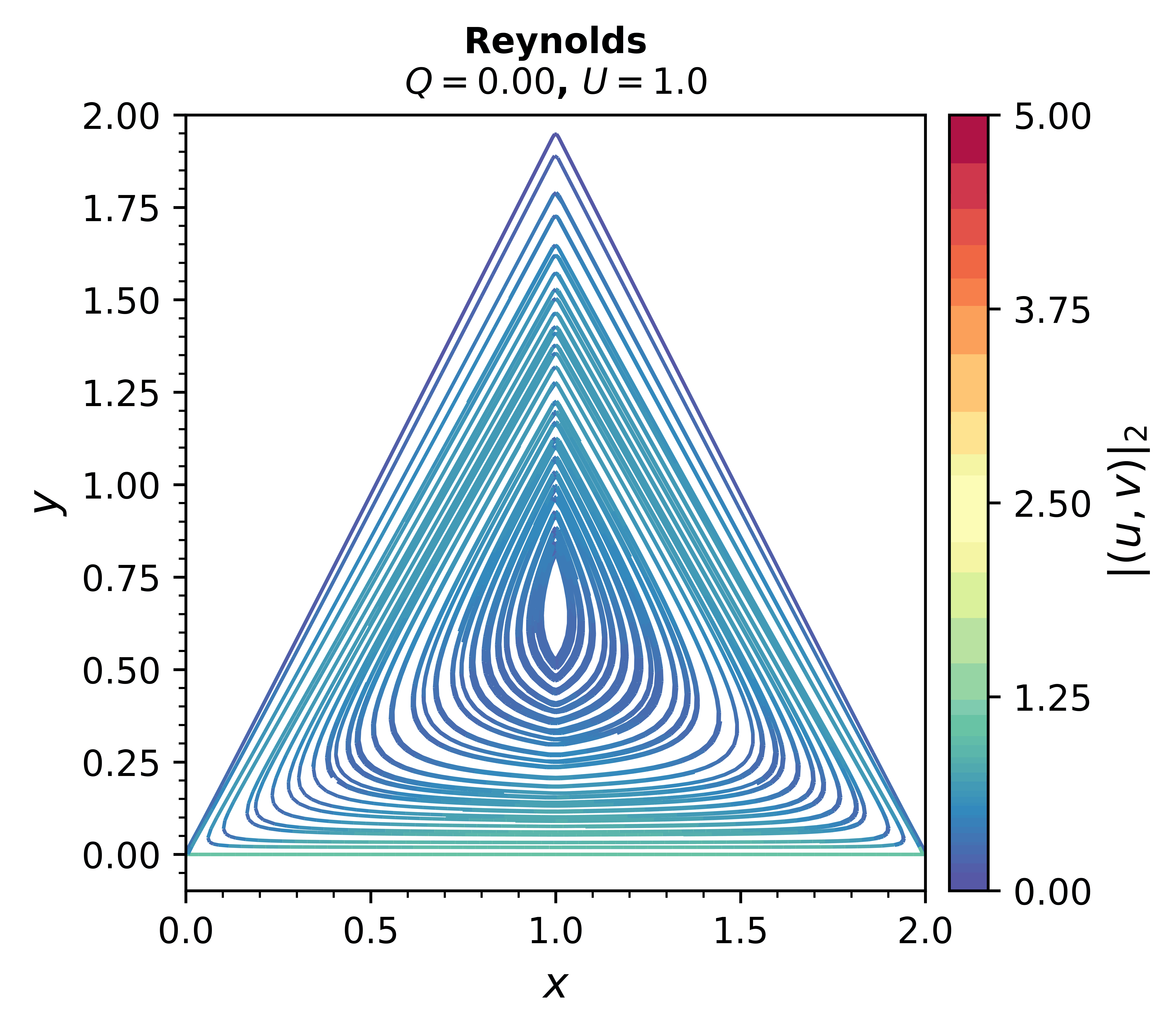}}
 \caption{Velocity streamlines for the lid-driven triangular cavity. In the Stokes solutions, a sequence of corner recirculation zones is observed; the points of flow separation lie closer to the triangle apex as the angle at the apex increases.}\label{tri_vel}
\end{figure}

\begin{figure}[h]
 \centering 
    \includegraphics[width=0.75\textwidth]{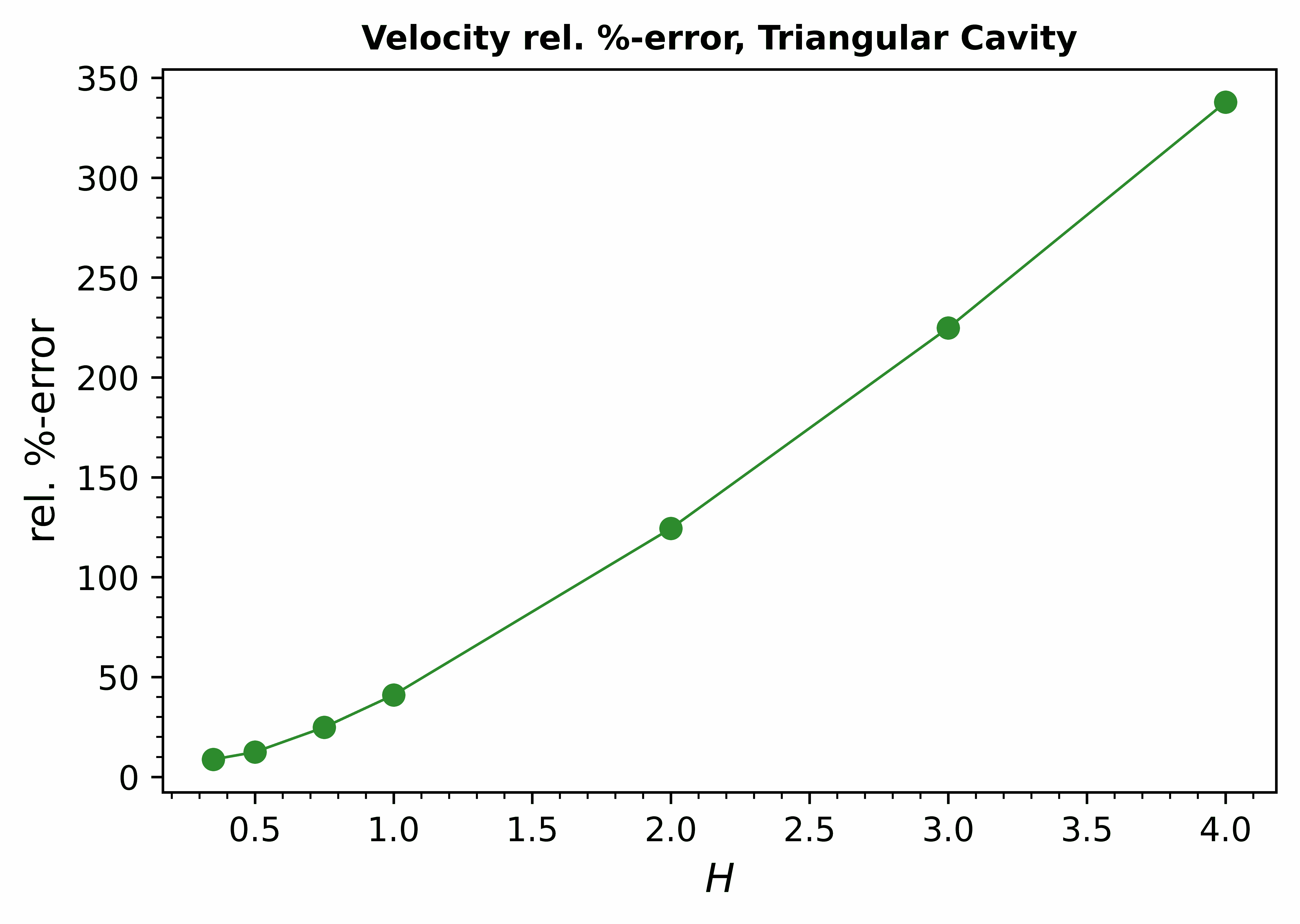}
 \caption{Relative percent error in $l_2$ norm of velocity for the triangular cavity, Stokes and Reynolds solutions. Error increases with increasing height of the triangle.}\label{tri_v_err}
\end{figure}

The velocity streamlines for the Reynolds and Stokes solutions to the triangular cavity are presented in \cref{tri_vel} for various $H$. The relative percent error in $l_2$ norm of velocity between the Reynolds and Stokes solutions is shown in \cref{tri_v_err}, the error increases rapidly with increasing height of the triangle. For sufficiently large $H$, the Stokes solution to the triangular cavity depicts a sequence of recirculation zones in the corner $(H, L/2)$. As with previous examples, the discontinuity in the surface gradient means that the Reynolds velocity is discontinuous. However, due to the surface gradient having continuous magnitude for the equilateral triangle, the Reynolds velocity has continuous magnitude, and the streamlines are continuous (although not smooth). Hence, the bulk flow recirculation seen in the Reynolds solution is not comparable to that of the Stokes solution. Moreover, the Reynolds solution does not depict flow separation or secondary recirculation.

The points of flow separation in the triangular cavity, Stokes solutions, are tabulated in \cref{tri_points} for various triangle heights $H$. In Stokes flow for the equilateral triangle, the flow separation points are symmetric; we list only the points $(x_k,y_k)$ on the left arm of the triangle $x_k < L/2$, and each has a partner point $(L-x_k, y_k)$ on the right arm of the triangle. Let $k=1$ denote the primary separation point, furthest from the triangle apex, separating the bulk recirculation from the secondary recirculation. At $H=4$ and $H=2$, we observe up to $k=5$ flow separation points, making 6 total recirculation regions. At $H=1$, we observe only $k=2$ flow separation points, for 3 total recirculation regions. In \cref{tri_points}, we include the flow separation points found by Biswas \& Kalita (2017) \cite{biswas_hoc_2017} for the triangular cavity at $\text{Re}=1$ with $H=4$, which show good agreement with our results at $\text{Re}=0$. In general, as the triangle increases in height (decreasing the angle at the triangle apex), the points of flow separation $(x_k,y_k)$ lie closer the triangle apex and the corner recirculation zones decrease in size.

\begin{table}[h]
\centering
\begin{tabular}{|c|c|c|c|c|}
\hline
& $H=4$ ($\text{Re}=1$, \cite{biswas_hoc_2017}) & $H=4$ & $H=2$ & $H=1$ \\ \hline\hline
$(x_1,y_1)$  & $(0.477, 1.919)$ & $(0.481,  1.925)$ & $(0.463,  0.925)$ & $(0.756, 0.756)$ \\ \hline
$(x_2,y_2)$  & $(0.739, 2.966)$ & $(0.744, 2.975)$  & $(0.494, 0.988)$  & $(0.963, 0.963)$ \\ \hline
$(x_3, y_3)$ & $(0.871, 3.490)$ & $(0.875, 3.50)$   & $(0.575, 1.150)$  & \\ \hline
$(x_4, y_4)$ & $(0.934, 3.748)$ & $(0.9375, 3.75)$  & $(0.750, 1.50)$   & \\ \hline
$(x_5, y_5)$ & $(0.969, 3.880)$ & $(0.975, 3.90)$   & $(0.95, 1.90)$    & \\ \hline
\end{tabular}
\vspace{.5em}
\caption{Flow separation points $(x_k, y_k)$ in the triangular cavity of varying heights $H$}\label{tri_points}
\end{table}

We observe flow separation and secondary recirculation in the Stokes solution to the triangular cavity for approximately $H > 0.7$ on length $L=2$. This corresponds to a critical angle of approximately $\theta < 110^\circ$, very close to the critical angle $\theta < 117^\circ$ we observed for the regularized BFS. We again highlight that for angles in the upper range of $\theta < 146^\circ$, (where $\theta = 146^\circ$ is the threshold for corner recirculation in Stokes flow determined analytically \cite{dean_steady_1949,moffatt_viscous_1964}), the corner recirculation region is very small, and cannot be observed with grid resolutions at our disposal. 

\section{Conclusions}
Here we have compared the solutions of lubrication theory and Stokes flow for various corner geometries, and examined corner flow recirculation in the Stokes solutions. The Reynolds equation was solved using an exact solution for piecewise linear domains, \cite{dennis_fast_2026}. The Stokes equation was solved using an iterative finite difference method \cite{sen_4oec_2015}, and we detailed a boundary approximation for this method in non-rectilinear domains. The Reynolds equation of lubrication theory is based on the assumption of a long and thin fluid domain and a small Reynolds number. As such, the lubrication approximation is highly sensitive to large surface gradients and sharp corners in the domain, even at low Reynolds numbers. Whereas, the Stokes approximation is not sensitive to the length scale ratio, and is based only in the assumption of zero Reynolds number. Hence the discrepancies between solutions to the Reynolds and Stokes equations characterize the extent to which the lubrication assumptions are applicable for each geometry. 

In the BFS example, the error between the Reynolds and Stokes solutions increases with increasing the expansion ratio. The sudden expansion at the step induces non-negligible cross-film pressure gradients not preserved in the lubrication approximation. As a result, the average pressure drop from the Stokes solutions is more extreme than from the Reynolds solution.  Moreover, the Stokes solution depicts flow separation and corner flow recirculation. We charted points of flow separation in the BFS and observed how the corner separated region increases in size with increasing expansion ratio. Flow separation may be regarded as a response to the significant cross film pressure gradients, and hence this phenomena is not observed in the lubrication approximation. 

We then considered the Stokes solutions to the wedged BFS; the concave corner of the step was replaced by a triangular wedge to fully or partially occlude the corner recirculation zone. The bulk flow structure and average pressure drop from the Stokes solutions were consistent for various wedge sizes. At low Reynolds numbers, regions of corner flow recirculation are very slow relative to the bulk flow; it may be desirable in applications to minimize such regions where fluid is stagnant. Our results indicate that corner geometries may be reshaped to minimize regions of flow recirculation without otherwise disrupting the bulk flow characteristics.  

As another example and variation on the BFS, we considered the regularized BFS. We observed that error between the Reynolds and Stokes solutions increased with increasing expansion ratio, and with increasing magnitude of surface gradient. Moreover, we observed corner flow recirculation in the Stokes solutions to the regularized BFS and charted the points of flow separation. For each expansion ratio, $\mathcal{H} = H_\text{in}/H_\text{out}$, the corner recirculation zone was significant for slopes $(H_\text{in}-H_\text{out})/\delta > 2$, increasing in size with increasing expansion ratio and magnitude of surface gradient. The results for the regularized BFS are consistent with those for the BFS, and highlight the importance of considering both the expansion ratio and the magnitude of surface gradient when selecting a fluid model.

Finally, we considered the lid-driven triangular cavity. The solutions of velocity from the Stokes and Reynolds equations to the triangular cavity are significant and increase with increase height on the triangle. For sufficiently tall triangles, the Stokes solution depicts a sequence of recirculation zones in the apex of the triangle. We charted the successive points of flow separation at various heights, and found that our results for $H=4$ and $\text{Re}=0$ are in good agreement with the results of \cite{biswas_hoc_2017} for $\text{Re}=1$. 

For the regularized BFS and the triangular cavity, we observed recirculation in corners with an angle approximately $\theta < 110$. This critical angle is slightly less than the critical angle $\theta < 146^\circ$ determined analytically for corner recirculation in the Stokes equations \cite{dean_steady_1949, moffatt_viscous_1964}. However, the size of the recirculation zone decreases as the angle at the corner increases, thus at these obtuse angles the corner recirculation is too small to capture with grid resolutions at our disposal. 

In all geometries considered here, the error in velocity between the Reynolds and Stokes solutions was significant. And for the BFS and regularized BFS, the error in velocity was consistently much greater than the error in pressure. Because the Reynolds velocity $u$ involves the height $h$, and $v$ involves the height gradient $\frac{dh}{dx}$, the Reynolds velocity has discontinuities at the discontinuities in $h$ or $\frac{dh}{dx}$. Not only does this prevent the Reynolds solution from capturing corner flow recirculation, as seen in the regularized BFS, the discontinuity in $\frac{dh}{dx}$ occurring at the minimum in $h$ caused significant blow-up in the velocity magnitude. In a previous work \cite{dennis_comparison_2026}, we considered models in extended lubrication theory, in which the lubrication assumptions are relaxed leading to the inclusion of higher order terms in the governing equations. While such models are not applicable to fully discontinuous geometries such as the BFS, they do provide significant improvements to the error in velocity and pressure for geometries with a continuous height profile, and can serve as a valuable intermediary model between the Reynolds and Stokes equations.

In conclusion, we find that large surface gradients are a primary contributor to error between the Reynolds and Stokes solutions. Moreover, the solutions to the Reynolds equation do not capture corner flow recirculation which is a significant feature of Stokes flow for corner geometries. In future work, the performance of the lubrication assumptions may be evaluated for various other geometries. In particular, it is not well understood how the lubrication assumptions perform for a sequence of texturings, particularly when we consider that successive textures share the length scale. Furthermore, given that the thin film assumption is often used to justify use of the Reynolds equation at non-zero Reynolds numbers, it would be a valuable avenue of future work to compare the Reynolds equation with the Navier-Stokes equations at low to moderate Reynolds numbers in corner domains. 
It may be that the geometry is long and thin, yet local surface discontinuities violate the lubrication assumptions, and thus contribute to significant errors in the lubrication model.

\section*{Acknowledgments}We thank Michael Fillon for providing feedback on an early version of this manuscript. This work was supported by National Science Foundation (NSF) grant DMS-2512565 to TGF. We also acknowledge use of the Brandeis High Performance Computing Cluster (HPCC) which is partially supported by the NSF through DMR-MRSEC 2011846 and OAC-1920147.

\appendix

\section{Exact solution to the Reynolds equation for piecewise linear geometries}
In our previous work \cite{dennis_fast_2026}, we presented a fast solver for the Reynolds equation, which is an exact solution for piecewise linear heights, like those we have considered here. A brief outline of the method is presented below. 

Suppose $h(x)$ defined on $[x_0,x_N]$ is piecewise linear with $N$ components. Let $\{x_k\}_0^N$ denote the critical points of $h(x)$ demarcating each piecewise linear component. Let $\{h_{k^\pm}\}_{k=0}^N$ denote the endpoints of $h(x)$ on each interval, $h_{k^\pm} = \lim_{x\to x_k^{\pm}}h(x)$. And let $\{\frac{\Delta h}{\Delta x}\big|_k\}_{k=0}^{N-1}$ denote the constant gradients of $h(x)$ for each interval, \begin{align}
    \frac{\Delta h}{\Delta x}\bigg|_k = \frac{h_{(k+1)^-}-h_{k^+}}{x_{k+1}-x_k} = \lim_{x\to x_k^+}\frac{dh}{dx}.
\end{align}
The solution to the Reynolds equation for a piecewise linear height is derived from coupling the solutions of each sub-interval. 

For a single interval $[x_k,x_{k+1}]$ on which $h(x)$ is piecewise linear, $p(x)$ satisfies,
\begin{equation}\label{pwl_p}
    p(x) = \begin{cases}
               -\Bigg(\tfrac{1}{2}\mathcal{C}_Q \Big[h(x)\Big]^{-2} + 6 \eta \mathcal{U} \Big[h(x)\Big]^{-1}\Bigg)\Big[\frac{\Delta h}{\Delta x}\Big|_k\,\Big]^{-1}  + \mathcal{C}_{P_k} &\frac{\Delta h}{ \Delta x}\Big|_k \ne 0\\
    \Big(\mathcal{C}_Q \Big[h(x)\Big]^{-3}+ 6 \eta \mathcal{U} \Big[h(x)\Big]^{-2}\Big)(x-x_k) + \mathcal{C}_{P_k}  & \frac{\Delta h}{\Delta x}\Big|_k = 0 \end{cases}.\end{equation}
The constant $C_Q$ arises in the first integration of the Reynolds equation and is directly proportional to the flux, $\mathcal{C}_Q =-12\eta \mathcal{Q}$. The constant $C_{P_k}$ arises in the second integration of the Reynolds equation and relates to the fixed endpoint pressure. 

For $h(x)$ with $N$ piecewise linear components, we couple solutions of the form \cref{pwl_p}. The fixed outlet pressure $p(x_L)=0$ determines $\mathcal{C}_{P_{N-1}}$,
\begin{equation}\label{boundary_cp}\mathcal{C}_{P_{N-1}} = \begin{cases}  
\mathcal{C}_Q\tfrac{1}{2}\frac{\Delta h}{\Delta x}\Big|_{N-1}^{-1}   h_{N^-}^{-2} +6 \eta \mathcal{U}\frac{\Delta h}{\Delta x} \Big|_{N-1}^{-1} h_{N^-}^{-1} &\frac{\Delta h}{ \Delta x}\Big|_{N-1}\ne 0\\
 -\mathcal{C}_Q  h_{N^-}^{-3}\Delta x|_{N-1} -6 \eta \mathcal{U}  h_{N^-}^{-2}\Delta x|_{N-1} & \frac{\Delta h}{\Delta x}\Big|_{N-1} = 0
\end{cases}.\end{equation} 
To solve for $\{\mathcal{C}_{P_k}\}_{k=0}^{N-2}$, we assume a continuous pressure and set $\lim_{x\to x_k^+} p(x) = \lim_{x\to x_k^-} p(x)$, where $p(x)$ is of the form \cref{pwl_p}.
Hence for each $0< k < N$, the integration constants $C_{P_k}$, $C_{P_{k-1}}$ and $C_Q$ satisfy,
\begin{multline} \label{cp cq cases}
    \begin{cases}
    \mathcal{C}_{P_{k}}- \mathcal{C}_{P_{k-1}} -\mathcal{C}_Q\bigg(\frac{1}{2}h_{k^+}^{-2}\frac{\Delta h}{\Delta x}\Big|_{k}^{-1} - \frac{1}{2} h_{k^-}^{-2} \frac{\Delta h}{\Delta x}\Big|_{k-1}^{-1} \bigg) \\
        \hspace{2em} = 6\eta \mathcal{U}\bigg(h_{k^+}^{-1}\frac{\Delta h}{\Delta x}\Big|_{k}^{-1} -h_{k^-}^{-1} \frac{\Delta h}{\Delta x}\Big|_{k-1}^{-1}   \bigg) 
            & \frac{\Delta h}{\Delta x}\big|_{k} \ne 0,   \frac{\Delta h}{\Delta x}\big|_{k-1} \ne 0  \\
   \mathcal{C}_{P_{k}}- \mathcal{C}_{P_{k-1}} -\mathcal{C}_Q\bigg(\frac{1}{2}h_{k^+}^{-2} \frac{\Delta h}{\Delta x}\Big|_{k}^{-1}  + h_{k^-}^{-3}\Delta x\Big|_{k-1}\bigg) \\
        \hspace{2em} = 6\eta \mathcal{U}\bigg( h_{k^+}^{-1}\frac{\Delta h}{\Delta x}\Big|_{k}^{-1}  + h_{k^-}^{-2}\Delta x\Big|_{k-1}  \bigg)
            & \frac{\Delta h}{\Delta x}\big|_{k} \ne 0,  \frac{\Delta h}{\Delta x}\big|_{k-1}= 0 \\
   \mathcal{C}_{P_{k}}- \mathcal{C}_{P_{k-1}} +\mathcal{C}_Q \frac{1}{2}h_{k^-}^{-2}\frac{\Delta h}{\Delta x}\Big|_{k-1}^{-1}\\
        \hspace{2em}= -6\eta \mathcal{U}h_{k^-}^{-1}\frac{\Delta h}{\Delta x}\big|_{k-1}^{-1}   
            &  \frac{\Delta h}{\Delta x}\big|_{k} = 0,\frac{\Delta h}{\Delta x}\big|_{k-1}\ne 0\vspace{.5em}\\
    \mathcal{C}_{P_{k}}- \mathcal{C}_{P_{k-1}} -\mathcal{C}_Q  h_{k^-}^{-3} \Delta x \big|_{k-1}\\
        \hspace{2em}= 6\eta \mathcal{U} h_{k^-}^{-2}\Delta x\big|_{k-1}  
            & \frac{\Delta h}{\Delta x}\big|_{k} = 0, \frac{\Delta h}{\Delta x}\big|_{k-1} = 0
    \end{cases}
\end{multline}

All together, $C_Q$ and $\{C_{P_k}\}_{k=0}^{N-1}$ constitute a size $N+1$ linear system of equations $\mathrm{M}{\bf x} = {\bf b}$ where ${\bf x} = [ \mathcal{C}_Q,\mathcal{C}_{P_0}, ...,  \mathcal{C}_{P_{N-1}} ]^T$. 
The matrix $\mathrm{M}$ has a block form,  \begin{equation}\mathrm{M} = \begin{bmatrix}
    1 & 0\\
   C & D
\end{bmatrix}.\end{equation} The identity row corresponds to $C_Q = -12\eta \mathcal{Q}$. The block $C$ is a size $N$ vector corresponding to the $C_Q$ coefficients from \cref{cp cq cases}. The block $D$ is a size $N$ bi-diagonal matrix with $1$ on the main diagonal and $-1$ on the upper diagonal, corresponding to the $C_{P_k}$ and $C_{P_{k-1}}$ coefficients from \cref{cp cq cases}. The inverse of $\mathrm{M}$ is, 
\begin{equation}\mathrm{M}^{-1} = \begin{bmatrix}
    1 & 0\\
    -D^{-1}C & D^{-1}
\end{bmatrix}.\end{equation} 
The matrix $D^{-1}$ is upper triangular with all $-1$ entries. The matrix vector product $-D^{-1}C$ is then the reversed partial sum of entries in $C$. Likewise, when evaluating $M^{-1}{\bf b}$, the lower block of ${\bf b}$ is multiplied by $D^{-1}$, resulting in the negative reversed partial sum of the entries in ${\bf b}$. These two steps constitute an $\mathcal{O}(N)$ time algorithm solving $\mathrm{M}{\bf x}={\bf b}$.
After solving for ${\bf x}$ comprising $C_Q$ and $\{C_{P_k}\}_{k=0}^{N-1}$, the pressure $p(x)$ on each region $\{[x_k,x_{k+1}]\}_{k=0}^{N-1}$ is obtained according to \cref{pwl_p}.

\section{Convergence}
\begin{figure}[h]
 \centering 
 \subfloat[BFS $\mathcal{H}=2$]{\includegraphics[width=.45\textwidth]{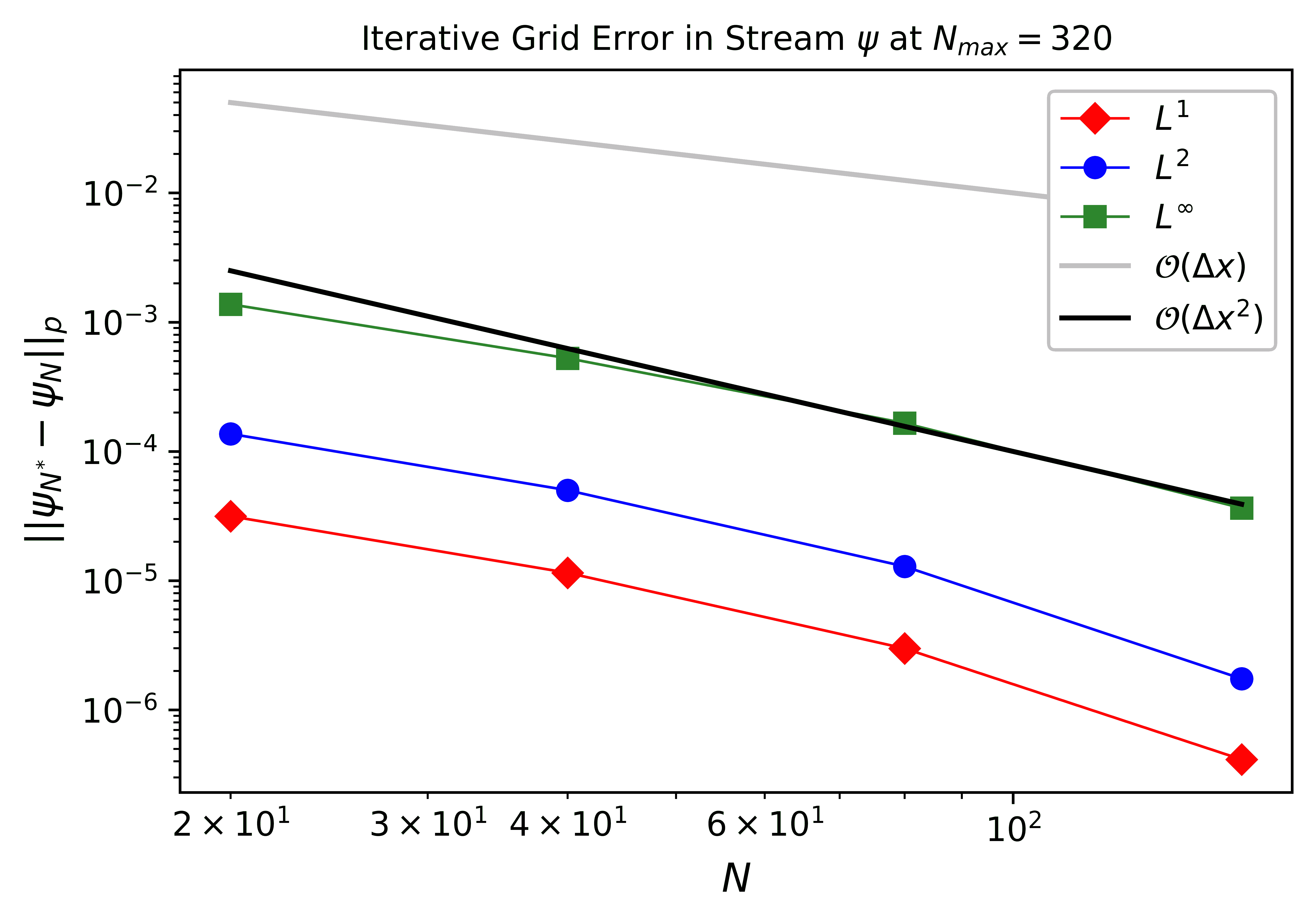}}
 \subfloat[Triangular cavity $H=4$]{\includegraphics[width=.45\textwidth]{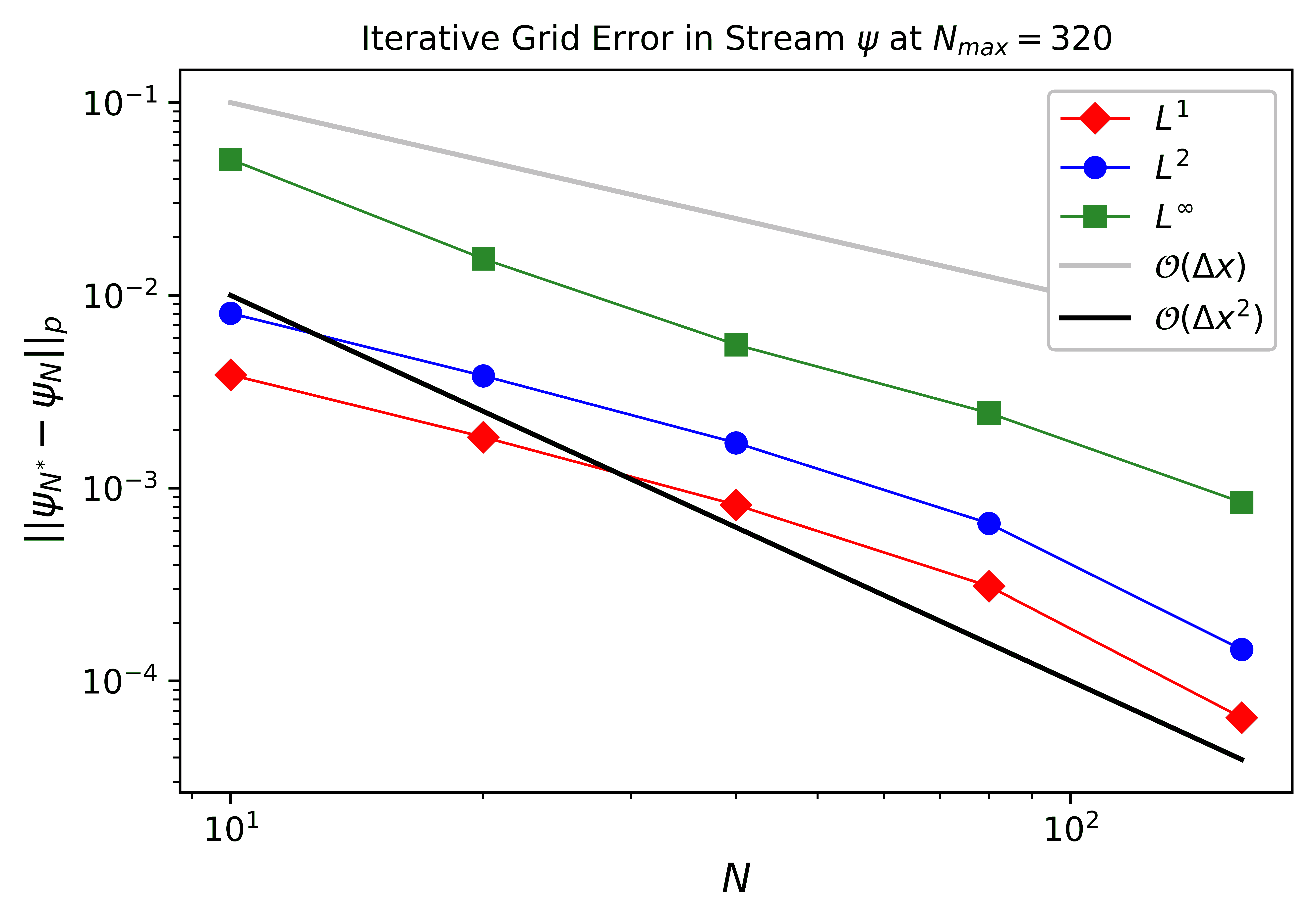}}
 \caption{Convergence of the numerical Stokes solution $\psi$ for the BFS and the triangular cavity. The BFS solution converges with order $\mathcal{O}(\Delta x^2)$ and the triangular cavity converges with order between $\mathcal{O}(\Delta x^2)$ and $\mathcal{O}(\Delta x)$}\label{convg}
\end{figure}

In the absence of an exact solution to the Stokes equation, we measure the grid convergence of the numerical solution by comparing to the solution on the finest grid resolution available.
We denote the grid scale $\mathcal{N} =\frac{1}{\Delta x} = \frac{1}{\Delta y}$. Convergence in the stream function $\psi$ for the BFS with $\mathcal{H}=2$ and for the triangular cavity with $H=4$ are shown in \cref{convg}. For the BFS, the convergence is order $\mathcal{O}(\Delta x^2)$. For the triangular cavity, the convergence in $\psi$ is between order $\mathcal{O}(\Delta x^2)$ and $\mathcal{O}(\Delta x)$. We expect a reduced order of convergence in the triangular cavity at $\mathcal{H}=4$ due to the boundary smoothing in the acute corners of the triangle.

\bibliography{ReynoldsStokes_bib}
\bibliographystyle{siamplain}
\end{document}